\documentclass[%
 reprint,
superscriptaddress,
 amsmath,amssymb,
 aps,physrev,
floatfix,
]{revtex4-1}

\usepackage{times}
\usepackage[pdftex]{graphicx}
\usepackage[bookmarksopen=true,bookmarksopenlevel=1]{hyperref}
\usepackage{epstopdf}
\usepackage{amsmath}
\usepackage{amsfonts}
\usepackage{amssymb}
\usepackage{graphicx}
\usepackage{sidecap}
\sidecaptionvpos{figure}{c}
\usepackage{color}
\usepackage{dcolumn}
\usepackage{bm}
\usepackage[compact]{titlesec}
\usepackage[normalem]{ulem}
\usepackage{bookmark}
\usepackage{natbib}
\usepackage{indentfirst}
\usepackage{appendix}
\usepackage{soul}

\usepackage{lineno}

\definecolor{mypink}{RGB}{219, 48, 122}
\definecolor{mygreen}{RGB}{51, 153, 102}
\definecolor{brown}{RGB}{165, 42, 42}

\newlength{\beforesection}
\newlength{\aftersection}
\newlength{\beforesubsection}
\newlength{\aftersubsection}
\titlespacing*{\section}{0pt}{\beforesection}{\aftersection}
\titlespacing*{\subsection}{0pt}{\beforesubsection}{\aftersubsection}

\def\root33{$\sqrt{3}\times\sqrt{3}$ {\it R}30$^\circ$}
\def\RT3{$\sqrt{3}$}

\begin{document}

\title{Dimensional band engineering in asymmetrical van der Waals heterostructures}

\author{Oliver J. Clark}
\email[Corresponding author. E-mail address: ] {oliver.clark@diamond.ac.uk}
\affiliation{School of Physics and Astronomy, Monash University, Clayton, VIC, Australia}

\author{Anugrah Azhar}
\affiliation{Department of Physics and Astronomy, University of Manchester, Oxford Road, Manchester M13 9PY, UK}
\affiliation{Physics Study Program, Faculty of Science and Technology, Syarif Hidayatullah State Islamic University Jakarta, Tangerang Selatan 15412, Indonesia}

\author{Ben A. Chambers}
\affiliation{Flinders Microscopy and Microanalysis, Flinders University, Bedford Park, South Australia 5042, Australia}

\author{Daniel McEwen}
\affiliation{School of Physics and Astronomy, Monash University, Clayton, VIC, Australia}
\affiliation{ARC Centre for Future Low Energy Electronics Technologies, Monash University, Clayton, VIC, Australia}

\author{Thi-Hai-Yen Vu}
\affiliation{School of Physics and Astronomy, Monash University, Clayton, VIC, Australia}

\author{M. Tofajjol H. Bhuiyan}
\affiliation{School of Physics and Astronomy, Monash University, Clayton, VIC, Australia}

\author{Rodion V. Belosludov}
\affiliation{Institute for Materials Research, Tohoku University, Sendai, 980-8577, Japan}

\author{Aaron Bostwick}
\affiliation{Advanced Light Source, Lawrence Berkeley National Laboratory, Berkeley, CA, 94720 USA}

\author{Chris Jozwiak}
\affiliation{Advanced Light Source, Lawrence Berkeley National Laboratory, Berkeley, CA, 94720 USA}

\author{Eli Rotenberg}
\affiliation{Advanced Light Source, Lawrence Berkeley National Laboratory, Berkeley, CA, 94720 USA}

\author{Seng Huat Lee}
\affiliation{Department of Physics, Pennsylvania State University, University Park, PA, 16802, USA}
\affiliation{2D Crystal Consortium, Materials Research Institute, Pennsylvania State University, University Park, PA,
16802, USA}

\author{Zhiqiang Mao}
\affiliation{Department of Physics, Pennsylvania State University, University Park, PA, 16802, USA}
\affiliation{2D Crystal Consortium, Materials Research Institute, Pennsylvania State University, University Park, PA,
16802, USA}

\author{Geetha Balakrishnan}
\affiliation{Department of Physics, University of Warwick, Coventry CV4 7AL, United Kingdom}

\author{Federico Mazzola}
\affiliation{Department of Molecular Sciences and Nanosystems,
Ca Foscari University of Venice, Venice IT-30172, Italy}
\affiliation{CNR-SPIN UOS Napoli, Complesso Universitario di
Monte Sant’Angelo, Via Cinthia 80126, Napoli, Italy}

\author{Sarah L. Harmer}
\affiliation{Institute for Nanoscale Science and Technology, Flinders University, Bedford Park, South Australia 5042, Australia}
\affiliation{Flinders Microscopy and Microanalysis, Flinders University, Bedford Park, South Australia 5042, Australia}

\author{Michael S. Fuhrer}
\affiliation{School of Physics and Astronomy, Monash University, Clayton, VIC, Australia}
\affiliation{ARC Centre for Future Low Energy Electronics Technologies, Monash University, Clayton, VIC, Australia}

\author{M. Saeed Bahramy}
\affiliation{Department of
Physics and Astronomy, University of Manchester, Oxford Road, Manchester M13 9PY, UK}

\author{Mark T. Edmonds}
\affiliation{School of Physics and Astronomy, Monash University, Clayton, VIC, Australia}
\affiliation{ARC Centre for Future Low Energy Electronics Technologies, Monash University, Clayton, VIC, Australia}
\affiliation{ANFF-VIC Technology Fellow, Melbourne Centre for Nanofabrication, Victorian Node of the Australian National Fabrication Facility, Clayton, VIC 3168, Australia}

\begin{abstract}
Van der Waals materials enable the construction of atomically sharp interfaces between compounds with distinct crystal and electronic properties. This is dramatically exploited in moir\'e systems, where a lattice mismatch or twist between monolayers generates an emergent in-plane periodicity, giving rise to electronic properties absent in the constituent materials. In contrast, vertical superlattices, formed by stacking dissimilar materials in the out-of-plane direction on the nanometer scale, have received far less attention despite their potential to realize analogous emergent phenomena in three dimensions. 
Through angle-resolved photoemission spectroscopy and density functional theory, we investigate six-to-eight-layer transition metal dichalcogenide (TMD) heterostructures constructed from pairs of stacked few-layer materials. {Counterintuitively, we find that even these single superlattice units can host fully-delocalised bands, evidencing a robust coherent interlayer coupling across lattice-mismatched interfaces over extended spatial scales.} We
show how uncompensated semimetallic phases and energetically-mismatched topological surface states are readily and exclusively stabilized within such asymmetrical architectures. These findings establish two-component  heterostructures in the intermediate layer-regime as platforms to invoke and control unprecedented combinations and instances of the diverse quantum phases native to many-layer TMDs.

\end{abstract}

 \maketitle

\section{\label{sec:intro}INTRODUCTION\protect\\ }

Transition metal dichalcogenides (TMDs) are popular components for functional devices due to the ease in which their vast array of intrinsic properties can be augmented~\cite{pham_2d_2022, chhowalla_chemistry_2013, ahn_2d_2020,  ciarrocchi_excitonic_2022, manzelli_2d_2017, regan_emerging_2022}. This is facilitated through anisotropic bonding, enabling both the isolation of TMD flakes of arbitrary thicknesses and their subsequent recombination into stacks with atomically sharp interfaces without the need for epitaxial registry~\cite{lisi_observation_2021, pei_observation_2022, nunn_arpes_2022, stansbury_visualising_2021, jones_visualising_2021}. These procedures can both alter the electronic properties inherent to the component compounds, and induce otherwise-inaccessible correlated phenomena through the fine tuning of the laboratory-defined relative orientations, or `twist angle'~\cite{devakul_magic_2021, guo_superconductivity_2024}. Despite van der Waals bonding, however, the interlayer hopping along the $c$-axis in TMDs is significant, driving orbital-dependent bandwidths of up to several electron volts in bulk systems~\cite{clark_fermiology_2018, clark_hidden_2022, riley_direct_2014, watson_novel_2024}. This is yet to be exploited in artificial heterostructures~\cite{wan_layered_2024}, with current research largely rooted to homo- and heterobilayer systems, and thus to TMD properties in the 2D limit. This precludes, for example, the fine-tuning of band gaps through band quantisation~\cite{splendiani_emerging_2010, mak_atomically_2010, sun_indirect_2016}, the induction of $k_z$-mediated topological phases~\cite{bahramy_ubiquitous_2018}, or the modification of charge density wave and magnetic transitions dependent on interlayer coupling~\cite{zhang_substantially_2024, watson_orbital_2019,feng_electronic_2018}. 

Here, through nano-focused angle-resolved photoemission (nano-ARPES) and first principles calculations, {we show how vertical 6-8 layer heterostructures, constructed simply by interfacing pairs of few-layer TMD flakes, possess emergent bands sensitive to the entire spatial extent of the hybrid system. Despite the absence of well-defined periodicity along the asymmetric $c$-axis, the fully delocalized bands mimic discretized three-dimensional states of ordered systems, but coexist with layer-locked states to produce an overall-spatially dependent electronic structure capable of supporting electronic phenomena usually found only in bulk compounds.}

We give special focus to $k_z$-mediated topological phase transitions, triggered by the interfacing of the few-layer TMD flakes, which produce pairs of distinct topologically non-trivial surfaces that are each defined largely by a single component material, unlike any naturally existing topological system. Similar phenomena should be expected for asymmetric heterostructures constructed from any pair of TMDs, and thus our findings unveil an enormous parameter space within {which 
customizable electronic frameworks, mixing  2- and 3D properties from across the TMD family, can be produced and fine-tuned.}

\section{\label{sec:methods}METHODS\protect\\ }

To construct the heterostructures characterised in this study, few-layer flakes of graphite, MoSe$_2$,  WSe$_2$ and NbSe$_2$ were exfoliated onto PDMS film and transferred  onto n-type Si substrates. After each stacking stage, the samples were rinsed with diisoproylamine, isopropanol and ethanol to remove contaminants, and annealed in a constant-flow Ar furnace at 200-250 degrees. {These procedures were performed in an Ar environment for samples containing NbSe$_2$, with a hot plate used in place of a furnace.}
The component flakes comprising the heterostructures {have} thicknesses determined through comparisons of photoluminescence spectra and optical images to those of known references. An additional, \textit{in situ} metric is provided by photoemission spectra obtained by probing the non-overlapping regions of each flake where available, {wherein the number of distinct quantized $k_z$-sub-bands within the $d_{z^2}$-manifold at  $\Gamma$ matches the number of monolayers along the c-axis~\cite{chang_thickness_2014,lefevre_two_2024}}. 

All ARPES data  were collected using either $p$-polarised or unpolarised  synchrotron light of energies between 60 and 144~eV at the MAESTRO beamline of the Advanced Light Source. Samples were cooled to cryogenic temperatures following a further 3+ hour UHV anneal at 270$^\circ$C. The base pressure of the setup was better than $\sim$1$\times$10$^{-10}$ mbar. All heterostructures were pre-characterised with NanoESCA III at FMMA (ROR: 04z91ja70)~\cite{FlindersNanoESCA} following a 3 hour UHV anneal at 270$^\circ$C, where an extraction voltage of 12kV and a HeI plasma discharge lamp was used for electronic structure measurements at room temperature. Electronic structure calculations were performed within Density Functional Theory using the Perdew-Burke-Ernzerhof exchange-correlation functional~\cite{pbe}, as implemented in the VASP programme package~\cite{Kresse_1996,Kresse_1999}. Due to the close similarity between the lattice parameters of 2H-WSe$_2$ and 2H-MoSe$_2$, the experimentally reported bulk structure of 2H-WSe$_2$~\cite{SCHUTTE1987207} used for both compounds in constructing (WSe$_2$)${_n}$/(MoSe$_2$)${_m}$ heterostructures. A supercell containing $n=3$ and $m=3$ or $5$ layers was employed, with a vacuum spacing of 15~\AA\ to prevent spurious interactions between periodic images. The plane-wave cut-off energy was set to 400 eV. The Brillouin zone was sampled using a $\mathbf{k}$-point mesh of $10 \times 10 \times 1$. Spin and orbital projections were obtained using Wannier functions~\cite{mostofi2008,Souza2001}, downfolded from the DFT Hamiltonian, with the valence $d$ orbitals of the transition metal atoms and the Se $4p$ orbitals chosen as projection centres.

\section{\label{sec:results}RESULTS\protect }

\subsection{\label{sec:f1}Theoretical Framework}

\begin{figure*}
	\centering
	\includegraphics[width=\textwidth]{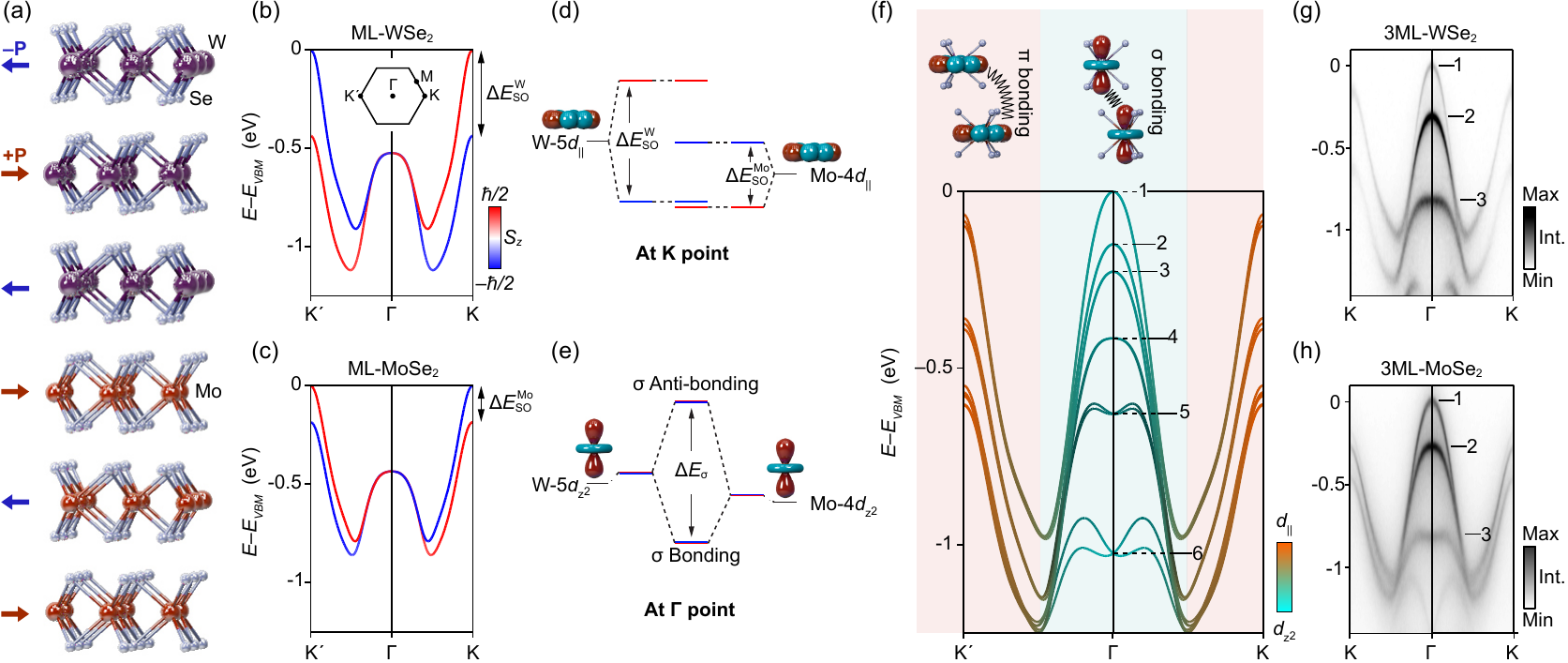}
	\caption{\label{Fig1} \text{Chemical nature of an anisotropic transition metal heterostructure.} (a) Schematic representation of a (WSe$_2$)$_3$/(MoSe$_2$)$_3$ heterostructure. Due to the $D_{3h}$ symmetry of each monolayer, a net dipole moment $\pm \mathbf{P}$ is induced, oriented perpendicular to the stacking axis, and oppositely aligned with respect to the adjacent layer. (b) and (c) Out-of-plane spin-projected electronic band structures for monolayers of WSe$_2$ and MoSe$_2$, respectively. The parameter $\Delta E_\text{SO}^\text{W/Mo}$ denotes the valley-dependent spin-orbit splitting at the K-point of the Brillouin zone.  (d) and (e) Schematic energy diagrams illustrating the bonding and anti-bonding interactions at the K and $\Gamma$ points, respectively, with their orbital character predominantly influenced by the in-plane $d_{||} =\{ {d_{xy}, d_{x^2 - y^2}}\}$ and out-of-plane $d_{z^2}$ orbitals of W/Mo. (f) Orbital-projected electronic structure of the (WSe$_2$)$_3$/(MoSe$_2$)$_3$ heterostructure, highlighting the distinct contributions from the constituent layers. The bands at $\Gamma$ are numbered from low energy to high. (g) and (h) Photoemission spectra collected from isolated 3ML thick flakes of WSe$_2$ and MoSe$_2$ (using 108 and 64~eV photons, respectively, and symmetrized about $\Gamma$). The bands at $\Gamma$ are numbered from low energy to high.} 
\end{figure*}

Figure~\ref{Fig1} overviews the mechanism underpinning our findings. Fig.\ref{Fig1}(a) illustrates the lattice structure of a 6-layer MoSe$_2$-WSe$_2$ heterostructure, composed of three WSe$_2$ layers stacked atop three MoSe$_2$ layers along the crystalline $c$-axis. Each monolayer possesses $D_{3h}$ symmetry, with an in-plane polarisation $\bf{P}$ arising from M–Se dipoles (where M = Mo or W). The 2H stacking sequence results in a {$\pi$} rotation of adjacent layers, aligning their polar axes anti-parallel. In this heterostructure, the polarisation $\bf{P}$ of MoSe$_2$ layers slightly exceeds that of WSe$_2$ layers, reflecting the greater electronegativity difference between Mo and Se compared to W and Se. However, the stronger atomic spin-orbit interaction (SOI) of W dominates the spin physics of the system. Coupled with $\bf{P}$, this SOI generates an out-of-plane spin-orbit field, $B_\text{SO} \propto (\bf{P} \times \bf{k})_z$, which induces a Zeeman-like spin splitting of energy bands. The effect is most pronounced at the K points of the Brillouin zone, as shown in Fig.~\ref{Fig1}(b) and (c), where the spin splitting in WSe$_2$ ($\Delta E^\text{W}_\text{SO}$) is more than twice that in MoSe$_2$ ($\Delta E^\text{Mo}_\text{SO}$). 

At the interface, the mismatch in spin-orbit coupling strengths gives rise to a hierarchical valley-spin polarisation. The MoSe$_2$ valence bands are entirely contained within the $\Delta E^\text{W}_\text{SO}$ gap of WSe$_2$, as shown in Fig.~\ref{Fig1}(d). This behaviour reflects the in-plane orbital nature of the K-point states, which are dominated by $d_\parallel = {d_{xy}, d_{x^2-y^2}}$ orbitals. These orbitals exhibit negligible out-of-plane coupling, suppressing interlayer hybridisation and preserving the monolayer-like electronic structure at K. By contrast, states at the $\Gamma$ point are dominated by $d_{z^2}$ orbitals, which extend into the van der Waals gap and exhibit substantial interlayer overlap. This leads to strong hybridisation at $\Gamma$, forming a broad bonding–anti-bonding energy window across the heterostructure, as depicted in Fig.\ref{Fig1}(e). While time-reversal symmetry suppresses spin splitting at $\Gamma$, the resulting band structure at this point is highly sensitive to interlayer coupling, in contrast to the robust monolayer-like dispersion retained at K. This dichotomy is evident in the full low-energy band structure of the 6-layer system (Fig.\ref{Fig1}(f)), where the K valleys retain their original characteristics while the $\Gamma$ valley forms a cascade of hybridised bands.

\begin{figure*}
	\centering
	\includegraphics[width=\textwidth]{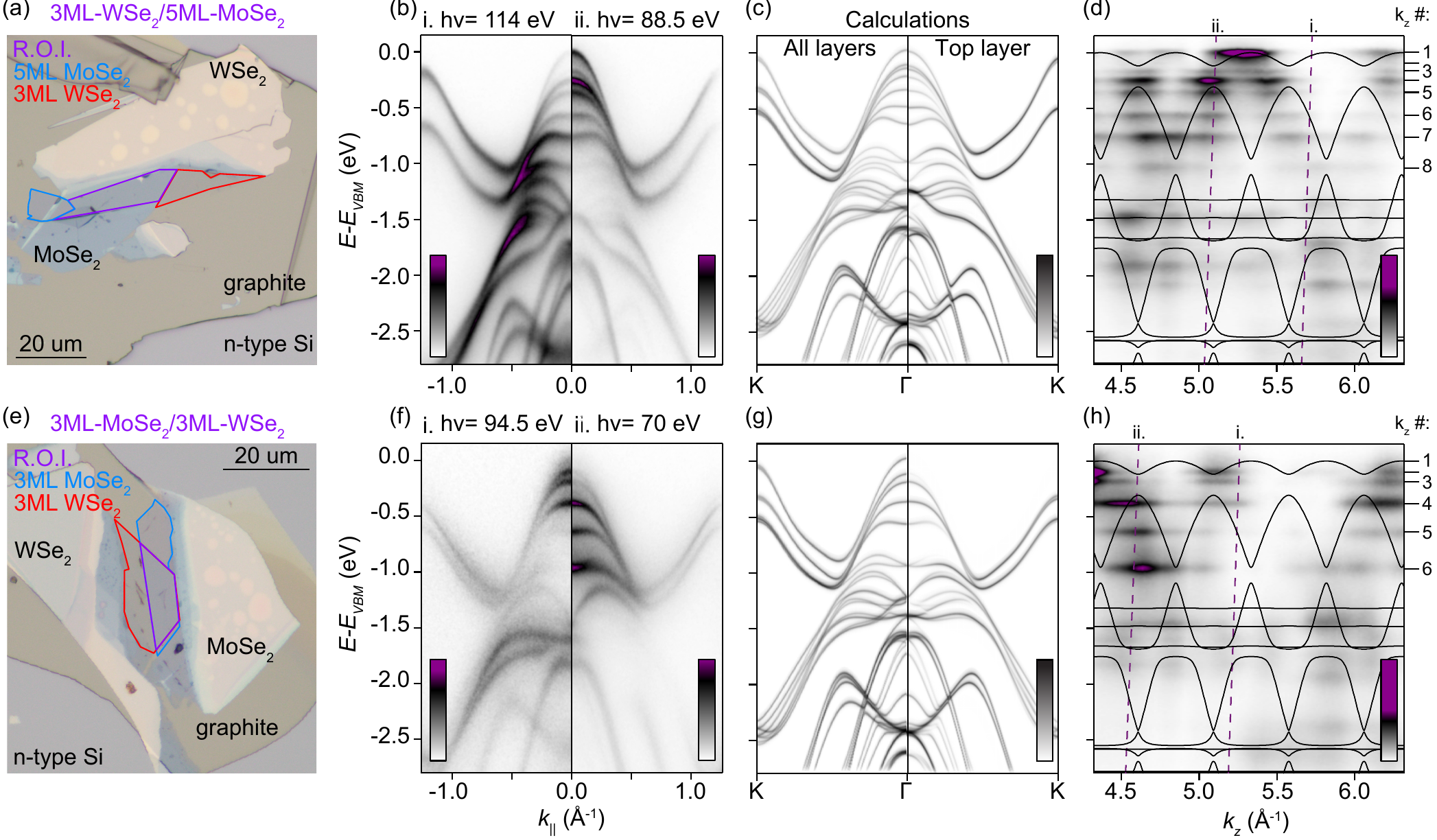}
	\caption{\label{Fig2} \text{Multi-layer asymmetric heterostructures constructed from MoSe$_2$ and WSe$_2$ flakes.} (a-d) Overview of a (WSe$_2$)$_3$/(MoSe$_2$)$_5$ heterostructure. (a) Optical image of the heterostructure region. The few-monolayer thick top and bottom layers are highlighted in red and blue respectively. The overlap region of interest (ROI) is highlighted in purple. (b) Band dispersions of the ROI along  axes close to the  K-$\Gamma$-K direction 
    for two stated photon energies. 
    (c) Density functional theory based slab calculations, projected through the full heterostructure (left) and to the top layer (right) along the K-$\Gamma$-K direction. (d) $k_z$ dispersion (h$v$=60-144~eV) at $k_{\parallel}=0$.  Dashed lines indicate the 
    slices in (b), and solid lines are a tight-binding model (TBM) of a bulk (WSe$_2$/MoSe$_2$)$_n$ system. {An inner potential of 16~eV is chosen to best match the periodicity of the calculated band structure.} $k_z$ sub-bands of the $d_{z^2}$ valence band maxima are numbered from low binding energy to high. {(e-h) Equivalent datasets for a (MoSe$_2$)$_3$/(WSe$_2$)$_3$ heterostructure.}}
\end{figure*}

This disparity highlights the importance of $\Gamma$-derived orbitals in band engineering, where artificial heterostructures can exploit interlayer hybridisation to tune the out-of-plane ($\parallel k_z$) dispersion by controlling layer composition, ordering, and alignment. Indeed, in the above simple case of a 6-layer MoSe$_2$-WSe$_2$ heterostructure,
though discretized, the bonding-antibonding formation of states at and around the $\Gamma$ point closely approximate a $k_z$-integrated bulk band structure of a bulk 2H-TMD system. This implies that, with only a few layers, the bulk band dispersion can be significantly restored and even controllably altered to exhibit a mixed characteristic between two or more materials. To emphasize the significance of this, in Fig.~\ref{Fig1}(g-h), we compare the calculated electronic structure of the interfaced system to experimental band structures of individual free-standing 3ML WSe$_2$ and MoSe$_2$ flakes. In each case, the number of $d_{z^2}$-derived bands at $\Gamma${, which differ only by their discrete $k_z$ quantum number,} is tripled relative to the respective monolayers as the three dimensional, $k_z$ dispersive, band continuum of the bulk limit begins to take form. The theoretical prediction in Fig.~\ref{Fig1}(f) demonstrates that, by simply combining these 3ML components, the evolution towards a 3D band continuum is entirely unhindered despite the interface between dissimilar compounds, and electrons at $\Gamma$ become fully delocalised through all six layers despite the enhanced spatial scales.

More obvious approaches to bulk-like band engineering would rely on the assembly of numerous monolayers with fixed relative orientations, stacked sequentially and periodically to replicate bulk periodicity in momentum space through the curation of a well-defined unit cell, similar to that of  the naturally occurring 4Hb polymorph of TaSe$_2$ with alternating 1T/1H layers~\cite{ribak_chiral_2020}. By bypassing this labor-intensive approach to instead employ much simpler, two-component heterostructures in the intermediate layer regime, the same essential three-dimensional electronic behavior can be accessed within an architecture that lends itself more naturally to fine tuning with e.g. layer thickness or twist angles. More generally, access to an extra dimension within which electronic properties can be tailored and controlled is of great relevance for nano-quantum device applications, where miniaturisation plays a major role.

{\subsection{\label{sec:f2}Delocalised band formation in two-component stacks }}

To demonstrate how this concept of achieving bulk-like band characteristics with mixed material properties can be realised in real systems, Figure~\ref{Fig2} overviews the electronic structure of a (WSe$_2$)$_3$/(MoSe$_2$)$_5$ (Fig.~\ref{Fig2}(a)) and a (MoSe$_2$)$_3$/(WSe$_2$)$_3$ (Fig.~\ref{Fig2}(e)) heterostructure constructed by interfacing two few-layer TMD flakes on graphite/n-type silicon substrates (see Methods). This pair of heterostructures thus closely approximate opposite surfaces of a single MoSe$_2$-WSe$_2$ system, {and the modest total layer numbers enable unambiguous comparisons of stacked regions relative to the two component flakes.} Photoemission spectra acquired from the stacked heterostructure regions are displayed in Fig.~\ref{Fig2}(b, f). 

As exemplified in Fig.~\ref{Fig1}(g-h), there is a one-to-one correspondence between the number of discrete sub-bands within the $d_{z^2}$-manifold at the $\Gamma$ point of 2H-structured TMDs and the number of monolayers within the system~\cite{chang_thickness_2014, lefevre_two_2024}. {These bands collectively describe a single three-dimensional $d_{z^2}$-derived band which is quantised in $k_z$ due to the finite size of the system along the $c-$axis~\cite{chiang_photoemission_2000, milun_quantum_2002, lefevre_two_2024}.} {Therefore}, as basic band theory stipulates that periodicity in momentum space cannot be achieved without periodicity in real space, the number of discrete $k_z$ sub-bands visible within the {overlapped} regions would not be expected to reflect the full extent of the heterostructures which lack a well-defined unit cell. Yet, the photoemission spectra in Fig.~\ref{Fig2}(b,f) demonstrate that the number of $k_z$ sub-bands is not only increased relative to equivalent datasets acquired from the constituent flakes (shown in full in Supplemental Fig. 1), but also matches the total layer number in real space (i.e. eight for (b) and six for (f)) of the entire heterostructure. The  electron mean free path of less than 0.7~nm, applicable to our photoemission experiments, precludes that the experimental result originates from a superposition of spectra from weakly-interacting top and bottom TMD flakes which each have thicknesses of $\sim$ 0.7~nm per monolayer~\cite{yang_ramen_2025}. This alternate scenario is further discarded through constant-energy $k_x$-$k_y$ contour mapping in Supplemental Figure 2. 
Instead, the six or eight distinct $d_{z^2}$ sub-bands at $\Gamma$ are delocalised throughout the structure, and are therefore accessible to a surface sensitive probe. 

While counterintuitive, these findings are entirely consistent with the discussion surrounding Figure~\ref{Fig1}: The array of hybridized bands at $\Gamma$ belonging to the collective band structure of the finite six or eight layer crystal structure closely mimics the discretized $k_z$-projected band manifold of ordered systems. For bands derived from orbitals with significant interlayer hopping potential, the $k_z$ dispersion is therefore effectively constructed without the need for a periodic repeating unit. To further test this description, the experimental band dispersions of both WSe$_2$-MoSe$_2$ heterostructures are compared to layer-projected, density functional theory (DFT)-based slab calculations in Fig.~\ref{Fig2}(c,g), (see Methods). 
{The electronic structure around the $\Gamma$ point is extremely well reproduced in each case, verifying the present interpretation while also demonstrating a negligible role of collective electron phases and  large twist angles (the latter chosen experimentally in order to decouple  Moiré effects, see Sup. Fig. 3), which are both neglected  by these calculations.}

To further scrutinize the observed three-dimensional behavior, in Figure~\ref{Fig2}(d,h) we track the evolution of the delocalised states at $\Gamma$ with changing photon energy, thus mapping their behavior with changing $k_z$~\cite{Dam2004}. 
These spectra can be compared to equivalent datasets for a three-layer MoSe$_2$ and WSe$_2$ flake (Sup. Fig.~1 (f,i)). For both the individual TMD flakes and the artificial heterostructures, each $d_{z^2}$-subband is two-dimensional (i.e. non-dispersive along $k_z$) in-line with their assignment as dimensionally confined states, though {each exhibits a photon energy-dependent spectral weight.} In the the eight- or six-layer hybrid systems (Fig.~\ref{Fig2}(d,h)), it is apparent that {these spectral weight variations are periodic in $k_z$, with each sub-band illuminated sequentially. This is consistent with their collective assignment as a singular $d_{z^2}$-orbital derived band, which remains quantized in $k_z$ due to the finite length scales of the total hybrid systems along the asymmetric $c-$axis. Though discrete, this periodic behavior is exactly as would be expected for three-dimensional bands in a bulk 2H-structured TMD system. Indeed, the overlaid solid lines, which well capture the observed periodicity, are a tight-binding model for an idealized bulk structure with alternating WSe$_2$ and MoSe$_2$ layers along the $c$-axis. We note that, due to significant $k_z$ broadening in photoemission generally~\cite{Dam2004}, the majority of these sub bands can be captured simultaneously when probed with a single photon energy, as evidenced by the in-plane band dispersions shown in Fig.~\ref{Fig2}(b,f). }

{In Supplemental Figure 4, we show how equivalent phenomena is observed for mismatched combinations of WSe$_2$ and MoSe$_2$ bilayers: the number of $k_z$ sub-bands within the $d_{z^2}$-manifold reflects the full $c$-axis extent of the system despite ill-defined unit cells, again demonstrating how the periodic $k_z$-dispersions expected for ordered solids are emulated when interfacing two-or-more TMD flakes in the intermediate layer regime. We stress that, while the observed three-dimensional band dispersions throughout this Section exhibit finite-size effects, the key properties of bulk TMDs are, in general, present and fully-developed in their six (or fewer)-monolayer variants despite $k_z$ quantisation~\cite{kim_thickness_2021,hlevyack2021,deng2019,lan_2d_2021,fridman_anomalous_2025,chang_thickness_2014,lefevre_two_2024}.
Nevertheless, we expect that the underpinning mechanism driving delocalised states along $k_z$ to hold for pairings of TMD flakes with larger total thicknesses, thus producing artificial 3D band properties closer to the true bulk limit. }

\subsection{\label{sec:f3}Orbital-dependent band localisation}

\begin{figure*}
	\centering
	\includegraphics[width=0.8\textwidth]{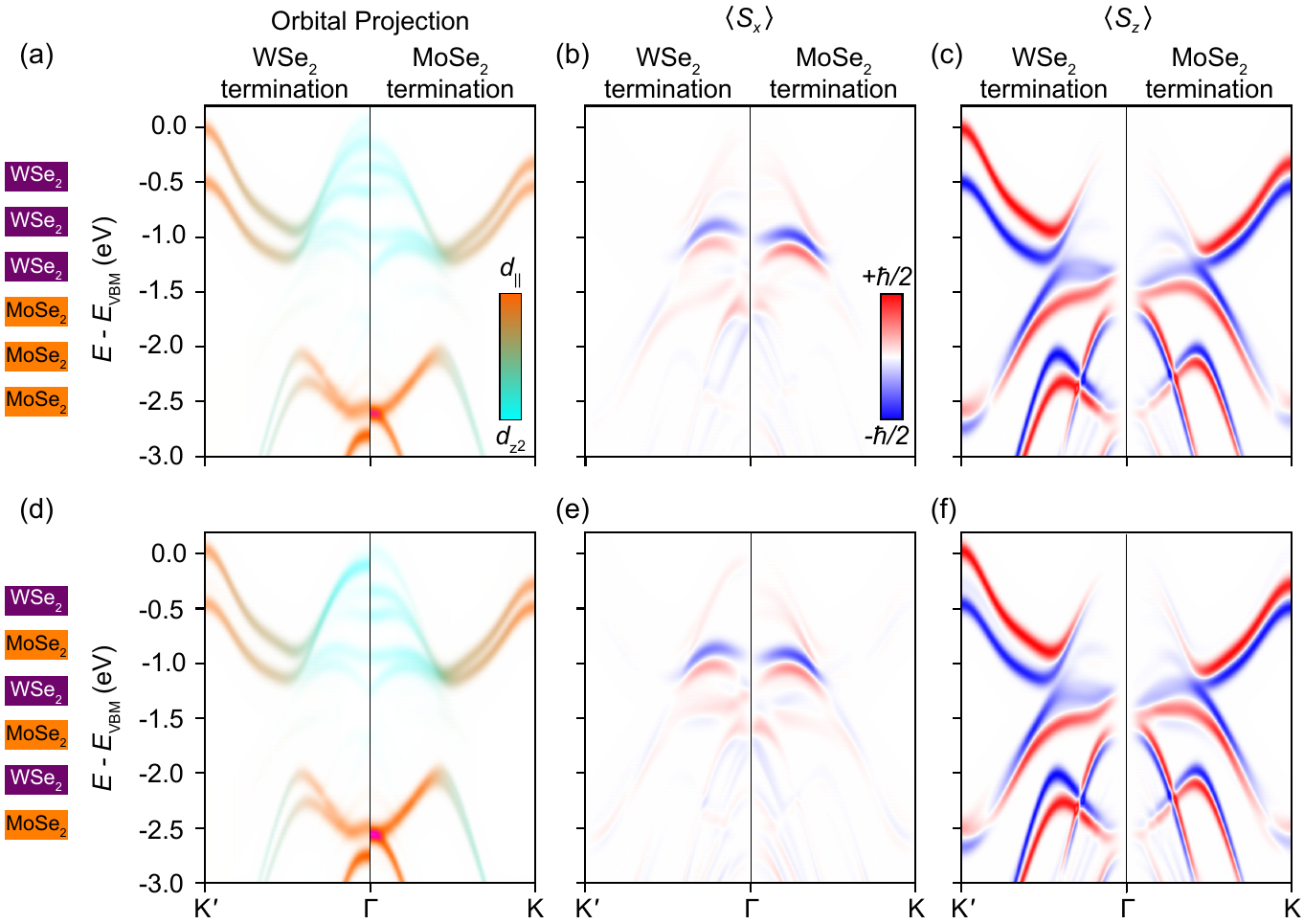}
	\caption{\label{Fig3} \text{Spin-orbital variation in the (MoSe$_2$)$_3$/(WSe$_2$)$_3$ heterostructure.} (a–c) Orbital- and spin-projected electronic structures calculated for both WSe$_2$- and MoSe$_2$-terminated configurations of the asymmetric (MoSe$_2$)$_3$/(WSe$_2$)$_3$ heterostructure. (d–f) Corresponding calculations for an idealised (MoSe$_2$)$_3$/(WSe$_2$)$_3$ heterostructure, where WSe$_2$ and MoSe$_2$ layers are alternately stacked along the $c$-axis.  
} 
\end{figure*}

While the presence of delocalised electron behavior within the $d_{z^2}$ orbital manifolds of the finite heterostructures is clear, the same is not true for regions of the electronic structure derived from in-plane orbitals. This stark contrast in orbital-dependent coupling is a hallmark of 2H-type transition metal dichalcogenides generally. The details of the $d_\parallel$ K-point valence bands are largely equivalent in monolayer and bulk 2H-TMD systems due to their minimal out-of-plane hopping potential~\cite{riley_negative_2015, mak_atomically_2010, xaio_coupled_2012}, whereas $d_{z^2}$-derived $\Gamma$-point states exhibit substantial out-of-plane coupling to drive significant $k_z$ dispersion in bulk materials ~\cite{riley_direct_2014, latzke_electronic_2015, kim_determination_2016, alarab_k_2023, lefevre_two_2024, chang_thickness_2014}, thus underpinning the indirect-to-direct bandgap transition observed in few-layer systems~\cite{mak_atomically_2010, sun_indirect_2016, ernandes_indirect_2021}. 
This natural orbital disparity interplays with the inherent asymmetry of heterostructures constructed by interfacing two dissimilar TMD flakes, ensuring a spatial dependence to the broader electronic structure along the $c$-axis between limits defined by the distinct component materials. The scale of this dependence is therefore dependent on orbital type.  

In Figure~\ref{Fig3} and Supplemental Figs. 5-6, this real space evolution is explored in full. In Fig.~\ref{Fig3}(a), the orbitally-resolved WSe$_2$ and MoSe$_2$ terminations of the experimentally realized (MoSe$_2$)$_3$/(WSe$_2$)$_3$ heterostructure are compared. The valence band maxima at K and $\Gamma$, derived exclusively from $d_{\parallel}$ and $d_{z^2}$-orbitals respectively, represent the two extremes of $c$-axis delocalisation: The details of the spin-orbit splitting at K is determined only by the local crystal structure, while the band structure at $\Gamma$ is equivalent between terminations and indeed between the top and bottom layers (shown explicitly in Supplemental Figure 5). The rest of the band structure behaves similarly, with the real-space electronic structure dependence directly determined by the $d_{\parallel}$/ $d_{z^2}$ orbital ratio, further influenced by  hybridization between bands of differing orbital origins. {In artificial, asymmetric heterostructures, this therefore produces an overall layer-dependent electronic structure despite the presence of fully delocalised band features. This is in stark contrast to naturally occurring TMDs, wherein the overall symmetry of the system precludes real space band structure variations despite a similar orbital-dependent localisation.} 

The asymmetry of our artificial heterostructures can also be expected to modify the spin structure relative to natural TMD systems. As discussed earlier, in a natural 2H-TMD, each constituent TMD layer, owing to its $D_{3h}$ symmetry, can only exhibit an out-of-plane spin-orbit field, whose direction is opposite to that of the neighbouring layers. Consequently, this implies that in-plane spin textures, including chiral spin-momentum locking types arising from the Rashba effect~\cite{Rashba84}, are effectively suppressed. Furthermore, the hidden out-of-plane spin textures of each layer can only become accessible at the surface, where translational symmetry breaking disrupts the natural cancellation of $B_\text{SO}$ along the stacking axis. 

In contrast, our artificial structures can naturally accommodate both in-plane Rashba-type and out-of-plane Zeeman-type spin textures due to their unique heterostructural properties. Specifically, the chiral Rashba spin polarisation can emerge naturally from the potential gradient formed at the MoSe$_2$/WSe$_2$ interface and propagate throughout the entire structure via $d_{z^2}$ and other out-of-plane orbitals. To elucidate the implications of this, in Fig.~\ref{Fig3}(b-f), we compare a (MoSe$_2$)$_3$/(WSe$_2$)$_3$ heterostructure to an idealised six-layer (MoSe$_2$/WSe$_2$)$_3$ system. The calculated layer-projected spin polarisation for the chiral ($x$) and out-of-plane ($z$) components is shown for both the asymmetric and idealised cases. As can be observed, in both configurations, strikingly similar patterns of in-plane and out-of-plane spin polarisation emerge, which remain valley-specific. Particularly, in both cases, the chiral ($x$) component is significantly enhanced around the $\Gamma$ point, where the $d_{z^2}$ contribution is dominant. As with the discussions surrounding the $k_z$ dispersions on show in Fig.~\ref{Fig2}, this again demonstrates how the key physics of ideally-stacked TMD heterostructures are very closely approximated by our simpler but asymmetric structures.

{\subsection{\label{sec:f4}Creation of layer-dependent semimetallic phases}}

\begin{figure*}
	\centering
	\includegraphics[width=\textwidth]{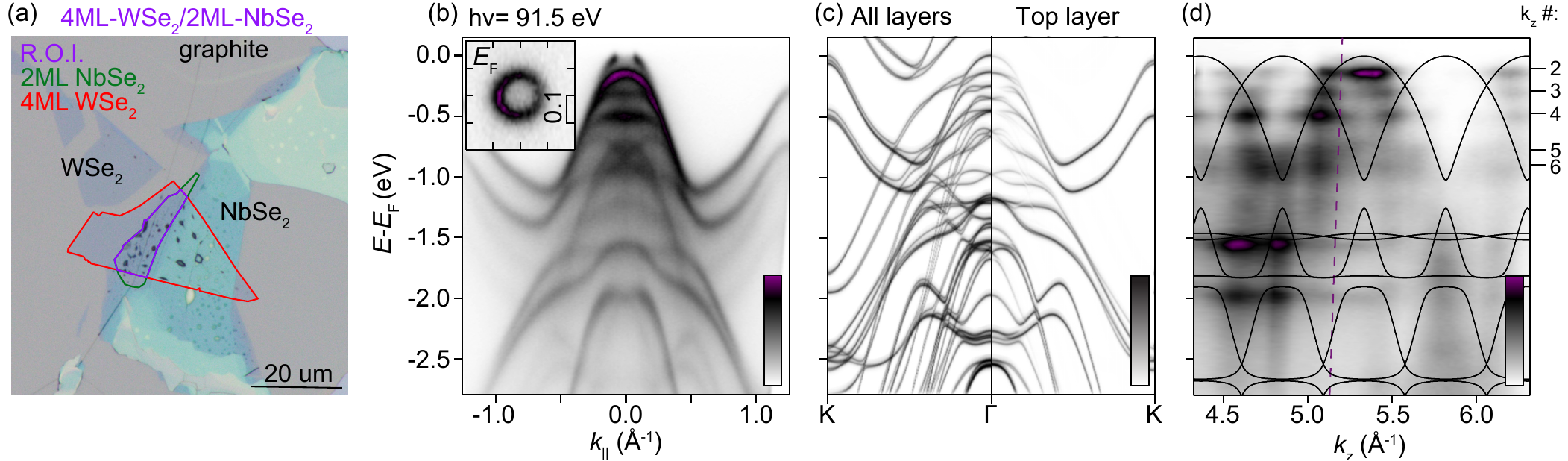}
	\caption{\label{Fig4} \text{Local semimetallic fermiology in an asymmetric WSe$_2$-NbSe$_2$ heterostructure} (a) Optical image of the heterostructure region. The few-monolayer thick top and bottom layers are highlighted in red and green respectively. The overlap region of interest (ROI) is highlighted in purple. (b) Band dispersions of the ROI along  axes close to the  K-$\Gamma$-K direction. The inset is a $k_x$-$k_y$ contour at the Fermi level, taken with 90~eV photons. 
    (c) Density functional theory based slab calculations, projected through the full heterostructure (left) and to the top layer (right) along the K-$\Gamma$-K direction. (d) $k_z$ dispersion (h$v$=60-144~eV) at $k_{\parallel}=0$. Dashed lines indicate the 
    slices in (b), and solid lines are a tight-binding model (TBM) of  bulk WSe$_2$ offset in energy to match the experimental scenario. {An inner potential of 16~eV is chosen to optimize the match to the periodicity of the calculated band structure.} $k_z$ sub-bands of the $d_{z^2}$ valence band maxima are numbered from low binding energy to high. We note that the spectrum in (b) is partially superimposed with that from pristine 4L-WSe$_2$, most likely due to the less-than-pristine overlap region. We note that, due to air sensitivity of air-exposed NbSe$_2$, atomic force microscopy (AFM) was employed to determine the thickness of the NbSe$_2$ bottom layer instead of \textit{in situ} photoemission. {The step height from the WSe$_2$ regions (red outline in (a)) to the overlap region (purple) was found to be 
    $\sim$2.25~nm consistent with 2 ML NbSe$_2$~\cite{wang_high_2017}}}
\end{figure*}

The asymmetric WSe$_2$-MoSe$_2$ heterostructures discussed thus far exhibit dissimilar top and bottom surfaces due to orbital-dependent band localisation. While the global symmetries of the asymmetric unit are fundamentally altered relative to naturally occurring TMDs or to an idealized stack, its layer-dependent electronic structure is derivative of the bulk forms of the component materials. For the specific case considered thus far, the similarities in the WSe$_2$ and MoSe$_2$ lattice constants~\cite{kang_band_2013} 
and in the energetics of the relevant W 5$d$ and Mo 4$d$ orbital manifolds precludes significant deviations of the formed hybrid electronic structure from those of the parent compounds. We stress however, that such a similarity in material components is not a necessary criterion to stabilize dimensionally-mixed frameworks: The mechanism underpinning our findings should apply generally to pairs of 2D materials with non-negligible out-of-plane hopping, even when the two components differ substantially. Such a mismatching could be used to engineer entirely unique quantum phases.

To exemplify the power of this eventuality, in Fig.~\ref{Fig4}, we showcase 
a (WSe$_2$)$_4$/(NbSe$_2$)$_2$ heterostructure. NbSe$_2$ is metallic, whereas WSe$_2$ is semiconducting. Consequently, the energy bands in WSe$_2$ are effectively lower than those in NbSe$_2$ by the energy gained from filling the valence bands with an additional electron, transferred from the W$^{+4}$ ion compared to the Nb$^{+3}$ ion. Despite this, not only do the number of $k_z$ subbands at $\Gamma$ continue to reflect the total layer number of the heterostructure thereby demonstrating the presence of fully delocalised bands, but the system becomes semimetallic, with the shallowest binding energy sub-band at the $\Gamma$ point forming an Fermi surface, shown inset in {Fig.~\ref{Fig4}b} and in Supplemental Fig. 7. This Fermi pocket at $\Gamma$ is the only Fermi band crossing on the top layers of the artificial heterostructure, unlike the electronic structure of any naturally occurring bulk or few-layer TMD. {In Fig.~\ref{Fig4}(c), the experimental picture is compared to DFT calculations. While, as with the previous heterostructures, there is strong quantitative agreement, we note the presence of additional fine structure to the bands  near the Fermi level at $\Gamma$ relative to the equivalent bands in the previous heterostructures. We attribute this to  hybridisation effects between the W $d_z^2$ manifold and the Se $p_z$ manifold from NbSe$_2$, which overlap in this hybrid heterostructure. Nevertheless, as verified by the layer-resolved calculations in Sup. Fig. 8, all sub-bands at $\Gamma$ are again present and at the same energy in each layer of the heterostructure, evidencing the presence of fully delocalised electrons along the $c$-axis and hence the formation {of a discretized three-dimensional band manifold. Indeed, the $k_z$ periodicity in the spectral weight of the experimentally observed sub-bands, shown in Fig.~\ref{Fig4}(d), is again well matched to the periodicity of bulk TMD calculations. }

The observed $c$-axis delocalisation, persisting even for this energetically-mismatched scenario, is facilitated by the lowering of the total energy of the system by partially lifting the constraints of 2D quantization. Along the $c$-axis, electrons form standing waves due to the finite thickness of the artificial crystals. This effect can be expected to vanish as one or both flakes approach the infinite-layer limit and electrons at the interface gain access to a bulk reservoir, ultimately preserving the quantized electronic structure of the remaining few-layer component. This is tested and shown explicitly in Supplemental Figure 9 through the characterization of a second overlap region within the sample in Fig.~\ref{Fig4}(a); between a 4ML-WSe$_2$ flake and a many-layer NbSe$_2$ crystal. While charge transfer from the semiconducting top-layer still occurs in addition to some normalization of band features relative to the isolated 4ML-WSe$_2$ top-layer, there are no additional well-defined $d_{z^2}$-subbands.

{\subsection{\label{sec:f5}$k_z$-mediated topological phases}}

\begin{figure*}
	\centering
	\includegraphics[width=\textwidth]{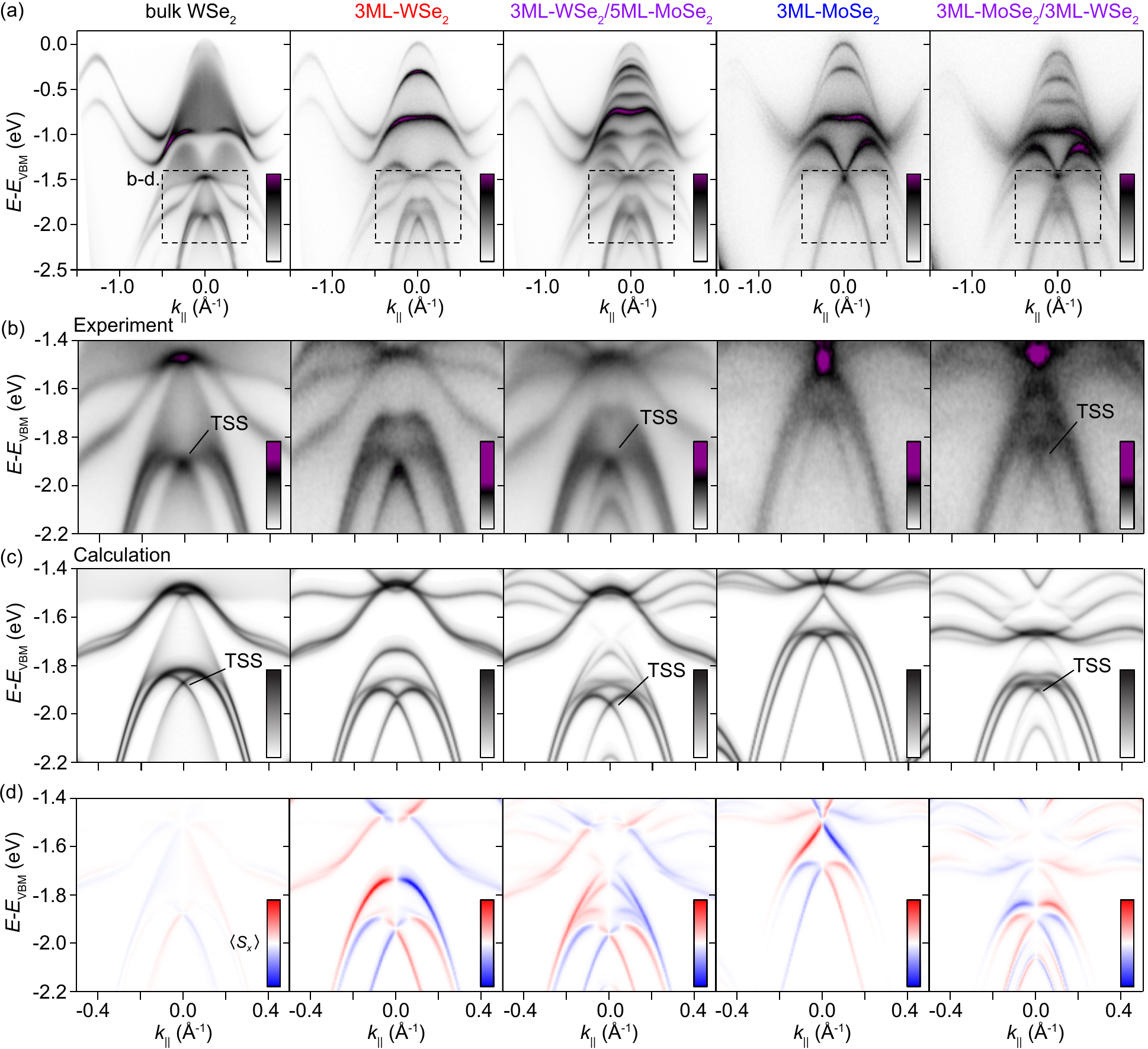}
	\caption{\label{Fig5} \text{$k_z$-mediated topological phases in asymmetric heterostructures} (a) Band dispersions along an axis close to K-$\Gamma$-K taken with 81~eV photons for (from left to right) bulk WSe$_2$, 3ML-WSe$_2$, a (WSe$_2$)$_3$/(MoSe$_2$)$_5$ heterostructure, 3ML MoSe$_2$ and a (MoSe$_2$)$_3$/(WSe$_2$)$_3$ heterostructure. Dashed rectangles show the energy-momentum region covered by the plots (b-d.)(b) Zoom ins of the spectra in (a) with enhanced colour contrast. (c) Top-layer projected DFT calculations for the equivalent regions (d) Equivalent plots with resolved chiral ($\langle S_x \rangle$) spin component. Fully-formed (i.e. ungapped) topological surface states (TSSs) are indicated in (b) and (c).} 
\end{figure*}

Though orbital-dependent, the presence of {delocalised} electronic behavior in asymmetric artificial heterostructures prompts the question of whether additional surface-localised states are produced by the truncation of  
global crystalline symmetries. While surface states, both topologically trivial and non-trivial, are commonplace in naturally occurring bulk TMDs, the disparate surfaces unique to asymmetric heterostructures would result in pairs of surface states with mismatched, surface-specific, properties.

To explore this potential, in Fig.~\ref{Fig5} we search for surface states enforced by topological phase transitions driven by {out-of-plane delocalised electron behavior.~\cite{bahramy_ubiquitous_2018,clark_fermiology_2018,ClarkGeneral2019}}. The families of three-fold symmetric 1T and 2H-structured TMDs possess a series of band inversions along their $k_z$ axes ($\parallel$ $\Gamma$-A)~\cite{huang_type_2016, yan_topological_2017, noh_experimental_2017, li_topological_2017, bahramy_ubiquitous_2018, clark_fermiology_2018, ClarkGeneral2019, ghosh_observation_2019, jiang_comprehensive_2020, mukherjee_fermi_2020, nurmamat_bulk_2021, clark_hidden_2022, chakraborty_observation_2023, kumar_electronic_2024}. These inversions, entirely contained to the chalcogen $p$-orbital manifold, generate up to five Dirac crossings (both bulk Dirac points and topological surface states (TSSs)) energetically stacked at the surface Brillouin centre. While the underpinning mechanism is equivalent for the dozen TMDs explicitly shown to host these states to date, the ladder of topological band crossings produced is delicately dependent on material-specific parameters, and thus are fingerprints of each individual system. The minimal criteria to generate topological ladders in TMDs are only that there exists a discrepancy in the $k_z$-bandwidths of in- and out-of-plane chalcogen $p$-orbital-derived states, and that $C_3$ symmetry is intact to maintain bulk Dirac crossings (though not crucial for the properties of surface Dirac states). The presence of bands well approximating a (discretized) $k_z$ dispersion along with orbital hopping asymmetries (discussed thus far for $d$-orbitals) imply that the above conditions are met in the experimental heterostructures considered here. {Previous ARPES studies have shown that 4-5 monolayers is sufficient to  stabilize the topologically non-trivial properties of  bulk TMD systems in full~\cite{hlevyack2021,deng2019}. Such physics is therefore relevant to the intermediate-layer regime considered here.}

Figure~\ref{Fig5} overviews the realisation of non-trivial band topology in our proof-of-principle MoSe$_2$-WSe$_2$ heterostructures. In {bulk} 2H-TMDs like WSe$_2$, there is one experimentally resolvable TSS centred $\sim$ 1.9~eV below the VBM~\cite{bahramy_ubiquitous_2018}, as observed in spectra from the bulk region of the flake used for device construction (Fig.~
\ref{Fig5}(a)). This is compared to band dispersions from the pair of WSe$_2$-MoSe$_2$ heterostructures and their respective 3ML top layers. Experimental zoom-ins  of the region wherein the bulk band inversion occurs and top-layer projected DFT calculations are provided in Fig.~\ref{Fig5}(b) and (c), respectively. While quasi-linear band dispersions are present in all cases, both 3ML TMD flakes undergo significant band renormalization when stacked atop a dissimilar flake. In 3ML-WSe$_2$, both experiment and theory demonstrate a small local SOC-mediated gap between a pair of bands at the binding energy where the Dirac crossing occurs in the bulk limit. When interfaced to form a 3ML-WSe$_2$/5ML-MoSe$_2$ heterostructure, however, the lower of these two bands shifts to higher binding energy to give way to a gapless TSS with well-defined upper and lower branches which merge into the surrounding bands at higher $k_{\parallel}$, as for the case of bulk WSe$_2$. Similarly contrasting behavior is evident for the second, thinner, heterostructure: While the smaller atomic SOC strength of Mo precludes resolving the finite band gaps for the isolated 3ML flake, a distinct crossing point emerges when stacked, differing from the states of 3L or 5ML-MoSe$_2$ (Sup. Fig.~1). With reference to Supplemental Figs.~5 and 6, one can verify that the spectral weight of the Dirac points are in each case localised to the top and bottom monolayers only, demonstrating  the surface states are a property of the collective lattice unit and not its individual components, in line with the $k_z$-mediated topological phase transition from which they derive~\cite{bahramy_ubiquitous_2018}.

In contrast to $k_z$-mediated TSS found in naturally occurring TMDs, the appearances, energetic positions and characteristic chiral momentum-locked spin textures of these Dirac crossings (Fig.~\ref{Fig5}(d)), can differ between the two oppositely-orientated surfaces, with a $\sim 55 $~meV discrepancy extracted for the specific (MoSe$_2$)$_3$/(WSe$_2$)$_5$ configuration from the layer-resolved calculations in Sup. Fig.~6, consistent with the experimental datasets from oppositely-orientated WSe$_2$-MoSe$_2$ structures in Fig.~\ref{Fig5}.  This physics should apply to any pair of TMDs which are susceptible to $k_z$-mediated topological phase transitions, each with a unique set of Dirac crossings with surface-dependent behavior. This could be utilized to drastically augment the five available Dirac cones in the 1T-structured TMD (semi)metals (Pd,Pt,Ni)(Se,Te)$_2$ which have high sensitivity to relative hopping parameters in the unit cell~\cite{bahramy_ubiquitous_2018, ClarkGeneral2019, ferreira_strain_2021, nicholson_uniaxial_2021}. {For example, compared to PtSe$_2$, PdTe$_2$ hosts an additional $k_z$-band crossing to produce a TSS at Fermi level~\cite{clark_fermiology_2018}. It follows that an asymmetric heterostructure formed from these two compounds would host a hybrid $k_z$-dispersion with a band inversion close to the quantum critical point, potentially enabling  \textit{in situ} switching of TSS contributions to transport.}

\section{\label{sec:concs}CONCLUSIONS\protect }

In conclusion, this work elucidates a new and straightforward method to  engineer designer electronic structures in artificial bulk-like systems of transition metal dichalcogenides. By interfacing two few-layer TMD flakes, emergent bands  with characteristics derived from both parent compounds are created, delocalised throughout the heterostructure and facilitated by charge transfer where necessary.}
This produces a scenario in which the bulk-like band dispersions of 3D systems and the layer-locked electron behavior of 2D systems coexist within the same asymmetric structure. {One particularly striking consequence of these contrasting behaviors is the production of $k_z$-mediated topological phases that stabilize surface layer-localised topological states with properties set by local crystal parameters, unlike any topologically non-trivial system existing in nature.}

When combined with controlled lattice-mismatching, this approach could offer opportunities to engineer a coexistence of strong electronic correlations and non-trivial band topology that, while extremely rare naturally, are pathways towards exotic phases such as topological superconductivity, quantum spin liquids and topological Mott insulators~\cite{tokura_quantum_2022}. Alternatively the non-equivalence of oppositely localized surface states, enforced by inherent $c$-axis structural asymmetry, could be exploited for 2D transistor-like components where the asymmetric surface electron transport can be controlled with band filling or layer-thickness.  While demonstrated using flakes with sizes of only tens of microns, new methodologies for creating wafer-scale 2D material flakes~\cite{kang_layer_2017} could be exploited for this purpose.  {We envision that combinations of any of the $>$ 30 stable TMD systems~\cite{chhowalla_chemistry_2013} could produce hybrid three-dimensional band properties analogous to those shown here{ with further applicability to other layered materials, including two-dimensional magnets.}} Together, the simplicity of this approach and the enormity of the phase space available together provide a new and accessible route towards a function-led construction of mixed-dimensional artificial systems.

\section{\label{sec:ack}ACKNOWLEDGMENTS\protect }
O. J. C. thanks Phil D. C. King for useful discussions.
O. J. C., M. S. F. and M. T. E. acknowledge funding support from ARC Discovery Project DP200101345. M. T. E. acknowledges funding support from ARC Future Fellowship FT2201000290. O. J. C., D. M. and M.T.E. acknowledge travel funding provided by the International Synchrotron Access Program (ISAP) managed by the Australian Synchrotron, part of ANSTO, and funded by the Australian Government. D. M. was supported by the FLEET Centre of Excellence (ARC grant CE170100039). A.A., R.V.B. and M.S.B. gratefully acknowledge the Research Infrastructures at the Center for Computational Materials Science at the Institute for Materials Research for allocations on the MASAMUNE-IMR supercomputer system (Project No. 202112-SCKXX-0510) and MAHAMERU BRIN HPC facility under the National Research and Innovation Agency of Indonesia. M.S.B. acknowledges support from Leverhulme Trust (Grant No. RPG-2023-253). A.A. had support from the Indonesia Endowment Fund for Education (LPDP), NIB/202209223311735. R.V.B. and M.S.B. are grateful to E-IMR center at the Institute for Materials Research, Tohoku University, for continuous support. B. C. and S. L. H.  acknowledge the facilities, and the scientific and technical assistance of Microscopy Australia (ROR: 042mm0k03) enabled by NCRIS and the Government of South Australia at Flinders Microscopy and Microanalysis (ROR: 04z91ja70), Flinders University (ROR: 01kpzv902). The work at the University of Warwick was supported by EPSRC, UK through Grant EP/T005963/1. This research used resources of the Advanced Light Source, which is a DOE Office of Science User Facility under contract no. DE-AC02-05CH11231. This work was performed in part at the Melbourne Centre for Nanofabrication (MCN) in the Victorian Node of the Australian National Fabrication Facility (ANFF)

\

\noindent {\bf Author Contributions}
\noindent {OJC, DM, T-H-YV, MTHB, FM and MTE performed the nano-ARPES measurements. OJC and BC performed the nano-ESCA measurements. OJC constructed the heterostructures shown throughout the manuscript, with the exception of the bilayer system shown in Sup. Fig. 4 which was constructed by T-H-YV. T-H-YV also performed atomic force microscopy on the WSe$_2$-NbSe$_2$ heterostructure. AA, RVB and MSB performed the theoretical calculations. CJ, AB and ER provided user support at BL 7 MAESTRO of the Advanced Light Source. BC and SLH provided access for the nano-ESCA instrument at Flinders University, South Australia. SHL and ZM supplied  semiconductors used in device construction GB supplied the NbSe$_2$ crystals used for device construction. OJC conceived the project and wrote the manuscript with significant contributions from MSB and MTE and further input from all other authors. OJC was responsible for the overall project planning and direction.  }

\

\bibliographystyle{apsrev4-1}

\bibliography{bib.bib}

\begin{thebibliography}{68}%
\makeatletter
\providecommand \@ifxundefined [1]{%
 \@ifx{#1\undefined}
}%
\providecommand \@ifnum [1]{%
 \ifnum #1\expandafter \@firstoftwo
 \else \expandafter \@secondoftwo
 \fi
}%
\providecommand \@ifx [1]{%
 \ifx #1\expandafter \@firstoftwo
 \else \expandafter \@secondoftwo
 \fi
}%
\providecommand \natexlab [1]{#1}%
\providecommand \enquote  [1]{``#1''}%
\providecommand \bibnamefont  [1]{#1}%
\providecommand \bibfnamefont [1]{#1}%
\providecommand \citenamefont [1]{#1}%
\providecommand \href@noop [0]{\@secondoftwo}%
\providecommand \href [0]{\begingroup \@sanitize@url \@href}%
\providecommand \@href[1]{\@@startlink{#1}\@@href}%
\providecommand \@@href[1]{\endgroup#1\@@endlink}%
\providecommand \@sanitize@url [0]{\catcode `\\12\catcode `\$12\catcode `\&12\catcode `\#12\catcode `\^12\catcode `\_12\catcode `\%12\relax}%
\providecommand \@@startlink[1]{}%
\providecommand \@@endlink[0]{}%
\providecommand \url  [0]{\begingroup\@sanitize@url \@url }%
\providecommand \@url [1]{\endgroup\@href {#1}{\urlprefix }}%
\providecommand \urlprefix  [0]{URL }%
\providecommand \Eprint [0]{\href }%
\providecommand \doibase [0]{http://dx.doi.org/}%
\providecommand \selectlanguage [0]{\@gobble}%
\providecommand \bibinfo  [0]{\@secondoftwo}%
\providecommand \bibfield  [0]{\@secondoftwo}%
\providecommand \translation [1]{[#1]}%
\providecommand \BibitemOpen [0]{}%
\providecommand \bibitemStop [0]{}%
\providecommand \bibitemNoStop [0]{.\EOS\space}%
\providecommand \EOS [0]{\spacefactor3000\relax}%
\providecommand \BibitemShut  [1]{\csname bibitem#1\endcsname}%
\let\auto@bib@innerbib\@empty
\bibitem [{\citenamefont {Pham}\ \emph {et~al.}(2022)\citenamefont {Pham}, \citenamefont {Bodepudi}, \citenamefont {Shehzad}, \citenamefont {Liu}, \citenamefont {Xu}, \citenamefont {Yu},\ and\ \citenamefont {Duan}}]{pham_2d_2022}%
  \BibitemOpen
  \bibfield  {author} {\bibinfo {author} {\bibfnamefont {P.~V.}\ \bibnamefont {Pham}}, \bibinfo {author} {\bibfnamefont {S.~C.}\ \bibnamefont {Bodepudi}}, \bibinfo {author} {\bibfnamefont {K.}~\bibnamefont {Shehzad}}, \bibinfo {author} {\bibfnamefont {Y.}~\bibnamefont {Liu}}, \bibinfo {author} {\bibfnamefont {Y.}~\bibnamefont {Xu}}, \bibinfo {author} {\bibfnamefont {B.}~\bibnamefont {Yu}}, \ and\ \bibinfo {author} {\bibfnamefont {X.}~\bibnamefont {Duan}},\ }\bibfield  {booktitle} {\emph {\bibinfo {booktitle} {Chemical Reviews}},\ }\href {\doibase 10.1021/acs.chemrev.1c00735} {\bibfield  {journal} {\bibinfo  {journal} {Chemical Reviews}\ }\textbf {\bibinfo {volume} {122}},\ \bibinfo {pages} {6514} (\bibinfo {year} {2022})}\BibitemShut {NoStop}%
\bibitem [{\citenamefont {Chhowalla}\ \emph {et~al.}(2013)\citenamefont {Chhowalla}, \citenamefont {Shin}, \citenamefont {Eda}, \citenamefont {Li}, \citenamefont {Loh},\ and\ \citenamefont {Zhang}}]{chhowalla_chemistry_2013}%
  \BibitemOpen
  \bibfield  {author} {\bibinfo {author} {\bibfnamefont {M.}~\bibnamefont {Chhowalla}}, \bibinfo {author} {\bibfnamefont {H.~S.}\ \bibnamefont {Shin}}, \bibinfo {author} {\bibfnamefont {G.}~\bibnamefont {Eda}}, \bibinfo {author} {\bibfnamefont {L.-J.}\ \bibnamefont {Li}}, \bibinfo {author} {\bibfnamefont {K.~P.}\ \bibnamefont {Loh}}, \ and\ \bibinfo {author} {\bibfnamefont {H.}~\bibnamefont {Zhang}},\ }\href {\doibase 10.1038/nchem.1589} {\bibfield  {journal} {\bibinfo  {journal} {Nature Chemistry}\ }\textbf {\bibinfo {volume} {5}},\ \bibinfo {pages} {263} (\bibinfo {year} {2013})}\BibitemShut {NoStop}%
\bibitem [{\citenamefont {Ahn}(2020)}]{ahn_2d_2020}%
  \BibitemOpen
  \bibfield  {author} {\bibinfo {author} {\bibfnamefont {E.~C.}\ \bibnamefont {Ahn}},\ }\href {\doibase 10.1038/s41699-020-0152-0} {\bibfield  {journal} {\bibinfo  {journal} {npj 2D Materials and Applications}\ }\textbf {\bibinfo {volume} {4}},\ \bibinfo {pages} {17} (\bibinfo {year} {2020})}\BibitemShut {NoStop}%
\bibitem [{\citenamefont {Ciarrocchi}\ \emph {et~al.}(2022)\citenamefont {Ciarrocchi}, \citenamefont {Tagarelli}, \citenamefont {Avsar},\ and\ \citenamefont {Kis}}]{ciarrocchi_excitonic_2022}%
  \BibitemOpen
  \bibfield  {author} {\bibinfo {author} {\bibfnamefont {A.}~\bibnamefont {Ciarrocchi}}, \bibinfo {author} {\bibfnamefont {F.}~\bibnamefont {Tagarelli}}, \bibinfo {author} {\bibfnamefont {A.}~\bibnamefont {Avsar}}, \ and\ \bibinfo {author} {\bibfnamefont {A.}~\bibnamefont {Kis}},\ }\href {\doibase 10.1038/s41578-021-00408-7} {\bibfield  {journal} {\bibinfo  {journal} {Nature Reviews Materials}\ }\textbf {\bibinfo {volume} {7}},\ \bibinfo {pages} {449} (\bibinfo {year} {2022})}\BibitemShut {NoStop}%
\bibitem [{\citenamefont {Manzeli}\ \emph {et~al.}(2017)\citenamefont {Manzeli}, \citenamefont {Ovchinnikov}, \citenamefont {Pasquier}, \citenamefont {Yazyev},\ and\ \citenamefont {Kis}}]{manzelli_2d_2017}%
  \BibitemOpen
  \bibfield  {author} {\bibinfo {author} {\bibfnamefont {S.}~\bibnamefont {Manzeli}}, \bibinfo {author} {\bibfnamefont {D.}~\bibnamefont {Ovchinnikov}}, \bibinfo {author} {\bibfnamefont {D.}~\bibnamefont {Pasquier}}, \bibinfo {author} {\bibfnamefont {O.~V.}\ \bibnamefont {Yazyev}}, \ and\ \bibinfo {author} {\bibfnamefont {A.}~\bibnamefont {Kis}},\ }\href {\doibase 10.1038/natrevmats.2017.33} {\bibfield  {journal} {\bibinfo  {journal} {Nature Reviews Materials}\ }\textbf {\bibinfo {volume} {2}},\ \bibinfo {pages} {17033} (\bibinfo {year} {2017})}\BibitemShut {NoStop}%
\bibitem [{\citenamefont {Regan}\ \emph {et~al.}(2022)\citenamefont {Regan}, \citenamefont {Wang}, \citenamefont {Paik}, \citenamefont {Zeng}, \citenamefont {Zhang}, \citenamefont {Zhu}, \citenamefont {MacDonald}, \citenamefont {Deng},\ and\ \citenamefont {Wang}}]{regan_emerging_2022}%
  \BibitemOpen
  \bibfield  {author} {\bibinfo {author} {\bibfnamefont {E.~C.}\ \bibnamefont {Regan}}, \bibinfo {author} {\bibfnamefont {D.}~\bibnamefont {Wang}}, \bibinfo {author} {\bibfnamefont {E.~Y.}\ \bibnamefont {Paik}}, \bibinfo {author} {\bibfnamefont {Y.}~\bibnamefont {Zeng}}, \bibinfo {author} {\bibfnamefont {L.}~\bibnamefont {Zhang}}, \bibinfo {author} {\bibfnamefont {J.}~\bibnamefont {Zhu}}, \bibinfo {author} {\bibfnamefont {A.~H.}\ \bibnamefont {MacDonald}}, \bibinfo {author} {\bibfnamefont {H.}~\bibnamefont {Deng}}, \ and\ \bibinfo {author} {\bibfnamefont {F.}~\bibnamefont {Wang}},\ }\href {\doibase 10.1038/s41578-022-00440-1} {\bibfield  {journal} {\bibinfo  {journal} {Nature Reviews Materials}\ }\textbf {\bibinfo {volume} {7}},\ \bibinfo {pages} {778} (\bibinfo {year} {2022})}\BibitemShut {NoStop}%
\bibitem [{\citenamefont {Lisi}\ \emph {et~al.}(2021)\citenamefont {Lisi}, \citenamefont {Lu}, \citenamefont {Benschop}, \citenamefont {de~Jong}, \citenamefont {Stepanov}, \citenamefont {Duran}, \citenamefont {Margot}, \citenamefont {Cucchi}, \citenamefont {Cappelli}, \citenamefont {Hunter}, \citenamefont {Tamai}, \citenamefont {Kandyba}, \citenamefont {Giampietri}, \citenamefont {Barinov}, \citenamefont {Jobst}, \citenamefont {Stalman}, \citenamefont {Leeuwenhoek}, \citenamefont {Watanabe}, \citenamefont {Taniguchi}, \citenamefont {Rademaker}, \citenamefont {van~der Molen}, \citenamefont {Allan}, \citenamefont {Efetov},\ and\ \citenamefont {Baumberger}}]{lisi_observation_2021}%
  \BibitemOpen
  \bibfield  {author} {\bibinfo {author} {\bibfnamefont {S.}~\bibnamefont {Lisi}}, \bibinfo {author} {\bibfnamefont {X.}~\bibnamefont {Lu}}, \bibinfo {author} {\bibfnamefont {T.}~\bibnamefont {Benschop}}, \bibinfo {author} {\bibfnamefont {T.~A.}\ \bibnamefont {de~Jong}}, \bibinfo {author} {\bibfnamefont {P.}~\bibnamefont {Stepanov}}, \bibinfo {author} {\bibfnamefont {J.~R.}\ \bibnamefont {Duran}}, \bibinfo {author} {\bibfnamefont {F.}~\bibnamefont {Margot}}, \bibinfo {author} {\bibfnamefont {I.}~\bibnamefont {Cucchi}}, \bibinfo {author} {\bibfnamefont {E.}~\bibnamefont {Cappelli}}, \bibinfo {author} {\bibfnamefont {A.}~\bibnamefont {Hunter}}, \bibinfo {author} {\bibfnamefont {A.}~\bibnamefont {Tamai}}, \bibinfo {author} {\bibfnamefont {V.}~\bibnamefont {Kandyba}}, \bibinfo {author} {\bibfnamefont {A.}~\bibnamefont {Giampietri}}, \bibinfo {author} {\bibfnamefont {A.}~\bibnamefont {Barinov}}, \bibinfo {author} {\bibfnamefont {J.}~\bibnamefont {Jobst}}, \bibinfo {author} {\bibfnamefont {V.}~\bibnamefont
  {Stalman}}, \bibinfo {author} {\bibfnamefont {M.}~\bibnamefont {Leeuwenhoek}}, \bibinfo {author} {\bibfnamefont {K.}~\bibnamefont {Watanabe}}, \bibinfo {author} {\bibfnamefont {T.}~\bibnamefont {Taniguchi}}, \bibinfo {author} {\bibfnamefont {L.}~\bibnamefont {Rademaker}}, \bibinfo {author} {\bibfnamefont {S.~J.}\ \bibnamefont {van~der Molen}}, \bibinfo {author} {\bibfnamefont {M.~P.}\ \bibnamefont {Allan}}, \bibinfo {author} {\bibfnamefont {D.~K.}\ \bibnamefont {Efetov}}, \ and\ \bibinfo {author} {\bibfnamefont {F.}~\bibnamefont {Baumberger}},\ }\href {\doibase 10.1038/s41567-020-01041-x} {\bibfield  {journal} {\bibinfo  {journal} {Nature Physics}\ }\textbf {\bibinfo {volume} {17}},\ \bibinfo {pages} {189} (\bibinfo {year} {2021})}\BibitemShut {NoStop}%
\bibitem [{\citenamefont {Pei}\ \emph {et~al.}(2022)\citenamefont {Pei}, \citenamefont {Wang}, \citenamefont {Zhou}, \citenamefont {He}, \citenamefont {An}, \citenamefont {He}, \citenamefont {Chen}, \citenamefont {Li}, \citenamefont {Wei}, \citenamefont {Liang}, \citenamefont {Avila}, \citenamefont {Dudin}, \citenamefont {Kandyba}, \citenamefont {Giampietri}, \citenamefont {Cattelan}, \citenamefont {Barinov}, \citenamefont {Liu}, \citenamefont {Liu}, \citenamefont {Weng}, \citenamefont {Wang}, \citenamefont {Xue},\ and\ \citenamefont {Chen}}]{pei_observation_2022}%
  \BibitemOpen
  \bibfield  {author} {\bibinfo {author} {\bibfnamefont {D.}~\bibnamefont {Pei}}, \bibinfo {author} {\bibfnamefont {B.}~\bibnamefont {Wang}}, \bibinfo {author} {\bibfnamefont {Z.}~\bibnamefont {Zhou}}, \bibinfo {author} {\bibfnamefont {Z.}~\bibnamefont {He}}, \bibinfo {author} {\bibfnamefont {L.}~\bibnamefont {An}}, \bibinfo {author} {\bibfnamefont {S.}~\bibnamefont {He}}, \bibinfo {author} {\bibfnamefont {C.}~\bibnamefont {Chen}}, \bibinfo {author} {\bibfnamefont {Y.}~\bibnamefont {Li}}, \bibinfo {author} {\bibfnamefont {L.}~\bibnamefont {Wei}}, \bibinfo {author} {\bibfnamefont {A.}~\bibnamefont {Liang}}, \bibinfo {author} {\bibfnamefont {J.}~\bibnamefont {Avila}}, \bibinfo {author} {\bibfnamefont {P.}~\bibnamefont {Dudin}}, \bibinfo {author} {\bibfnamefont {V.}~\bibnamefont {Kandyba}}, \bibinfo {author} {\bibfnamefont {A.}~\bibnamefont {Giampietri}}, \bibinfo {author} {\bibfnamefont {M.}~\bibnamefont {Cattelan}}, \bibinfo {author} {\bibfnamefont {A.}~\bibnamefont {Barinov}}, \bibinfo {author} {\bibfnamefont
  {Z.}~\bibnamefont {Liu}}, \bibinfo {author} {\bibfnamefont {J.}~\bibnamefont {Liu}}, \bibinfo {author} {\bibfnamefont {H.}~\bibnamefont {Weng}}, \bibinfo {author} {\bibfnamefont {N.}~\bibnamefont {Wang}}, \bibinfo {author} {\bibfnamefont {J.}~\bibnamefont {Xue}}, \ and\ \bibinfo {author} {\bibfnamefont {Y.}~\bibnamefont {Chen}},\ }\href {\doibase 10.1103/PhysRevX.12.021065} {\bibfield  {journal} {\bibinfo  {journal} {Phys. Rev. X}\ }\textbf {\bibinfo {volume} {12}},\ \bibinfo {pages} {021065} (\bibinfo {year} {2022})}\BibitemShut {NoStop}%
\bibitem [{\citenamefont {Nunn}\ \emph {et~al.}(2023)\citenamefont {Nunn}, \citenamefont {McEllistrim}, \citenamefont {Weston}, \citenamefont {Garcia-Ruiz}, \citenamefont {Watson}, \citenamefont {Mucha-Kruczynski}, \citenamefont {Cacho}, \citenamefont {Gorbachev}, \citenamefont {Fal'ko},\ and\ \citenamefont {Wilson}}]{nunn_arpes_2022}%
  \BibitemOpen
  \bibfield  {author} {\bibinfo {author} {\bibfnamefont {J.~E.}\ \bibnamefont {Nunn}}, \bibinfo {author} {\bibfnamefont {A.}~\bibnamefont {McEllistrim}}, \bibinfo {author} {\bibfnamefont {A.}~\bibnamefont {Weston}}, \bibinfo {author} {\bibfnamefont {A.}~\bibnamefont {Garcia-Ruiz}}, \bibinfo {author} {\bibfnamefont {M.~D.}\ \bibnamefont {Watson}}, \bibinfo {author} {\bibfnamefont {M.}~\bibnamefont {Mucha-Kruczynski}}, \bibinfo {author} {\bibfnamefont {C.}~\bibnamefont {Cacho}}, \bibinfo {author} {\bibfnamefont {R.~V.}\ \bibnamefont {Gorbachev}}, \bibinfo {author} {\bibfnamefont {V.~I.}\ \bibnamefont {Fal'ko}}, \ and\ \bibinfo {author} {\bibfnamefont {N.~R.}\ \bibnamefont {Wilson}},\ }\bibfield  {booktitle} {\emph {\bibinfo {booktitle} {Nano Letters}},\ }\href {\doibase 10.1021/acs.nanolett.3c01173} {\bibfield  {journal} {\bibinfo  {journal} {Nano Letters}\ }\textbf {\bibinfo {volume} {23}},\ \bibinfo {pages} {5201} (\bibinfo {year} {2023})}\BibitemShut {NoStop}%
\bibitem [{\citenamefont {Stansbury}\ \emph {et~al.}()\citenamefont {Stansbury}, \citenamefont {Utama}, \citenamefont {Fatuzzo}, \citenamefont {Regan}, \citenamefont {Wang}, \citenamefont {Xiang}, \citenamefont {Ding}, \citenamefont {Watanabe}, \citenamefont {Taniguchi}, \citenamefont {Blei}, \citenamefont {Shen}, \citenamefont {Lorcy}, \citenamefont {Bostwick}, \citenamefont {Jozwiak}, \citenamefont {Koch}, \citenamefont {Tongay}, \citenamefont {Avila}, \citenamefont {Rotenberg}, \citenamefont {Wang},\ and\ \citenamefont {Lanzara}}]{stansbury_visualising_2021}%
  \BibitemOpen
  \bibfield  {author} {\bibinfo {author} {\bibfnamefont {C.~H.}\ \bibnamefont {Stansbury}}, \bibinfo {author} {\bibfnamefont {M.~I.~B.}\ \bibnamefont {Utama}}, \bibinfo {author} {\bibfnamefont {C.~G.}\ \bibnamefont {Fatuzzo}}, \bibinfo {author} {\bibfnamefont {E.~C.}\ \bibnamefont {Regan}}, \bibinfo {author} {\bibfnamefont {D.}~\bibnamefont {Wang}}, \bibinfo {author} {\bibfnamefont {Z.}~\bibnamefont {Xiang}}, \bibinfo {author} {\bibfnamefont {M.}~\bibnamefont {Ding}}, \bibinfo {author} {\bibfnamefont {K.}~\bibnamefont {Watanabe}}, \bibinfo {author} {\bibfnamefont {T.}~\bibnamefont {Taniguchi}}, \bibinfo {author} {\bibfnamefont {M.}~\bibnamefont {Blei}}, \bibinfo {author} {\bibfnamefont {Y.}~\bibnamefont {Shen}}, \bibinfo {author} {\bibfnamefont {S.}~\bibnamefont {Lorcy}}, \bibinfo {author} {\bibfnamefont {A.}~\bibnamefont {Bostwick}}, \bibinfo {author} {\bibfnamefont {C.}~\bibnamefont {Jozwiak}}, \bibinfo {author} {\bibfnamefont {R.}~\bibnamefont {Koch}}, \bibinfo {author} {\bibfnamefont {S.}~\bibnamefont
  {Tongay}}, \bibinfo {author} {\bibfnamefont {J.}~\bibnamefont {Avila}}, \bibinfo {author} {\bibfnamefont {E.}~\bibnamefont {Rotenberg}}, \bibinfo {author} {\bibfnamefont {F.}~\bibnamefont {Wang}}, \ and\ \bibinfo {author} {\bibfnamefont {A.}~\bibnamefont {Lanzara}},\ }\bibfield  {booktitle} {\emph {\bibinfo {booktitle} {Science Advances}},\ }\href {\doibase 10.1126/sciadv.abf4387} {\bibfield  {journal} {\bibinfo  {journal} {Science Advances}\ }\textbf {\bibinfo {volume} {7}},\ \bibinfo {pages} {eabf4387}}\BibitemShut {NoStop}%
\bibitem [{\citenamefont {Jones}\ \emph {et~al.}(2022)\citenamefont {Jones}, \citenamefont {Muzzio}, \citenamefont {Pakdel}, \citenamefont {Biswas}, \citenamefont {Curcio}, \citenamefont {Lanat{\`a}}, \citenamefont {Hofmann}, \citenamefont {McCreary}, \citenamefont {Jonker}, \citenamefont {Watanabe}, \citenamefont {Taniguchi}, \citenamefont {Singh}, \citenamefont {Koch}, \citenamefont {Jozwiak}, \citenamefont {Rotenberg}, \citenamefont {Bostwick}, \citenamefont {Miwa}, \citenamefont {Katoch},\ and\ \citenamefont {Ulstrup}}]{jones_visualising_2021}%
  \BibitemOpen
  \bibfield  {author} {\bibinfo {author} {\bibfnamefont {A.~J.~H.}\ \bibnamefont {Jones}}, \bibinfo {author} {\bibfnamefont {R.}~\bibnamefont {Muzzio}}, \bibinfo {author} {\bibfnamefont {S.}~\bibnamefont {Pakdel}}, \bibinfo {author} {\bibfnamefont {D.}~\bibnamefont {Biswas}}, \bibinfo {author} {\bibfnamefont {D.}~\bibnamefont {Curcio}}, \bibinfo {author} {\bibfnamefont {N.}~\bibnamefont {Lanat{\`a}}}, \bibinfo {author} {\bibfnamefont {P.}~\bibnamefont {Hofmann}}, \bibinfo {author} {\bibfnamefont {K.~M.}\ \bibnamefont {McCreary}}, \bibinfo {author} {\bibfnamefont {B.~T.}\ \bibnamefont {Jonker}}, \bibinfo {author} {\bibfnamefont {K.}~\bibnamefont {Watanabe}}, \bibinfo {author} {\bibfnamefont {T.}~\bibnamefont {Taniguchi}}, \bibinfo {author} {\bibfnamefont {S.}~\bibnamefont {Singh}}, \bibinfo {author} {\bibfnamefont {R.~J.}\ \bibnamefont {Koch}}, \bibinfo {author} {\bibfnamefont {C.}~\bibnamefont {Jozwiak}}, \bibinfo {author} {\bibfnamefont {E.}~\bibnamefont {Rotenberg}}, \bibinfo {author} {\bibfnamefont
  {A.}~\bibnamefont {Bostwick}}, \bibinfo {author} {\bibfnamefont {J.~A.}\ \bibnamefont {Miwa}}, \bibinfo {author} {\bibfnamefont {J.}~\bibnamefont {Katoch}}, \ and\ \bibinfo {author} {\bibfnamefont {S.}~\bibnamefont {Ulstrup}},\ }\bibfield  {booktitle} {\emph {\bibinfo {booktitle} {2D Materials}},\ }\href {\doibase 10.1088/2053-1583/ac3feb} {\ \textbf {\bibinfo {volume} {9}},\ \bibinfo {pages} {015032} (\bibinfo {year} {2022})}\BibitemShut {NoStop}%
\bibitem [{\citenamefont {Devakul}\ \emph {et~al.}(2021)\citenamefont {Devakul}, \citenamefont {Cr{\'e}pel}, \citenamefont {Zhang},\ and\ \citenamefont {Fu}}]{devakul_magic_2021}%
  \BibitemOpen
  \bibfield  {author} {\bibinfo {author} {\bibfnamefont {T.}~\bibnamefont {Devakul}}, \bibinfo {author} {\bibfnamefont {V.}~\bibnamefont {Cr{\'e}pel}}, \bibinfo {author} {\bibfnamefont {Y.}~\bibnamefont {Zhang}}, \ and\ \bibinfo {author} {\bibfnamefont {L.}~\bibnamefont {Fu}},\ }\href {\doibase 10.1038/s41467-021-27042-9} {\bibfield  {journal} {\bibinfo  {journal} {Nature Communications}\ }\textbf {\bibinfo {volume} {12}},\ \bibinfo {pages} {6730} (\bibinfo {year} {2021})}\BibitemShut {NoStop}%
\bibitem [{\citenamefont {Guo}\ \emph {et~al.}(2023)\citenamefont {Guo}, \citenamefont {Pack}, \citenamefont {Swann}, \citenamefont {Holtzman}, \citenamefont {Cothrine}, \citenamefont {Watanabe}, \citenamefont {Taniguchi}, \citenamefont {Mandrus}, \citenamefont {Barmak}, \citenamefont {Hone}, \citenamefont {Millis}, \citenamefont {Pasupathy},\ and\ \citenamefont {Dean}}]{guo_superconductivity_2024}%
  \BibitemOpen
  \bibfield  {author} {\bibinfo {author} {\bibfnamefont {Y.}~\bibnamefont {Guo}}, \bibinfo {author} {\bibfnamefont {J.}~\bibnamefont {Pack}}, \bibinfo {author} {\bibfnamefont {J.}~\bibnamefont {Swann}}, \bibinfo {author} {\bibfnamefont {L.}~\bibnamefont {Holtzman}}, \bibinfo {author} {\bibfnamefont {M.}~\bibnamefont {Cothrine}}, \bibinfo {author} {\bibfnamefont {K.}~\bibnamefont {Watanabe}}, \bibinfo {author} {\bibfnamefont {T.}~\bibnamefont {Taniguchi}}, \bibinfo {author} {\bibfnamefont {D.}~\bibnamefont {Mandrus}}, \bibinfo {author} {\bibfnamefont {K.}~\bibnamefont {Barmak}}, \bibinfo {author} {\bibfnamefont {J.}~\bibnamefont {Hone}}, \bibinfo {author} {\bibfnamefont {A.~J.}\ \bibnamefont {Millis}}, \bibinfo {author} {\bibfnamefont {A.~N.}\ \bibnamefont {Pasupathy}}, \ and\ \bibinfo {author} {\bibfnamefont {C.~R.}\ \bibnamefont {Dean}},\ }\href@noop {} {\bibfield  {journal} {\bibinfo  {journal} {arXiv:2310.11317}\ } (\bibinfo {year} {2023})}\BibitemShut {NoStop}%
\bibitem [{\citenamefont {Clark}\ \emph {et~al.}(2018)\citenamefont {Clark}, \citenamefont {Neat}, \citenamefont {Okawa}, \citenamefont {Bawden}, \citenamefont {Markovi\ifmmode~\acute{c}\else \'{c}\fi{}}, \citenamefont {Mazzola}, \citenamefont {Feng}, \citenamefont {Sunko}, \citenamefont {Riley}, \citenamefont {Meevasana}, \citenamefont {Fujii}, \citenamefont {Vobornik}, \citenamefont {Kim}, \citenamefont {Hoesch}, \citenamefont {Sasagawa}, \citenamefont {Wahl}, \citenamefont {Bahramy},\ and\ \citenamefont {King}}]{clark_fermiology_2018}%
  \BibitemOpen
  \bibfield  {author} {\bibinfo {author} {\bibfnamefont {O.~J.}\ \bibnamefont {Clark}}, \bibinfo {author} {\bibfnamefont {M.~J.}\ \bibnamefont {Neat}}, \bibinfo {author} {\bibfnamefont {K.}~\bibnamefont {Okawa}}, \bibinfo {author} {\bibfnamefont {L.}~\bibnamefont {Bawden}}, \bibinfo {author} {\bibfnamefont {I.}~\bibnamefont {Markovi\ifmmode~\acute{c}\else \'{c}\fi{}}}, \bibinfo {author} {\bibfnamefont {F.}~\bibnamefont {Mazzola}}, \bibinfo {author} {\bibfnamefont {J.}~\bibnamefont {Feng}}, \bibinfo {author} {\bibfnamefont {V.}~\bibnamefont {Sunko}}, \bibinfo {author} {\bibfnamefont {J.~M.}\ \bibnamefont {Riley}}, \bibinfo {author} {\bibfnamefont {W.}~\bibnamefont {Meevasana}}, \bibinfo {author} {\bibfnamefont {J.}~\bibnamefont {Fujii}}, \bibinfo {author} {\bibfnamefont {I.}~\bibnamefont {Vobornik}}, \bibinfo {author} {\bibfnamefont {T.~K.}\ \bibnamefont {Kim}}, \bibinfo {author} {\bibfnamefont {M.}~\bibnamefont {Hoesch}}, \bibinfo {author} {\bibfnamefont {T.}~\bibnamefont {Sasagawa}}, \bibinfo {author}
  {\bibfnamefont {P.}~\bibnamefont {Wahl}}, \bibinfo {author} {\bibfnamefont {M.~S.}\ \bibnamefont {Bahramy}}, \ and\ \bibinfo {author} {\bibfnamefont {P.~D.~C.}\ \bibnamefont {King}},\ }\href {\doibase 10.1103/PhysRevLett.120.156401} {\bibfield  {journal} {\bibinfo  {journal} {Phys. Rev. Lett.}\ }\textbf {\bibinfo {volume} {120}},\ \bibinfo {pages} {156401} (\bibinfo {year} {2018})}\BibitemShut {NoStop}%
\bibitem [{\citenamefont {Clark}\ \emph {et~al.}(2022)\citenamefont {Clark}, \citenamefont {Dowinton}, \citenamefont {Bahramy},\ and\ \citenamefont {S{\'a}nchez-Barriga}}]{clark_hidden_2022}%
  \BibitemOpen
  \bibfield  {author} {\bibinfo {author} {\bibfnamefont {O.~J.}\ \bibnamefont {Clark}}, \bibinfo {author} {\bibfnamefont {O.}~\bibnamefont {Dowinton}}, \bibinfo {author} {\bibfnamefont {M.~S.}\ \bibnamefont {Bahramy}}, \ and\ \bibinfo {author} {\bibfnamefont {J.}~\bibnamefont {S{\'a}nchez-Barriga}},\ }\href {\doibase 10.1038/s41467-022-31539-2} {\bibfield  {journal} {\bibinfo  {journal} {Nature Communications}\ }\textbf {\bibinfo {volume} {13}},\ \bibinfo {pages} {4147} (\bibinfo {year} {2022})}\BibitemShut {NoStop}%
\bibitem [{\citenamefont {Riley}\ \emph {et~al.}(2014)\citenamefont {Riley}, \citenamefont {Mazzola}, \citenamefont {Dendzik}, \citenamefont {Michiardi}, \citenamefont {Takayama}, \citenamefont {Bawden}, \citenamefont {Graner{\o}d}, \citenamefont {Leandersson}, \citenamefont {Balasubramanian}, \citenamefont {Hoesch}, \citenamefont {Kim}, \citenamefont {Takagi}, \citenamefont {Meevasana}, \citenamefont {Hofmann}, \citenamefont {Bahramy}, \citenamefont {Wells},\ and\ \citenamefont {King}}]{riley_direct_2014}%
  \BibitemOpen
  \bibfield  {author} {\bibinfo {author} {\bibfnamefont {J.~M.}\ \bibnamefont {Riley}}, \bibinfo {author} {\bibfnamefont {F.}~\bibnamefont {Mazzola}}, \bibinfo {author} {\bibfnamefont {M.}~\bibnamefont {Dendzik}}, \bibinfo {author} {\bibfnamefont {M.}~\bibnamefont {Michiardi}}, \bibinfo {author} {\bibfnamefont {T.}~\bibnamefont {Takayama}}, \bibinfo {author} {\bibfnamefont {L.}~\bibnamefont {Bawden}}, \bibinfo {author} {\bibfnamefont {C.}~\bibnamefont {Graner{\o}d}}, \bibinfo {author} {\bibfnamefont {M.}~\bibnamefont {Leandersson}}, \bibinfo {author} {\bibfnamefont {T.}~\bibnamefont {Balasubramanian}}, \bibinfo {author} {\bibfnamefont {M.}~\bibnamefont {Hoesch}}, \bibinfo {author} {\bibfnamefont {T.~K.}\ \bibnamefont {Kim}}, \bibinfo {author} {\bibfnamefont {H.}~\bibnamefont {Takagi}}, \bibinfo {author} {\bibfnamefont {W.}~\bibnamefont {Meevasana}}, \bibinfo {author} {\bibfnamefont {P.}~\bibnamefont {Hofmann}}, \bibinfo {author} {\bibfnamefont {M.~S.}\ \bibnamefont {Bahramy}}, \bibinfo {author} {\bibfnamefont
  {J.~W.}\ \bibnamefont {Wells}}, \ and\ \bibinfo {author} {\bibfnamefont {P.~D.~C.}\ \bibnamefont {King}},\ }\href {\doibase 10.1038/nphys3105} {\bibfield  {journal} {\bibinfo  {journal} {Nature Physics}\ }\textbf {\bibinfo {volume} {10}},\ \bibinfo {pages} {835} (\bibinfo {year} {2014})}\BibitemShut {NoStop}%
\bibitem [{\citenamefont {Watson}\ \emph {et~al.}(2024)\citenamefont {Watson}, \citenamefont {Date}, \citenamefont {Louat},\ and\ \citenamefont {Schr\"oter}}]{watson_novel_2024}%
  \BibitemOpen
  \bibfield  {author} {\bibinfo {author} {\bibfnamefont {M.~D.}\ \bibnamefont {Watson}}, \bibinfo {author} {\bibfnamefont {M.}~\bibnamefont {Date}}, \bibinfo {author} {\bibfnamefont {A.}~\bibnamefont {Louat}}, \ and\ \bibinfo {author} {\bibfnamefont {N.~B.~M.}\ \bibnamefont {Schr\"oter}},\ }\href {\doibase 10.1103/PhysRevB.110.L121121} {\bibfield  {journal} {\bibinfo  {journal} {Phys. Rev. B}\ }\textbf {\bibinfo {volume} {110}},\ \bibinfo {pages} {L121121} (\bibinfo {year} {2024})}\BibitemShut {NoStop}%
\bibitem [{\citenamefont {Wan}\ \emph {et~al.}(2024)\citenamefont {Wan}, \citenamefont {Qian}, \citenamefont {Huang},\ and\ \citenamefont {Duan}}]{wan_layered_2024}%
  \BibitemOpen
  \bibfield  {author} {\bibinfo {author} {\bibfnamefont {Z.}~\bibnamefont {Wan}}, \bibinfo {author} {\bibfnamefont {Q.}~\bibnamefont {Qian}}, \bibinfo {author} {\bibfnamefont {Y.}~\bibnamefont {Huang}}, \ and\ \bibinfo {author} {\bibfnamefont {X.}~\bibnamefont {Duan}},\ }\href {\doibase 10.1038/s41586-024-07858-3} {\bibfield  {journal} {\bibinfo  {journal} {Nature}\ }\textbf {\bibinfo {volume} {635}},\ \bibinfo {pages} {49} (\bibinfo {year} {2024})}\BibitemShut {NoStop}%
\bibitem [{\citenamefont {Splendiani}\ \emph {et~al.}(2010)\citenamefont {Splendiani}, \citenamefont {Sun}, \citenamefont {Zhang}, \citenamefont {Li}, \citenamefont {Kim}, \citenamefont {Chim}, \citenamefont {Galli},\ and\ \citenamefont {Wang}}]{splendiani_emerging_2010}%
  \BibitemOpen
  \bibfield  {author} {\bibinfo {author} {\bibfnamefont {A.}~\bibnamefont {Splendiani}}, \bibinfo {author} {\bibfnamefont {L.}~\bibnamefont {Sun}}, \bibinfo {author} {\bibfnamefont {Y.}~\bibnamefont {Zhang}}, \bibinfo {author} {\bibfnamefont {T.}~\bibnamefont {Li}}, \bibinfo {author} {\bibfnamefont {J.}~\bibnamefont {Kim}}, \bibinfo {author} {\bibfnamefont {C.-Y.}\ \bibnamefont {Chim}}, \bibinfo {author} {\bibfnamefont {G.}~\bibnamefont {Galli}}, \ and\ \bibinfo {author} {\bibfnamefont {F.}~\bibnamefont {Wang}},\ }\bibfield  {booktitle} {\emph {\bibinfo {booktitle} {Nano Letters}},\ }\href {\doibase 10.1021/nl903868w} {\bibfield  {journal} {\bibinfo  {journal} {Nano Letters}\ }\textbf {\bibinfo {volume} {10}},\ \bibinfo {pages} {1271} (\bibinfo {year} {2010})}\BibitemShut {NoStop}%
\bibitem [{\citenamefont {Mak}\ \emph {et~al.}(2010)\citenamefont {Mak}, \citenamefont {Lee}, \citenamefont {Hone}, \citenamefont {Shan},\ and\ \citenamefont {Heinz}}]{mak_atomically_2010}%
  \BibitemOpen
  \bibfield  {author} {\bibinfo {author} {\bibfnamefont {K.~F.}\ \bibnamefont {Mak}}, \bibinfo {author} {\bibfnamefont {C.}~\bibnamefont {Lee}}, \bibinfo {author} {\bibfnamefont {J.}~\bibnamefont {Hone}}, \bibinfo {author} {\bibfnamefont {J.}~\bibnamefont {Shan}}, \ and\ \bibinfo {author} {\bibfnamefont {T.~F.}\ \bibnamefont {Heinz}},\ }\href {\doibase 10.1103/PhysRevLett.105.136805} {\bibfield  {journal} {\bibinfo  {journal} {Phys. Rev. Lett.}\ }\textbf {\bibinfo {volume} {105}},\ \bibinfo {pages} {136805} (\bibinfo {year} {2010})}\BibitemShut {NoStop}%
\bibitem [{\citenamefont {Sun}\ \emph {et~al.}(2016)\citenamefont {Sun}, \citenamefont {Wang},\ and\ \citenamefont {Shuai}}]{sun_indirect_2016}%
  \BibitemOpen
  \bibfield  {author} {\bibinfo {author} {\bibfnamefont {Y.}~\bibnamefont {Sun}}, \bibinfo {author} {\bibfnamefont {D.}~\bibnamefont {Wang}}, \ and\ \bibinfo {author} {\bibfnamefont {Z.}~\bibnamefont {Shuai}},\ }\bibfield  {booktitle} {\emph {\bibinfo {booktitle} {The Journal of Physical Chemistry C}},\ }\href {\doibase 10.1021/acs.jpcc.6b08748} {\bibfield  {journal} {\bibinfo  {journal} {The Journal of Physical Chemistry C}\ }\textbf {\bibinfo {volume} {120}},\ \bibinfo {pages} {21866} (\bibinfo {year} {2016})}\BibitemShut {NoStop}%
\bibitem [{\citenamefont {Bahramy}\ \emph {et~al.}(2018)\citenamefont {Bahramy}, \citenamefont {Clark}, \citenamefont {Yang}, \citenamefont {Feng}, \citenamefont {Bawden}, \citenamefont {Riley}, \citenamefont {I.}, \citenamefont {Mazzola}, \citenamefont {Sunko}, \citenamefont {Biswas}, \citenamefont {Cooil}, \citenamefont {Jorge}, \citenamefont {Wells}, \citenamefont {Leandersson}, \citenamefont {Balasubramanian}, \citenamefont {Fujii}, \citenamefont {Vobornik}, \citenamefont {Rault}, \citenamefont {Kim}, \citenamefont {Hoesch}, \citenamefont {Okawa}, \citenamefont {Asakawa}, \citenamefont {Sasagawa}, \citenamefont {Eknapakul}, \citenamefont {Meevasana},\ and\ \citenamefont {King}}]{bahramy_ubiquitous_2018}%
  \BibitemOpen
  \bibfield  {author} {\bibinfo {author} {\bibfnamefont {M.~S.}\ \bibnamefont {Bahramy}}, \bibinfo {author} {\bibfnamefont {O.~J.}\ \bibnamefont {Clark}}, \bibinfo {author} {\bibfnamefont {B.-J.}\ \bibnamefont {Yang}}, \bibinfo {author} {\bibfnamefont {J.}~\bibnamefont {Feng}}, \bibinfo {author} {\bibfnamefont {L.}~\bibnamefont {Bawden}}, \bibinfo {author} {\bibfnamefont {J.~M.}\ \bibnamefont {Riley}}, \bibinfo {author} {\bibfnamefont {M.}~\bibnamefont {I.}}, \bibinfo {author} {\bibfnamefont {F.}~\bibnamefont {Mazzola}}, \bibinfo {author} {\bibfnamefont {V.}~\bibnamefont {Sunko}}, \bibinfo {author} {\bibfnamefont {D.}~\bibnamefont {Biswas}}, \bibinfo {author} {\bibfnamefont {S.~P.}\ \bibnamefont {Cooil}}, \bibinfo {author} {\bibfnamefont {M.}~\bibnamefont {Jorge}}, \bibinfo {author} {\bibfnamefont {J.~W.}\ \bibnamefont {Wells}}, \bibinfo {author} {\bibfnamefont {M.}~\bibnamefont {Leandersson}}, \bibinfo {author} {\bibfnamefont {T.}~\bibnamefont {Balasubramanian}}, \bibinfo {author} {\bibfnamefont
  {J.}~\bibnamefont {Fujii}}, \bibinfo {author} {\bibfnamefont {I.}~\bibnamefont {Vobornik}}, \bibinfo {author} {\bibfnamefont {J.}~\bibnamefont {Rault}}, \bibinfo {author} {\bibfnamefont {T.~K.}\ \bibnamefont {Kim}}, \bibinfo {author} {\bibfnamefont {M.}~\bibnamefont {Hoesch}}, \bibinfo {author} {\bibfnamefont {K.}~\bibnamefont {Okawa}}, \bibinfo {author} {\bibfnamefont {M.}~\bibnamefont {Asakawa}}, \bibinfo {author} {\bibfnamefont {T.}~\bibnamefont {Sasagawa}}, \bibinfo {author} {\bibfnamefont {T.}~\bibnamefont {Eknapakul}}, \bibinfo {author} {\bibfnamefont {W.}~\bibnamefont {Meevasana}}, \ and\ \bibinfo {author} {\bibfnamefont {P.~D.~C.}\ \bibnamefont {King}},\ }\href@noop {} {\bibfield  {journal} {\bibinfo  {journal} {Nature Materials}\ }\textbf {\bibinfo {volume} {17}} (\bibinfo {year} {2018})}\BibitemShut {NoStop}%
\bibitem [{\citenamefont {Zhang}\ \emph {et~al.}(2025)\citenamefont {Zhang}, \citenamefont {Lu}, \citenamefont {Shen}, \citenamefont {Niu}, \citenamefont {Liu}, \citenamefont {Lu}, \citenamefont {Lin}, \citenamefont {Han}, \citenamefont {Weng}, \citenamefont {Shao}, \citenamefont {Yan}, \citenamefont {Ren}, \citenamefont {Li}, \citenamefont {Chang}, \citenamefont {Singh}, \citenamefont {He}, \citenamefont {He}, \citenamefont {Liu}, \citenamefont {Bian}, \citenamefont {Miao},\ and\ \citenamefont {Xu}}]{zhang_substantially_2024}%
  \BibitemOpen
  \bibfield  {author} {\bibinfo {author} {\bibfnamefont {X.}~\bibnamefont {Zhang}}, \bibinfo {author} {\bibfnamefont {Q.}~\bibnamefont {Lu}}, \bibinfo {author} {\bibfnamefont {Z.-X.}\ \bibnamefont {Shen}}, \bibinfo {author} {\bibfnamefont {W.}~\bibnamefont {Niu}}, \bibinfo {author} {\bibfnamefont {X.}~\bibnamefont {Liu}}, \bibinfo {author} {\bibfnamefont {J.}~\bibnamefont {Lu}}, \bibinfo {author} {\bibfnamefont {W.}~\bibnamefont {Lin}}, \bibinfo {author} {\bibfnamefont {L.}~\bibnamefont {Han}}, \bibinfo {author} {\bibfnamefont {Y.}~\bibnamefont {Weng}}, \bibinfo {author} {\bibfnamefont {T.}~\bibnamefont {Shao}}, \bibinfo {author} {\bibfnamefont {P.}~\bibnamefont {Yan}}, \bibinfo {author} {\bibfnamefont {Q.}~\bibnamefont {Ren}}, \bibinfo {author} {\bibfnamefont {H.}~\bibnamefont {Li}}, \bibinfo {author} {\bibfnamefont {T.-R.}\ \bibnamefont {Chang}}, \bibinfo {author} {\bibfnamefont {D.~J.}\ \bibnamefont {Singh}}, \bibinfo {author} {\bibfnamefont {L.}~\bibnamefont {He}}, \bibinfo {author} {\bibfnamefont
  {L.}~\bibnamefont {He}}, \bibinfo {author} {\bibfnamefont {C.}~\bibnamefont {Liu}}, \bibinfo {author} {\bibfnamefont {G.}~\bibnamefont {Bian}}, \bibinfo {author} {\bibfnamefont {L.}~\bibnamefont {Miao}}, \ and\ \bibinfo {author} {\bibfnamefont {Y.}~\bibnamefont {Xu}},\ }\bibfield  {booktitle} {\emph {\bibinfo {booktitle} {Advanced Materials}},\ }\href {\doibase https://doi.org/10.1002/adma.202411137} {\bibfield  {journal} {\bibinfo  {journal} {Advanced Materials}\ }\textbf {\bibinfo {volume} {37}},\ \bibinfo {pages} {2411137} (\bibinfo {year} {2025})}\BibitemShut {NoStop}%
\bibitem [{\citenamefont {Watson}\ \emph {et~al.}(2019)\citenamefont {Watson}, \citenamefont {Clark}, \citenamefont {Mazzola}, \citenamefont {Markovi\ifmmode~\acute{c}\else \'{c}\fi{}}, \citenamefont {Sunko}, \citenamefont {Kim}, \citenamefont {Rossnagel},\ and\ \citenamefont {King}}]{watson_orbital_2019}%
  \BibitemOpen
  \bibfield  {author} {\bibinfo {author} {\bibfnamefont {M.~D.}\ \bibnamefont {Watson}}, \bibinfo {author} {\bibfnamefont {O.~J.}\ \bibnamefont {Clark}}, \bibinfo {author} {\bibfnamefont {F.}~\bibnamefont {Mazzola}}, \bibinfo {author} {\bibfnamefont {I.}~\bibnamefont {Markovi\ifmmode~\acute{c}\else \'{c}\fi{}}}, \bibinfo {author} {\bibfnamefont {V.}~\bibnamefont {Sunko}}, \bibinfo {author} {\bibfnamefont {T.~K.}\ \bibnamefont {Kim}}, \bibinfo {author} {\bibfnamefont {K.}~\bibnamefont {Rossnagel}}, \ and\ \bibinfo {author} {\bibfnamefont {P.~D.~C.}\ \bibnamefont {King}},\ }\href {\doibase 10.1103/PhysRevLett.122.076404} {\bibfield  {journal} {\bibinfo  {journal} {Phys. Rev. Lett.}\ }\textbf {\bibinfo {volume} {122}},\ \bibinfo {pages} {076404} (\bibinfo {year} {2019})}\BibitemShut {NoStop}%
\bibitem [{\citenamefont {Feng}\ \emph {et~al.}(2018)\citenamefont {Feng}, \citenamefont {Biswas}, \citenamefont {Rajan}, \citenamefont {Watson}, \citenamefont {Mazzola}, \citenamefont {Clark}, \citenamefont {Underwood}, \citenamefont {Markovi{\'c}}, \citenamefont {McLaren}, \citenamefont {Hunter}, \citenamefont {Burn}, \citenamefont {Duffy}, \citenamefont {Barua}, \citenamefont {Balakrishnan}, \citenamefont {Bertran}, \citenamefont {Le~F{\`e}vre}, \citenamefont {Kim}, \citenamefont {van~der Laan}, \citenamefont {Hesjedal}, \citenamefont {Wahl},\ and\ \citenamefont {King}}]{feng_electronic_2018}%
  \BibitemOpen
  \bibfield  {author} {\bibinfo {author} {\bibfnamefont {J.}~\bibnamefont {Feng}}, \bibinfo {author} {\bibfnamefont {D.}~\bibnamefont {Biswas}}, \bibinfo {author} {\bibfnamefont {A.}~\bibnamefont {Rajan}}, \bibinfo {author} {\bibfnamefont {M.~D.}\ \bibnamefont {Watson}}, \bibinfo {author} {\bibfnamefont {F.}~\bibnamefont {Mazzola}}, \bibinfo {author} {\bibfnamefont {O.~J.}\ \bibnamefont {Clark}}, \bibinfo {author} {\bibfnamefont {K.}~\bibnamefont {Underwood}}, \bibinfo {author} {\bibfnamefont {I.}~\bibnamefont {Markovi{\'c}}}, \bibinfo {author} {\bibfnamefont {M.}~\bibnamefont {McLaren}}, \bibinfo {author} {\bibfnamefont {A.}~\bibnamefont {Hunter}}, \bibinfo {author} {\bibfnamefont {D.~M.}\ \bibnamefont {Burn}}, \bibinfo {author} {\bibfnamefont {L.~B.}\ \bibnamefont {Duffy}}, \bibinfo {author} {\bibfnamefont {S.}~\bibnamefont {Barua}}, \bibinfo {author} {\bibfnamefont {G.}~\bibnamefont {Balakrishnan}}, \bibinfo {author} {\bibfnamefont {F.}~\bibnamefont {Bertran}}, \bibinfo {author} {\bibfnamefont
  {P.}~\bibnamefont {Le~F{\`e}vre}}, \bibinfo {author} {\bibfnamefont {T.~K.}\ \bibnamefont {Kim}}, \bibinfo {author} {\bibfnamefont {G.}~\bibnamefont {van~der Laan}}, \bibinfo {author} {\bibfnamefont {T.}~\bibnamefont {Hesjedal}}, \bibinfo {author} {\bibfnamefont {P.}~\bibnamefont {Wahl}}, \ and\ \bibinfo {author} {\bibfnamefont {P.~D.~C.}\ \bibnamefont {King}},\ }\bibfield  {booktitle} {\emph {\bibinfo {booktitle} {Nano Letters}},\ }\href {\doibase 10.1021/acs.nanolett.8b01649} {\bibfield  {journal} {\bibinfo  {journal} {Nano Letters}\ }\textbf {\bibinfo {volume} {18}},\ \bibinfo {pages} {4493} (\bibinfo {year} {2018})}\BibitemShut {NoStop}%
\bibitem [{\citenamefont {Chang}\ \emph {et~al.}(2014)\citenamefont {Chang}, \citenamefont {Lin}, \citenamefont {Jeng},\ and\ \citenamefont {Bansil}}]{chang_thickness_2014}%
  \BibitemOpen
  \bibfield  {author} {\bibinfo {author} {\bibfnamefont {T.-R.}\ \bibnamefont {Chang}}, \bibinfo {author} {\bibfnamefont {H.}~\bibnamefont {Lin}}, \bibinfo {author} {\bibfnamefont {H.-T.}\ \bibnamefont {Jeng}}, \ and\ \bibinfo {author} {\bibfnamefont {A.}~\bibnamefont {Bansil}},\ }\href {\doibase 10.1038/srep06270} {\bibfield  {journal} {\bibinfo  {journal} {Scientific Reports}\ }\textbf {\bibinfo {volume} {4}},\ \bibinfo {pages} {6270} (\bibinfo {year} {2014})}\BibitemShut {NoStop}%
\bibitem [{\citenamefont {Le~F\'evre}\ \emph {et~al.}(2024)\citenamefont {Le~F\'evre}, \citenamefont {Salazar}, \citenamefont {Jamet}, \citenamefont {Bertran}, \citenamefont {Bigi}, \citenamefont {Ourghi}, \citenamefont {Vergnaud}, \citenamefont {Pulkkinen}, \citenamefont {Minar}, \citenamefont {Jaouen},\ and\ \citenamefont {Rault}}]{lefevre_two_2024}%
  \BibitemOpen
  \bibfield  {author} {\bibinfo {author} {\bibfnamefont {P.}~\bibnamefont {Le~F\'evre}}, \bibinfo {author} {\bibfnamefont {R.}~\bibnamefont {Salazar}}, \bibinfo {author} {\bibfnamefont {M.}~\bibnamefont {Jamet}}, \bibinfo {author} {\bibfnamefont {F.}~\bibnamefont {Bertran}}, \bibinfo {author} {\bibfnamefont {C.}~\bibnamefont {Bigi}}, \bibinfo {author} {\bibfnamefont {A.}~\bibnamefont {Ourghi}}, \bibinfo {author} {\bibfnamefont {C.}~\bibnamefont {Vergnaud}}, \bibinfo {author} {\bibfnamefont {A.}~\bibnamefont {Pulkkinen}}, \bibinfo {author} {\bibfnamefont {J.}~\bibnamefont {Minar}}, \bibinfo {author} {\bibfnamefont {T.}~\bibnamefont {Jaouen}}, \ and\ \bibinfo {author} {\bibfnamefont {J.}~\bibnamefont {Rault}},\ }\href@noop {} {\bibfield  {journal} {\bibinfo  {journal} {arXiv:2407.03768}\ } (\bibinfo {year} {2024})}\BibitemShut {NoStop}%
\bibitem [{Fli()}]{FlindersNanoESCA}%
  \BibitemOpen
  \href {\doibase 10.25957/flinders.nanoesca3} {\ 10.25957/flinders.nanoesca3}\BibitemShut {NoStop}%
\bibitem [{\citenamefont {Perdew}\ \emph {et~al.}(1996)\citenamefont {Perdew}, \citenamefont {Burke},\ and\ \citenamefont {Ernzerhof}}]{pbe}%
  \BibitemOpen
  \bibfield  {author} {\bibinfo {author} {\bibfnamefont {J.~P.}\ \bibnamefont {Perdew}}, \bibinfo {author} {\bibfnamefont {K.}~\bibnamefont {Burke}}, \ and\ \bibinfo {author} {\bibfnamefont {M.}~\bibnamefont {Ernzerhof}},\ }\href {\doibase 10.1103/PhysRevLett.77.3865} {\bibfield  {journal} {\bibinfo  {journal} {Phys. Rev. Lett.}\ }\textbf {\bibinfo {volume} {77}},\ \bibinfo {pages} {3865} (\bibinfo {year} {1996})}\BibitemShut {NoStop}%
\bibitem [{\citenamefont {Kresse}\ and\ \citenamefont {Furthm\"uller}(1996)}]{Kresse_1996}%
  \BibitemOpen
  \bibfield  {author} {\bibinfo {author} {\bibfnamefont {G.}~\bibnamefont {Kresse}}\ and\ \bibinfo {author} {\bibfnamefont {J.}~\bibnamefont {Furthm\"uller}},\ }\href {\doibase 10.1103/PhysRevB.54.11169} {\bibfield  {journal} {\bibinfo  {journal} {Phys. Rev. B}\ }\textbf {\bibinfo {volume} {54}},\ \bibinfo {pages} {11169} (\bibinfo {year} {1996})}\BibitemShut {NoStop}%
\bibitem [{\citenamefont {Kresse}\ and\ \citenamefont {Joubert}(1999)}]{Kresse_1999}%
  \BibitemOpen
  \bibfield  {author} {\bibinfo {author} {\bibfnamefont {G.}~\bibnamefont {Kresse}}\ and\ \bibinfo {author} {\bibfnamefont {D.}~\bibnamefont {Joubert}},\ }\href {\doibase 10.1103/PhysRevB.59.1758} {\bibfield  {journal} {\bibinfo  {journal} {Phys. Rev. B}\ }\textbf {\bibinfo {volume} {59}},\ \bibinfo {pages} {1758} (\bibinfo {year} {1999})}\BibitemShut {NoStop}%
\bibitem [{\citenamefont {Schutte}\ \emph {et~al.}(1987)\citenamefont {Schutte}, \citenamefont {{De Boer}},\ and\ \citenamefont {Jellinek}}]{SCHUTTE1987207}%
  \BibitemOpen
  \bibfield  {author} {\bibinfo {author} {\bibfnamefont {W.}~\bibnamefont {Schutte}}, \bibinfo {author} {\bibfnamefont {J.}~\bibnamefont {{De Boer}}}, \ and\ \bibinfo {author} {\bibfnamefont {F.}~\bibnamefont {Jellinek}},\ }\href {\doibase https://doi.org/10.1016/0022-4596(87)90057-0} {\bibfield  {journal} {\bibinfo  {journal} {Journal of Solid State Chemistry}\ }\textbf {\bibinfo {volume} {70}},\ \bibinfo {pages} {207} (\bibinfo {year} {1987})}\BibitemShut {NoStop}%
\bibitem [{\citenamefont {Mostofi}\ \emph {et~al.}(2008)\citenamefont {Mostofi}, \citenamefont {Yates}, \citenamefont {Lee}, \citenamefont {Souza}, \citenamefont {Vanderbilt},\ and\ \citenamefont {Marzari}}]{mostofi2008}%
  \BibitemOpen
  \bibfield  {author} {\bibinfo {author} {\bibfnamefont {A.~A.}\ \bibnamefont {Mostofi}}, \bibinfo {author} {\bibfnamefont {J.~R.}\ \bibnamefont {Yates}}, \bibinfo {author} {\bibfnamefont {Y.-S.}\ \bibnamefont {Lee}}, \bibinfo {author} {\bibfnamefont {I.}~\bibnamefont {Souza}}, \bibinfo {author} {\bibfnamefont {D.}~\bibnamefont {Vanderbilt}}, \ and\ \bibinfo {author} {\bibfnamefont {N.}~\bibnamefont {Marzari}},\ }\href {\doibase https://doi.org/10.1016/j.cpc.2007.11.016} {\bibfield  {journal} {\bibinfo  {journal} {Comput. Phys. Commun.}\ }\textbf {\bibinfo {volume} {178}},\ \bibinfo {pages} {685} (\bibinfo {year} {2008})}\BibitemShut {NoStop}%
\bibitem [{\citenamefont {Souza}\ \emph {et~al.}(2001)\citenamefont {Souza}, \citenamefont {Marzari},\ and\ \citenamefont {Vanderbilt}}]{Souza2001}%
  \BibitemOpen
  \bibfield  {author} {\bibinfo {author} {\bibfnamefont {I.}~\bibnamefont {Souza}}, \bibinfo {author} {\bibfnamefont {N.}~\bibnamefont {Marzari}}, \ and\ \bibinfo {author} {\bibfnamefont {D.}~\bibnamefont {Vanderbilt}},\ }\href@noop {} {\bibfield  {journal} {\bibinfo  {journal} {Phys. Rev. B}\ }\textbf {\bibinfo {volume} {65}},\ \bibinfo {pages} {035109} (\bibinfo {year} {2001})}\BibitemShut {NoStop}%
\bibitem [{\citenamefont {Ribak}\ \emph {et~al.}()\citenamefont {Ribak}, \citenamefont {Skiff}, \citenamefont {Mograbi}, \citenamefont {Rout}, \citenamefont {Fischer}, \citenamefont {Ruhman}, \citenamefont {Chashka}, \citenamefont {Dagan},\ and\ \citenamefont {Kanigel}}]{ribak_chiral_2020}%
  \BibitemOpen
  \bibfield  {author} {\bibinfo {author} {\bibfnamefont {A.}~\bibnamefont {Ribak}}, \bibinfo {author} {\bibfnamefont {R.~M.}\ \bibnamefont {Skiff}}, \bibinfo {author} {\bibfnamefont {M.}~\bibnamefont {Mograbi}}, \bibinfo {author} {\bibfnamefont {P.~K.}\ \bibnamefont {Rout}}, \bibinfo {author} {\bibfnamefont {M.~H.}\ \bibnamefont {Fischer}}, \bibinfo {author} {\bibfnamefont {J.}~\bibnamefont {Ruhman}}, \bibinfo {author} {\bibfnamefont {K.}~\bibnamefont {Chashka}}, \bibinfo {author} {\bibfnamefont {Y.}~\bibnamefont {Dagan}}, \ and\ \bibinfo {author} {\bibfnamefont {A.}~\bibnamefont {Kanigel}},\ }\bibfield  {booktitle} {\emph {\bibinfo {booktitle} {Science Advances}},\ }\href {\doibase 10.1126/sciadv.aax9480} {\bibfield  {journal} {\bibinfo  {journal} {Science Advances}\ }\textbf {\bibinfo {volume} {6}},\ \bibinfo {pages} {eaax9480}}\BibitemShut {NoStop}%
\bibitem [{\citenamefont {Chiang}(2000)}]{chiang_photoemission_2000}%
  \BibitemOpen
  \bibfield  {author} {\bibinfo {author} {\bibfnamefont {T.~C.}\ \bibnamefont {Chiang}},\ }\href {\doibase https://doi.org/10.1016/S0167-5729(00)00006-6} {\bibfield  {journal} {\bibinfo  {journal} {Surface Science Reports}\ }\textbf {\bibinfo {volume} {39}},\ \bibinfo {pages} {181} (\bibinfo {year} {2000})}\BibitemShut {NoStop}%
\bibitem [{\citenamefont {Milun}\ \emph {et~al.}(2002)\citenamefont {Milun}, \citenamefont {Pervan},\ and\ \citenamefont {Woodruff}}]{milun_quantum_2002}%
  \BibitemOpen
  \bibfield  {author} {\bibinfo {author} {\bibfnamefont {M.}~\bibnamefont {Milun}}, \bibinfo {author} {\bibfnamefont {P.}~\bibnamefont {Pervan}}, \ and\ \bibinfo {author} {\bibfnamefont {D.~P.}\ \bibnamefont {Woodruff}},\ }\href {\doibase 10.1088/0034-4885/65/2/201} {\bibfield  {journal} {\bibinfo  {journal} {Reports on Progress in Physics}\ }\textbf {\bibinfo {volume} {65}},\ \bibinfo {pages} {99} (\bibinfo {year} {2002})}\BibitemShut {NoStop}%
\bibitem [{\citenamefont {Pan}\ and\ \citenamefont {Zahn}(2022)}]{yang_ramen_2025}%
  \BibitemOpen
  \bibfield  {author} {\bibinfo {author} {\bibfnamefont {Y.}~\bibnamefont {Pan}}\ and\ \bibinfo {author} {\bibfnamefont {D.~R.~T.}\ \bibnamefont {Zahn}},\ }\bibfield  {booktitle} {\emph {\bibinfo {booktitle} {Nanomaterials}},\ }\href {\doibase 10.3390/nano12223949} {\ \textbf {\bibinfo {volume} {12}} (\bibinfo {year} {2022}),\ 10.3390/nano12223949}\BibitemShut {NoStop}%
\bibitem [{\citenamefont {Damascelli}(2004)}]{Dam2004}%
  \BibitemOpen
  \bibfield  {author} {\bibinfo {author} {\bibfnamefont {A.}~\bibnamefont {Damascelli}},\ }\href@noop {} {\bibfield  {journal} {\bibinfo  {journal} {Physica Scripta}\ }\textbf {\bibinfo {volume} {2004}},\ \bibinfo {pages} {T109 61} (\bibinfo {year} {2004})}\BibitemShut {NoStop}%
\bibitem [{\citenamefont {Kim}\ and\ \citenamefont {Choi}(2021)}]{kim_thickness_2021}%
  \BibitemOpen
  \bibfield  {author} {\bibinfo {author} {\bibfnamefont {H.-g.}\ \bibnamefont {Kim}}\ and\ \bibinfo {author} {\bibfnamefont {H.~J.}\ \bibnamefont {Choi}},\ }\href {\doibase 10.1103/PhysRevB.103.085404} {\bibfield  {journal} {\bibinfo  {journal} {Phys. Rev. B}\ }\textbf {\bibinfo {volume} {103}},\ \bibinfo {pages} {085404} (\bibinfo {year} {2021})}\BibitemShut {NoStop}%
\bibitem [{\citenamefont {Hlevyack}\ \emph {et~al.}(2021)\citenamefont {Hlevyack}, \citenamefont {Feng}, \citenamefont {Lin}, \citenamefont {Villaos}, \citenamefont {Liu}, \citenamefont {Chen}, \citenamefont {Li}, \citenamefont {Mo}, \citenamefont {Chuang},\ and\ \citenamefont {Chiang}}]{hlevyack2021}%
  \BibitemOpen
  \bibfield  {author} {\bibinfo {author} {\bibfnamefont {J.~A.}\ \bibnamefont {Hlevyack}}, \bibinfo {author} {\bibfnamefont {L.-Y.}\ \bibnamefont {Feng}}, \bibinfo {author} {\bibfnamefont {M.-K.}\ \bibnamefont {Lin}}, \bibinfo {author} {\bibfnamefont {R.~A.~B.}\ \bibnamefont {Villaos}}, \bibinfo {author} {\bibfnamefont {R.-Y.}\ \bibnamefont {Liu}}, \bibinfo {author} {\bibfnamefont {P.}~\bibnamefont {Chen}}, \bibinfo {author} {\bibfnamefont {Y.}~\bibnamefont {Li}}, \bibinfo {author} {\bibfnamefont {S.-K.}\ \bibnamefont {Mo}}, \bibinfo {author} {\bibfnamefont {F.-C.}\ \bibnamefont {Chuang}}, \ and\ \bibinfo {author} {\bibfnamefont {T.~C.}\ \bibnamefont {Chiang}},\ }\href {\doibase 10.1038/s41699-021-00218-z} {\bibfield  {journal} {\bibinfo  {journal} {npj 2D Materials and Applications}\ }\textbf {\bibinfo {volume} {5}},\ \bibinfo {pages} {40} (\bibinfo {year} {2021})}\BibitemShut {NoStop}%
\bibitem [{\citenamefont {Deng}\ \emph {et~al.}(2019)\citenamefont {Deng}, \citenamefont {Yan}, \citenamefont {Yu}, \citenamefont {Li}, \citenamefont {Zhou}, \citenamefont {Zhang}, \citenamefont {Zhao}, \citenamefont {Miyamoto}, \citenamefont {Okuda}, \citenamefont {Duan}, \citenamefont {Wu}, \citenamefont {Zhong},\ and\ \citenamefont {Zhou}}]{deng2019}%
  \BibitemOpen
  \bibfield  {author} {\bibinfo {author} {\bibfnamefont {K.}~\bibnamefont {Deng}}, \bibinfo {author} {\bibfnamefont {M.}~\bibnamefont {Yan}}, \bibinfo {author} {\bibfnamefont {C.-P.}\ \bibnamefont {Yu}}, \bibinfo {author} {\bibfnamefont {J.}~\bibnamefont {Li}}, \bibinfo {author} {\bibfnamefont {X.}~\bibnamefont {Zhou}}, \bibinfo {author} {\bibfnamefont {K.}~\bibnamefont {Zhang}}, \bibinfo {author} {\bibfnamefont {Y.}~\bibnamefont {Zhao}}, \bibinfo {author} {\bibfnamefont {K.}~\bibnamefont {Miyamoto}}, \bibinfo {author} {\bibfnamefont {T.}~\bibnamefont {Okuda}}, \bibinfo {author} {\bibfnamefont {W.}~\bibnamefont {Duan}}, \bibinfo {author} {\bibfnamefont {Y.}~\bibnamefont {Wu}}, \bibinfo {author} {\bibfnamefont {X.}~\bibnamefont {Zhong}}, \ and\ \bibinfo {author} {\bibfnamefont {S.}~\bibnamefont {Zhou}},\ }\bibfield  {booktitle} {\emph {\bibinfo {booktitle} {Crossover from 2D metal to 3D Dirac semimetal in metallic PtTe2 films with local Rashba effect}},\ }\href {\doibase
  https://doi.org/10.1016/j.scib.2019.05.023} {\bibfield  {journal} {\bibinfo  {journal} {Science Bulletin}\ }\textbf {\bibinfo {volume} {64}},\ \bibinfo {pages} {1044} (\bibinfo {year} {2019})}\BibitemShut {NoStop}%
\bibitem [{\citenamefont {Lan}\ \emph {et~al.}(2021)\citenamefont {Lan}, \citenamefont {Li}, \citenamefont {Ho},\ and\ \citenamefont {Liu}}]{lan_2d_2021}%
  \BibitemOpen
  \bibfield  {author} {\bibinfo {author} {\bibfnamefont {C.}~\bibnamefont {Lan}}, \bibinfo {author} {\bibfnamefont {C.}~\bibnamefont {Li}}, \bibinfo {author} {\bibfnamefont {J.~C.}\ \bibnamefont {Ho}}, \ and\ \bibinfo {author} {\bibfnamefont {Y.}~\bibnamefont {Liu}},\ }\bibfield  {booktitle} {\emph {\bibinfo {booktitle} {Advanced Electronic Materials}},\ }\href {\doibase https://doi.org/10.1002/aelm.202000688} {\bibfield  {journal} {\bibinfo  {journal} {Advanced Electronic Materials}\ }\textbf {\bibinfo {volume} {7}},\ \bibinfo {pages} {2000688} (\bibinfo {year} {2021})}\BibitemShut {NoStop}%
\bibitem [{\citenamefont {Fridman}\ \emph {et~al.}(2025)\citenamefont {Fridman}, \citenamefont {Feld}, \citenamefont {Noah}, \citenamefont {Zalic}, \citenamefont {Markman}, \citenamefont {Devidas}, \citenamefont {Zur}, \citenamefont {Grynszpan}, \citenamefont {Gutfreund}, \citenamefont {Keren}, \citenamefont {Vakahi}, \citenamefont {Remennik}, \citenamefont {Watanabe}, \citenamefont {Taniguchi}, \citenamefont {Huber}, \citenamefont {Aleiner}, \citenamefont {Steinberg}, \citenamefont {Agam},\ and\ \citenamefont {Anahory}}]{fridman_anomalous_2025}%
  \BibitemOpen
  \bibfield  {author} {\bibinfo {author} {\bibfnamefont {N.}~\bibnamefont {Fridman}}, \bibinfo {author} {\bibfnamefont {T.~D.}\ \bibnamefont {Feld}}, \bibinfo {author} {\bibfnamefont {A.}~\bibnamefont {Noah}}, \bibinfo {author} {\bibfnamefont {A.}~\bibnamefont {Zalic}}, \bibinfo {author} {\bibfnamefont {M.}~\bibnamefont {Markman}}, \bibinfo {author} {\bibfnamefont {T.~R.}\ \bibnamefont {Devidas}}, \bibinfo {author} {\bibfnamefont {Y.}~\bibnamefont {Zur}}, \bibinfo {author} {\bibfnamefont {E.}~\bibnamefont {Grynszpan}}, \bibinfo {author} {\bibfnamefont {A.}~\bibnamefont {Gutfreund}}, \bibinfo {author} {\bibfnamefont {I.}~\bibnamefont {Keren}}, \bibinfo {author} {\bibfnamefont {A.}~\bibnamefont {Vakahi}}, \bibinfo {author} {\bibfnamefont {S.}~\bibnamefont {Remennik}}, \bibinfo {author} {\bibfnamefont {K.}~\bibnamefont {Watanabe}}, \bibinfo {author} {\bibfnamefont {T.}~\bibnamefont {Taniguchi}}, \bibinfo {author} {\bibfnamefont {M.~E.}\ \bibnamefont {Huber}}, \bibinfo {author} {\bibfnamefont {I.}~\bibnamefont
  {Aleiner}}, \bibinfo {author} {\bibfnamefont {H.}~\bibnamefont {Steinberg}}, \bibinfo {author} {\bibfnamefont {O.}~\bibnamefont {Agam}}, \ and\ \bibinfo {author} {\bibfnamefont {Y.}~\bibnamefont {Anahory}},\ }\href {\doibase 10.1038/s41467-025-57817-3} {\bibfield  {journal} {\bibinfo  {journal} {Nature Communications}\ }\textbf {\bibinfo {volume} {16}},\ \bibinfo {pages} {2696} (\bibinfo {year} {2025})}\BibitemShut {NoStop}%
\bibitem [{\citenamefont {Riley}\ \emph {et~al.}(2015)\citenamefont {Riley}, \citenamefont {Meevasana}, \citenamefont {Bawden}, \citenamefont {Asakawa}, \citenamefont {Takayama}, \citenamefont {Eknapakul}, \citenamefont {Kim}, \citenamefont {Hoesch}, \citenamefont {Mo}, \citenamefont {Takagi}, \citenamefont {Sasagawa}, \citenamefont {Bahramy},\ and\ \citenamefont {King}}]{riley_negative_2015}%
  \BibitemOpen
  \bibfield  {author} {\bibinfo {author} {\bibfnamefont {J.~M.}\ \bibnamefont {Riley}}, \bibinfo {author} {\bibfnamefont {W.}~\bibnamefont {Meevasana}}, \bibinfo {author} {\bibfnamefont {L.}~\bibnamefont {Bawden}}, \bibinfo {author} {\bibfnamefont {M.}~\bibnamefont {Asakawa}}, \bibinfo {author} {\bibfnamefont {T.}~\bibnamefont {Takayama}}, \bibinfo {author} {\bibfnamefont {T.}~\bibnamefont {Eknapakul}}, \bibinfo {author} {\bibfnamefont {T.~K.}\ \bibnamefont {Kim}}, \bibinfo {author} {\bibfnamefont {M.}~\bibnamefont {Hoesch}}, \bibinfo {author} {\bibfnamefont {S.~K.}\ \bibnamefont {Mo}}, \bibinfo {author} {\bibfnamefont {H.}~\bibnamefont {Takagi}}, \bibinfo {author} {\bibfnamefont {T.}~\bibnamefont {Sasagawa}}, \bibinfo {author} {\bibfnamefont {M.~S.}\ \bibnamefont {Bahramy}}, \ and\ \bibinfo {author} {\bibfnamefont {P.~D.~C.}\ \bibnamefont {King}},\ }\href {\doibase 10.1038/nnano.2015.217} {\bibfield  {journal} {\bibinfo  {journal} {Nature Nanotechnology}\ }\textbf {\bibinfo {volume} {10}},\ \bibinfo {pages}
  {1043} (\bibinfo {year} {2015})}\BibitemShut {NoStop}%
\bibitem [{\citenamefont {Xiao}\ \emph {et~al.}(2012)\citenamefont {Xiao}, \citenamefont {Liu}, \citenamefont {Feng}, \citenamefont {Xu},\ and\ \citenamefont {Yao}}]{xaio_coupled_2012}%
  \BibitemOpen
  \bibfield  {author} {\bibinfo {author} {\bibfnamefont {D.}~\bibnamefont {Xiao}}, \bibinfo {author} {\bibfnamefont {G.-B.}\ \bibnamefont {Liu}}, \bibinfo {author} {\bibfnamefont {W.}~\bibnamefont {Feng}}, \bibinfo {author} {\bibfnamefont {X.}~\bibnamefont {Xu}}, \ and\ \bibinfo {author} {\bibfnamefont {W.}~\bibnamefont {Yao}},\ }\href {\doibase 10.1103/PhysRevLett.108.196802} {\bibfield  {journal} {\bibinfo  {journal} {Phys. Rev. Lett.}\ }\textbf {\bibinfo {volume} {108}},\ \bibinfo {pages} {196802} (\bibinfo {year} {2012})}\BibitemShut {NoStop}%
\bibitem [{\citenamefont {Latzke}\ \emph {et~al.}(2015)\citenamefont {Latzke}, \citenamefont {Zhang}, \citenamefont {Suslu}, \citenamefont {Chang}, \citenamefont {Lin}, \citenamefont {Jeng}, \citenamefont {Tongay}, \citenamefont {Wu}, \citenamefont {Bansil},\ and\ \citenamefont {Lanzara}}]{latzke_electronic_2015}%
  \BibitemOpen
  \bibfield  {author} {\bibinfo {author} {\bibfnamefont {D.~W.}\ \bibnamefont {Latzke}}, \bibinfo {author} {\bibfnamefont {W.}~\bibnamefont {Zhang}}, \bibinfo {author} {\bibfnamefont {A.}~\bibnamefont {Suslu}}, \bibinfo {author} {\bibfnamefont {T.-R.}\ \bibnamefont {Chang}}, \bibinfo {author} {\bibfnamefont {H.}~\bibnamefont {Lin}}, \bibinfo {author} {\bibfnamefont {H.-T.}\ \bibnamefont {Jeng}}, \bibinfo {author} {\bibfnamefont {S.}~\bibnamefont {Tongay}}, \bibinfo {author} {\bibfnamefont {J.}~\bibnamefont {Wu}}, \bibinfo {author} {\bibfnamefont {A.}~\bibnamefont {Bansil}}, \ and\ \bibinfo {author} {\bibfnamefont {A.}~\bibnamefont {Lanzara}},\ }\href {\doibase 10.1103/PhysRevB.91.235202} {\bibfield  {journal} {\bibinfo  {journal} {Phys. Rev. B}\ }\textbf {\bibinfo {volume} {91}},\ \bibinfo {pages} {235202} (\bibinfo {year} {2015})}\BibitemShut {NoStop}%
\bibitem [{\citenamefont {Kim}\ \emph {et~al.}(2016)\citenamefont {Kim}, \citenamefont {Rhim}, \citenamefont {Kim}, \citenamefont {Kim},\ and\ \citenamefont {Park}}]{kim_determination_2016}%
  \BibitemOpen
  \bibfield  {author} {\bibinfo {author} {\bibfnamefont {B.~S.}\ \bibnamefont {Kim}}, \bibinfo {author} {\bibfnamefont {J.-W.}\ \bibnamefont {Rhim}}, \bibinfo {author} {\bibfnamefont {B.}~\bibnamefont {Kim}}, \bibinfo {author} {\bibfnamefont {C.}~\bibnamefont {Kim}}, \ and\ \bibinfo {author} {\bibfnamefont {S.~R.}\ \bibnamefont {Park}},\ }\href {\doibase 10.1038/srep36389} {\bibfield  {journal} {\bibinfo  {journal} {Scientific Reports}\ }\textbf {\bibinfo {volume} {6}},\ \bibinfo {pages} {36389} (\bibinfo {year} {2016})}\BibitemShut {NoStop}%
\bibitem [{\citenamefont {Alarab}\ \emph {et~al.}(2023)\citenamefont {Alarab}, \citenamefont {Minar}, \citenamefont {Constantinou}, \citenamefont {Nafday}, \citenamefont {Schmitt}, \citenamefont {Wang},\ and\ \citenamefont {Strocov}}]{alarab_k_2023}%
  \BibitemOpen
  \bibfield  {author} {\bibinfo {author} {\bibfnamefont {F.}~\bibnamefont {Alarab}}, \bibinfo {author} {\bibfnamefont {J.}~\bibnamefont {Minar}}, \bibinfo {author} {\bibfnamefont {P.}~\bibnamefont {Constantinou}}, \bibinfo {author} {\bibfnamefont {D.}~\bibnamefont {Nafday}}, \bibinfo {author} {\bibfnamefont {T.}~\bibnamefont {Schmitt}}, \bibinfo {author} {\bibfnamefont {X.}~\bibnamefont {Wang}}, \ and\ \bibinfo {author} {\bibfnamefont {V.~N.}\ \bibnamefont {Strocov}},\ }\href@noop {} {\bibfield  {journal} {\bibinfo  {journal} {arXiv:2310.11317}\ } (\bibinfo {year} {2023})}\BibitemShut {NoStop}%
\bibitem [{\citenamefont {Ernandes}\ \emph {et~al.}(2021)\citenamefont {Ernandes}, \citenamefont {Khalil}, \citenamefont {Almabrouk}, \citenamefont {Pierucci}, \citenamefont {Zheng}, \citenamefont {Avila}, \citenamefont {Dudin}, \citenamefont {Chaste}, \citenamefont {Oehler}, \citenamefont {Pala}, \citenamefont {Bisti}, \citenamefont {Brul{\'e}}, \citenamefont {Lhuillier}, \citenamefont {Pan},\ and\ \citenamefont {Ouerghi}}]{ernandes_indirect_2021}%
  \BibitemOpen
  \bibfield  {author} {\bibinfo {author} {\bibfnamefont {C.}~\bibnamefont {Ernandes}}, \bibinfo {author} {\bibfnamefont {L.}~\bibnamefont {Khalil}}, \bibinfo {author} {\bibfnamefont {H.}~\bibnamefont {Almabrouk}}, \bibinfo {author} {\bibfnamefont {D.}~\bibnamefont {Pierucci}}, \bibinfo {author} {\bibfnamefont {B.}~\bibnamefont {Zheng}}, \bibinfo {author} {\bibfnamefont {J.}~\bibnamefont {Avila}}, \bibinfo {author} {\bibfnamefont {P.}~\bibnamefont {Dudin}}, \bibinfo {author} {\bibfnamefont {J.}~\bibnamefont {Chaste}}, \bibinfo {author} {\bibfnamefont {F.}~\bibnamefont {Oehler}}, \bibinfo {author} {\bibfnamefont {M.}~\bibnamefont {Pala}}, \bibinfo {author} {\bibfnamefont {F.}~\bibnamefont {Bisti}}, \bibinfo {author} {\bibfnamefont {T.}~\bibnamefont {Brul{\'e}}}, \bibinfo {author} {\bibfnamefont {E.}~\bibnamefont {Lhuillier}}, \bibinfo {author} {\bibfnamefont {A.}~\bibnamefont {Pan}}, \ and\ \bibinfo {author} {\bibfnamefont {A.}~\bibnamefont {Ouerghi}},\ }\href {\doibase 10.1038/s41699-020-00187-9} {\bibfield
  {journal} {\bibinfo  {journal} {npj 2D Materials and Applications}\ }\textbf {\bibinfo {volume} {5}},\ \bibinfo {pages} {7} (\bibinfo {year} {2021})}\BibitemShut {NoStop}%
\bibitem [{\citenamefont {Bychkov}\ and\ \citenamefont {Rashba}(1984)}]{Rashba84}%
  \BibitemOpen
  \bibfield  {author} {\bibinfo {author} {\bibfnamefont {Y.~A.}\ \bibnamefont {Bychkov}}\ and\ \bibinfo {author} {\bibfnamefont {E.~I.}\ \bibnamefont {Rashba}},\ }\bibfield  {booktitle} {\emph {\bibinfo {booktitle} {Journal of Physics C: Solid State Physics}},\ }\href {\doibase 10.1088/0022-3719/17/33/015} {\ \textbf {\bibinfo {volume} {17}},\ \bibinfo {pages} {6039} (\bibinfo {year} {1984})}\BibitemShut {NoStop}%
\bibitem [{\citenamefont {Wang}\ \emph {et~al.}(2017)\citenamefont {Wang}, \citenamefont {Huang}, \citenamefont {Lin}, \citenamefont {Cui}, \citenamefont {Chen}, \citenamefont {Zhu}, \citenamefont {Liu}, \citenamefont {Zeng}, \citenamefont {Zhou}, \citenamefont {Yu}, \citenamefont {Wang}, \citenamefont {He}, \citenamefont {Tsang}, \citenamefont {Gao}, \citenamefont {Suenaga}, \citenamefont {Ma}, \citenamefont {Yang}, \citenamefont {Lu}, \citenamefont {Yu}, \citenamefont {Teo}, \citenamefont {Liu},\ and\ \citenamefont {Liu}}]{wang_high_2017}%
  \BibitemOpen
  \bibfield  {author} {\bibinfo {author} {\bibfnamefont {H.}~\bibnamefont {Wang}}, \bibinfo {author} {\bibfnamefont {X.}~\bibnamefont {Huang}}, \bibinfo {author} {\bibfnamefont {J.}~\bibnamefont {Lin}}, \bibinfo {author} {\bibfnamefont {J.}~\bibnamefont {Cui}}, \bibinfo {author} {\bibfnamefont {Y.}~\bibnamefont {Chen}}, \bibinfo {author} {\bibfnamefont {C.}~\bibnamefont {Zhu}}, \bibinfo {author} {\bibfnamefont {F.}~\bibnamefont {Liu}}, \bibinfo {author} {\bibfnamefont {Q.}~\bibnamefont {Zeng}}, \bibinfo {author} {\bibfnamefont {J.}~\bibnamefont {Zhou}}, \bibinfo {author} {\bibfnamefont {P.}~\bibnamefont {Yu}}, \bibinfo {author} {\bibfnamefont {X.}~\bibnamefont {Wang}}, \bibinfo {author} {\bibfnamefont {H.}~\bibnamefont {He}}, \bibinfo {author} {\bibfnamefont {S.~H.}\ \bibnamefont {Tsang}}, \bibinfo {author} {\bibfnamefont {W.}~\bibnamefont {Gao}}, \bibinfo {author} {\bibfnamefont {K.}~\bibnamefont {Suenaga}}, \bibinfo {author} {\bibfnamefont {F.}~\bibnamefont {Ma}}, \bibinfo {author} {\bibfnamefont
  {C.}~\bibnamefont {Yang}}, \bibinfo {author} {\bibfnamefont {L.}~\bibnamefont {Lu}}, \bibinfo {author} {\bibfnamefont {T.}~\bibnamefont {Yu}}, \bibinfo {author} {\bibfnamefont {E.~H.~T.}\ \bibnamefont {Teo}}, \bibinfo {author} {\bibfnamefont {G.}~\bibnamefont {Liu}}, \ and\ \bibinfo {author} {\bibfnamefont {Z.}~\bibnamefont {Liu}},\ }\href {\doibase 10.1038/s41467-017-00427-5} {\bibfield  {journal} {\bibinfo  {journal} {Nature Communications}\ }\textbf {\bibinfo {volume} {8}},\ \bibinfo {pages} {394} (\bibinfo {year} {2017})}\BibitemShut {NoStop}%
\bibitem [{\citenamefont {Kang}\ \emph {et~al.}(2013)\citenamefont {Kang}, \citenamefont {Tongay}, \citenamefont {Zhou}, \citenamefont {Li},\ and\ \citenamefont {Wu}}]{kang_band_2013}%
  \BibitemOpen
  \bibfield  {author} {\bibinfo {author} {\bibfnamefont {J.}~\bibnamefont {Kang}}, \bibinfo {author} {\bibfnamefont {S.}~\bibnamefont {Tongay}}, \bibinfo {author} {\bibfnamefont {J.}~\bibnamefont {Zhou}}, \bibinfo {author} {\bibfnamefont {J.}~\bibnamefont {Li}}, \ and\ \bibinfo {author} {\bibfnamefont {J.}~\bibnamefont {Wu}},\ }\href {\doibase 10.1063/1.4774090} {\bibfield  {journal} {\bibinfo  {journal} {Applied Physics Letters}\ }\textbf {\bibinfo {volume} {102}},\ \bibinfo {pages} {012111} (\bibinfo {year} {2013})},\ \Eprint {http://arxiv.org/abs/https://pubs.aip.org/aip/apl/article-pdf/doi/10.1063/1.4774090/14263519/012111\_1\_online.pdf} {https://pubs.aip.org/aip/apl/article-pdf/doi/10.1063/1.4774090/14263519/012111\_1\_online.pdf} \BibitemShut {NoStop}%
\bibitem [{\citenamefont {Clark}\ \emph {et~al.}(2019)\citenamefont {Clark}, \citenamefont {Mazzola}, \citenamefont {Markovi\'c}, \citenamefont {Riley}, \citenamefont {Feng}, \citenamefont {Yang}, \citenamefont {Sumida}, \citenamefont {Okuda}, \citenamefont {Fujii}, \citenamefont {Vobornik}, \citenamefont {Kim}, \citenamefont {Okawa}, \citenamefont {Sasagawa}, \citenamefont {Bahramy},\ and\ \citenamefont {King}}]{ClarkGeneral2019}%
  \BibitemOpen
  \bibfield  {author} {\bibinfo {author} {\bibfnamefont {O.~J.}\ \bibnamefont {Clark}}, \bibinfo {author} {\bibfnamefont {F.}~\bibnamefont {Mazzola}}, \bibinfo {author} {\bibfnamefont {I.}~\bibnamefont {Markovi\'c}}, \bibinfo {author} {\bibfnamefont {J.~M.}\ \bibnamefont {Riley}}, \bibinfo {author} {\bibfnamefont {J.}~\bibnamefont {Feng}}, \bibinfo {author} {\bibfnamefont {B.-J.}\ \bibnamefont {Yang}}, \bibinfo {author} {\bibfnamefont {K.}~\bibnamefont {Sumida}}, \bibinfo {author} {\bibfnamefont {T.}~\bibnamefont {Okuda}}, \bibinfo {author} {\bibfnamefont {J.}~\bibnamefont {Fujii}}, \bibinfo {author} {\bibfnamefont {I.}~\bibnamefont {Vobornik}}, \bibinfo {author} {\bibfnamefont {T.~K.}\ \bibnamefont {Kim}}, \bibinfo {author} {\bibfnamefont {K.}~\bibnamefont {Okawa}}, \bibinfo {author} {\bibfnamefont {T.}~\bibnamefont {Sasagawa}}, \bibinfo {author} {\bibfnamefont {M.~S.}\ \bibnamefont {Bahramy}}, \ and\ \bibinfo {author} {\bibfnamefont {P.~D.~C.}\ \bibnamefont {King}},\ }\href@noop {} {\bibfield  {journal}
  {\bibinfo  {journal} {Electronic Structure}\ }\textbf {\bibinfo {volume} {1}},\ \bibinfo {pages} {014002} (\bibinfo {year} {2019})}\BibitemShut {NoStop}%
\bibitem [{\citenamefont {Huang}\ \emph {et~al.}(2016)\citenamefont {Huang}, \citenamefont {Zhou},\ and\ \citenamefont {Duan}}]{huang_type_2016}%
  \BibitemOpen
  \bibfield  {author} {\bibinfo {author} {\bibfnamefont {H.}~\bibnamefont {Huang}}, \bibinfo {author} {\bibfnamefont {S.}~\bibnamefont {Zhou}}, \ and\ \bibinfo {author} {\bibfnamefont {W.}~\bibnamefont {Duan}},\ }\href {\doibase 10.1103/PhysRevB.94.121117} {\bibfield  {journal} {\bibinfo  {journal} {Phys. Rev. B}\ }\textbf {\bibinfo {volume} {94}},\ \bibinfo {pages} {121117} (\bibinfo {year} {2016})}\BibitemShut {NoStop}%
\bibitem [{\citenamefont {Yan}\ \emph {et~al.}(2017)\citenamefont {Yan}, \citenamefont {Huang}, \citenamefont {Zhang}, \citenamefont {Wang}, \citenamefont {Yao}, \citenamefont {Deng}, \citenamefont {Wan}, \citenamefont {Zhang}, \citenamefont {Arita}, \citenamefont {Yang}, \citenamefont {Sun}, \citenamefont {Yao}, \citenamefont {Wu}, \citenamefont {Fan}, \citenamefont {Duan},\ and\ \citenamefont {Zhou}}]{yan_topological_2017}%
  \BibitemOpen
  \bibfield  {author} {\bibinfo {author} {\bibfnamefont {M.}~\bibnamefont {Yan}}, \bibinfo {author} {\bibfnamefont {H.}~\bibnamefont {Huang}}, \bibinfo {author} {\bibfnamefont {K.}~\bibnamefont {Zhang}}, \bibinfo {author} {\bibfnamefont {E.}~\bibnamefont {Wang}}, \bibinfo {author} {\bibfnamefont {W.}~\bibnamefont {Yao}}, \bibinfo {author} {\bibfnamefont {K.}~\bibnamefont {Deng}}, \bibinfo {author} {\bibfnamefont {G.}~\bibnamefont {Wan}}, \bibinfo {author} {\bibfnamefont {H.}~\bibnamefont {Zhang}}, \bibinfo {author} {\bibfnamefont {M.}~\bibnamefont {Arita}}, \bibinfo {author} {\bibfnamefont {H.}~\bibnamefont {Yang}}, \bibinfo {author} {\bibfnamefont {Z.}~\bibnamefont {Sun}}, \bibinfo {author} {\bibfnamefont {H.}~\bibnamefont {Yao}}, \bibinfo {author} {\bibfnamefont {Y.}~\bibnamefont {Wu}}, \bibinfo {author} {\bibfnamefont {S.}~\bibnamefont {Fan}}, \bibinfo {author} {\bibfnamefont {W.}~\bibnamefont {Duan}}, \ and\ \bibinfo {author} {\bibfnamefont {S.}~\bibnamefont {Zhou}},\ }\href {\doibase
  10.1038/s41467-017-00280-6} {\bibfield  {journal} {\bibinfo  {journal} {Nature Communications}\ }\textbf {\bibinfo {volume} {8}},\ \bibinfo {pages} {257} (\bibinfo {year} {2017})}\BibitemShut {NoStop}%
\bibitem [{\citenamefont {Noh}\ \emph {et~al.}(2017)\citenamefont {Noh}, \citenamefont {Jeong}, \citenamefont {Cho}, \citenamefont {Kim}, \citenamefont {Min},\ and\ \citenamefont {Park}}]{noh_experimental_2017}%
  \BibitemOpen
  \bibfield  {author} {\bibinfo {author} {\bibfnamefont {H.-J.}\ \bibnamefont {Noh}}, \bibinfo {author} {\bibfnamefont {J.}~\bibnamefont {Jeong}}, \bibinfo {author} {\bibfnamefont {E.-J.}\ \bibnamefont {Cho}}, \bibinfo {author} {\bibfnamefont {K.}~\bibnamefont {Kim}}, \bibinfo {author} {\bibfnamefont {B.~I.}\ \bibnamefont {Min}}, \ and\ \bibinfo {author} {\bibfnamefont {B.-G.}\ \bibnamefont {Park}},\ }\href {\doibase 10.1103/PhysRevLett.119.016401} {\bibfield  {journal} {\bibinfo  {journal} {Phys. Rev. Lett.}\ }\textbf {\bibinfo {volume} {119}},\ \bibinfo {pages} {016401} (\bibinfo {year} {2017})}\BibitemShut {NoStop}%
\bibitem [{\citenamefont {Li}\ \emph {et~al.}(2017)\citenamefont {Li}, \citenamefont {Xia}, \citenamefont {Ekahana}, \citenamefont {Kumar}, \citenamefont {Jiang}, \citenamefont {Yang}, \citenamefont {Chen}, \citenamefont {Liu}, \citenamefont {Yan}, \citenamefont {Felser}, \citenamefont {Li}, \citenamefont {Liu},\ and\ \citenamefont {Chen}}]{li_topological_2017}%
  \BibitemOpen
  \bibfield  {author} {\bibinfo {author} {\bibfnamefont {Y.}~\bibnamefont {Li}}, \bibinfo {author} {\bibfnamefont {Y.}~\bibnamefont {Xia}}, \bibinfo {author} {\bibfnamefont {S.~A.}\ \bibnamefont {Ekahana}}, \bibinfo {author} {\bibfnamefont {N.}~\bibnamefont {Kumar}}, \bibinfo {author} {\bibfnamefont {J.}~\bibnamefont {Jiang}}, \bibinfo {author} {\bibfnamefont {L.}~\bibnamefont {Yang}}, \bibinfo {author} {\bibfnamefont {C.}~\bibnamefont {Chen}}, \bibinfo {author} {\bibfnamefont {C.}~\bibnamefont {Liu}}, \bibinfo {author} {\bibfnamefont {B.}~\bibnamefont {Yan}}, \bibinfo {author} {\bibfnamefont {C.}~\bibnamefont {Felser}}, \bibinfo {author} {\bibfnamefont {G.}~\bibnamefont {Li}}, \bibinfo {author} {\bibfnamefont {Z.}~\bibnamefont {Liu}}, \ and\ \bibinfo {author} {\bibfnamefont {Y.}~\bibnamefont {Chen}},\ }\href {\doibase 10.1103/PhysRevMaterials.1.074202} {\bibfield  {journal} {\bibinfo  {journal} {Phys. Rev. Materials}\ }\textbf {\bibinfo {volume} {1}},\ \bibinfo {pages} {074202} (\bibinfo {year}
  {2017})}\BibitemShut {NoStop}%
\bibitem [{\citenamefont {Ghosh}\ \emph {et~al.}(2019)\citenamefont {Ghosh}, \citenamefont {Mondal}, \citenamefont {Kuo}, \citenamefont {Lue}, \citenamefont {Nayak}, \citenamefont {Fujii}, \citenamefont {Vobornik}, \citenamefont {Politano},\ and\ \citenamefont {Agarwal}}]{ghosh_observation_2019}%
  \BibitemOpen
  \bibfield  {author} {\bibinfo {author} {\bibfnamefont {B.}~\bibnamefont {Ghosh}}, \bibinfo {author} {\bibfnamefont {D.}~\bibnamefont {Mondal}}, \bibinfo {author} {\bibfnamefont {C.-N.}\ \bibnamefont {Kuo}}, \bibinfo {author} {\bibfnamefont {C.~S.}\ \bibnamefont {Lue}}, \bibinfo {author} {\bibfnamefont {J.}~\bibnamefont {Nayak}}, \bibinfo {author} {\bibfnamefont {J.}~\bibnamefont {Fujii}}, \bibinfo {author} {\bibfnamefont {I.}~\bibnamefont {Vobornik}}, \bibinfo {author} {\bibfnamefont {A.}~\bibnamefont {Politano}}, \ and\ \bibinfo {author} {\bibfnamefont {A.}~\bibnamefont {Agarwal}},\ }\href {\doibase 10.1103/PhysRevB.100.195134} {\bibfield  {journal} {\bibinfo  {journal} {Phys. Rev. B}\ }\textbf {\bibinfo {volume} {100}},\ \bibinfo {pages} {195134} (\bibinfo {year} {2019})}\BibitemShut {NoStop}%
\bibitem [{\citenamefont {Jiang}\ \emph {et~al.}(2020)\citenamefont {Jiang}, \citenamefont {Lee}, \citenamefont {Fei}, \citenamefont {Song}, \citenamefont {Vescovo}, \citenamefont {Kaznatcheev}, \citenamefont {Walker},\ and\ \citenamefont {Ahn}}]{jiang_comprehensive_2020}%
  \BibitemOpen
  \bibfield  {author} {\bibinfo {author} {\bibfnamefont {J.}~\bibnamefont {Jiang}}, \bibinfo {author} {\bibfnamefont {S.}~\bibnamefont {Lee}}, \bibinfo {author} {\bibfnamefont {F.}~\bibnamefont {Fei}}, \bibinfo {author} {\bibfnamefont {F.}~\bibnamefont {Song}}, \bibinfo {author} {\bibfnamefont {E.}~\bibnamefont {Vescovo}}, \bibinfo {author} {\bibfnamefont {K.}~\bibnamefont {Kaznatcheev}}, \bibinfo {author} {\bibfnamefont {F.~J.}\ \bibnamefont {Walker}}, \ and\ \bibinfo {author} {\bibfnamefont {C.~H.}\ \bibnamefont {Ahn}},\ }\href {\doibase 10.1063/5.0011549} {\bibfield  {journal} {\bibinfo  {journal} {APL Materials}\ }\textbf {\bibinfo {volume} {8}},\ \bibinfo {pages} {061106} (\bibinfo {year} {2020})},\ \Eprint {http://arxiv.org/abs/https://pubs.aip.org/aip/apm/article-pdf/doi/10.1063/5.0011549/14564647/061106\_1\_online.pdf} {https://pubs.aip.org/aip/apm/article-pdf/doi/10.1063/5.0011549/14564647/061106\_1\_online.pdf} \BibitemShut {NoStop}%
\bibitem [{\citenamefont {Mukherjee}\ \emph {et~al.}(2020)\citenamefont {Mukherjee}, \citenamefont {Jung}, \citenamefont {Weber}, \citenamefont {Xu}, \citenamefont {Qian}, \citenamefont {Xu}, \citenamefont {Biswas}, \citenamefont {Kim}, \citenamefont {Chapon}, \citenamefont {Watson}, \citenamefont {Neaton},\ and\ \citenamefont {Cacho}}]{mukherjee_fermi_2020}%
  \BibitemOpen
  \bibfield  {author} {\bibinfo {author} {\bibfnamefont {S.}~\bibnamefont {Mukherjee}}, \bibinfo {author} {\bibfnamefont {S.~W.}\ \bibnamefont {Jung}}, \bibinfo {author} {\bibfnamefont {S.~F.}\ \bibnamefont {Weber}}, \bibinfo {author} {\bibfnamefont {C.}~\bibnamefont {Xu}}, \bibinfo {author} {\bibfnamefont {D.}~\bibnamefont {Qian}}, \bibinfo {author} {\bibfnamefont {X.}~\bibnamefont {Xu}}, \bibinfo {author} {\bibfnamefont {P.~K.}\ \bibnamefont {Biswas}}, \bibinfo {author} {\bibfnamefont {T.~K.}\ \bibnamefont {Kim}}, \bibinfo {author} {\bibfnamefont {L.~C.}\ \bibnamefont {Chapon}}, \bibinfo {author} {\bibfnamefont {M.~D.}\ \bibnamefont {Watson}}, \bibinfo {author} {\bibfnamefont {J.~B.}\ \bibnamefont {Neaton}}, \ and\ \bibinfo {author} {\bibfnamefont {C.}~\bibnamefont {Cacho}},\ }\href {\doibase 10.1038/s41598-020-69926-8} {\bibfield  {journal} {\bibinfo  {journal} {Scientific Reports}\ }\textbf {\bibinfo {volume} {10}},\ \bibinfo {pages} {12957} (\bibinfo {year} {2020})}\BibitemShut {NoStop}%
\bibitem [{\citenamefont {Nurmamat}\ \emph {et~al.}(2021)\citenamefont {Nurmamat}, \citenamefont {Eremeev}, \citenamefont {Wang}, \citenamefont {Yoshikawa}, \citenamefont {Kono}, \citenamefont {Kakoki}, \citenamefont {Muro}, \citenamefont {Jiang}, \citenamefont {Sun}, \citenamefont {Ye},\ and\ \citenamefont {Kimura}}]{nurmamat_bulk_2021}%
  \BibitemOpen
  \bibfield  {author} {\bibinfo {author} {\bibfnamefont {M.}~\bibnamefont {Nurmamat}}, \bibinfo {author} {\bibfnamefont {S.~V.}\ \bibnamefont {Eremeev}}, \bibinfo {author} {\bibfnamefont {X.}~\bibnamefont {Wang}}, \bibinfo {author} {\bibfnamefont {T.}~\bibnamefont {Yoshikawa}}, \bibinfo {author} {\bibfnamefont {T.}~\bibnamefont {Kono}}, \bibinfo {author} {\bibfnamefont {M.}~\bibnamefont {Kakoki}}, \bibinfo {author} {\bibfnamefont {T.}~\bibnamefont {Muro}}, \bibinfo {author} {\bibfnamefont {Q.}~\bibnamefont {Jiang}}, \bibinfo {author} {\bibfnamefont {Z.}~\bibnamefont {Sun}}, \bibinfo {author} {\bibfnamefont {M.}~\bibnamefont {Ye}}, \ and\ \bibinfo {author} {\bibfnamefont {A.}~\bibnamefont {Kimura}},\ }\href {\doibase 10.1103/PhysRevB.104.155133} {\bibfield  {journal} {\bibinfo  {journal} {Phys. Rev. B}\ }\textbf {\bibinfo {volume} {104}},\ \bibinfo {pages} {155133} (\bibinfo {year} {2021})}\BibitemShut {NoStop}%
\bibitem [{\citenamefont {Chakraborty}\ \emph {et~al.}(2023)\citenamefont {Chakraborty}, \citenamefont {Fujii}, \citenamefont {Kuo}, \citenamefont {Lue}, \citenamefont {Politano}, \citenamefont {Vobornik},\ and\ \citenamefont {Agarwal}}]{chakraborty_observation_2023}%
  \BibitemOpen
  \bibfield  {author} {\bibinfo {author} {\bibfnamefont {A.}~\bibnamefont {Chakraborty}}, \bibinfo {author} {\bibfnamefont {J.}~\bibnamefont {Fujii}}, \bibinfo {author} {\bibfnamefont {C.-N.}\ \bibnamefont {Kuo}}, \bibinfo {author} {\bibfnamefont {C.~S.}\ \bibnamefont {Lue}}, \bibinfo {author} {\bibfnamefont {A.}~\bibnamefont {Politano}}, \bibinfo {author} {\bibfnamefont {I.}~\bibnamefont {Vobornik}}, \ and\ \bibinfo {author} {\bibfnamefont {A.}~\bibnamefont {Agarwal}},\ }\href {\doibase 10.1103/PhysRevB.107.085406} {\bibfield  {journal} {\bibinfo  {journal} {Phys. Rev. B}\ }\textbf {\bibinfo {volume} {107}},\ \bibinfo {pages} {085406} (\bibinfo {year} {2023})}\BibitemShut {NoStop}%
\bibitem [{\citenamefont {Kimar}\ \emph {et~al.}()\citenamefont {Kimar}, \citenamefont {Kumar}, \citenamefont {Yenugonda}, \citenamefont {Oishi}, \citenamefont {Nayak}, \citenamefont {Chen}, \citenamefont {Singh}, \citenamefont {Onimaru}, \citenamefont {Shimura}, \citenamefont {Ideta},\ and\ \citenamefont {Shimada}}]{kumar_electronic_2024}%
  \BibitemOpen
  \bibfield  {author} {\bibinfo {author} {\bibfnamefont {Y.}~\bibnamefont {Kimar}}, \bibinfo {author} {\bibfnamefont {S.}~\bibnamefont {Kumar}}, \bibinfo {author} {\bibfnamefont {V.}~\bibnamefont {Yenugonda}}, \bibinfo {author} {\bibfnamefont {R.}~\bibnamefont {Oishi}}, \bibinfo {author} {\bibfnamefont {J.}~\bibnamefont {Nayak}}, \bibinfo {author} {\bibfnamefont {C.}~\bibnamefont {Chen}}, \bibinfo {author} {\bibfnamefont {R.~P.}\ \bibnamefont {Singh}}, \bibinfo {author} {\bibfnamefont {T.}~\bibnamefont {Onimaru}}, \bibinfo {author} {\bibfnamefont {Y.}~\bibnamefont {Shimura}}, \bibinfo {author} {\bibfnamefont {S.}~\bibnamefont {Ideta}}, \ and\ \bibinfo {author} {\bibfnamefont {K.}~\bibnamefont {Shimada}},\ }\href@noop {} {\bibinfo  {journal} {arXiv:2408.15500}\ }\BibitemShut {NoStop}%
\bibitem [{\citenamefont {Ferreira}\ \emph {et~al.}(2021)\citenamefont {Ferreira}, \citenamefont {Manesco}, \citenamefont {Dorini}, \citenamefont {Correa}, \citenamefont {Weber}, \citenamefont {Machado},\ and\ \citenamefont {Eleno}}]{ferreira_strain_2021}%
  \BibitemOpen
\bibfield  {journal} {  }\bibfield  {author} {\bibinfo {author} {\bibfnamefont {P.~P.}\ \bibnamefont {Ferreira}}, \bibinfo {author} {\bibfnamefont {A.~L.~R.}\ \bibnamefont {Manesco}}, \bibinfo {author} {\bibfnamefont {T.~T.}\ \bibnamefont {Dorini}}, \bibinfo {author} {\bibfnamefont {L.~E.}\ \bibnamefont {Correa}}, \bibinfo {author} {\bibfnamefont {G.}~\bibnamefont {Weber}}, \bibinfo {author} {\bibfnamefont {A.~J.~S.}\ \bibnamefont {Machado}}, \ and\ \bibinfo {author} {\bibfnamefont {L.~T.~F.}\ \bibnamefont {Eleno}},\ }\href {\doibase 10.1103/PhysRevB.103.125134} {\bibfield  {journal} {\bibinfo  {journal} {Phys. Rev. B}\ }\textbf {\bibinfo {volume} {103}},\ \bibinfo {pages} {125134} (\bibinfo {year} {2021})}\BibitemShut {NoStop}%
\bibitem [{\citenamefont {Nicholson}\ \emph {et~al.}(2021)\citenamefont {Nicholson}, \citenamefont {Rumo}, \citenamefont {Pulkkinen}, \citenamefont {Kremer}, \citenamefont {Salzmann}, \citenamefont {Mottas}, \citenamefont {Hildebrand}, \citenamefont {Jaouen}, \citenamefont {Kim}, \citenamefont {Mukherjee}, \citenamefont {Ma}, \citenamefont {Muntwiler}, \citenamefont {von Rohr}, \citenamefont {Cacho},\ and\ \citenamefont {Monney}}]{nicholson_uniaxial_2021}%
  \BibitemOpen
  \bibfield  {author} {\bibinfo {author} {\bibfnamefont {C.}~\bibnamefont {Nicholson}}, \bibinfo {author} {\bibfnamefont {M.}~\bibnamefont {Rumo}}, \bibinfo {author} {\bibfnamefont {A.}~\bibnamefont {Pulkkinen}}, \bibinfo {author} {\bibfnamefont {G.}~\bibnamefont {Kremer}}, \bibinfo {author} {\bibfnamefont {B.}~\bibnamefont {Salzmann}}, \bibinfo {author} {\bibfnamefont {M.-L.}\ \bibnamefont {Mottas}}, \bibinfo {author} {\bibfnamefont {B.}~\bibnamefont {Hildebrand}}, \bibinfo {author} {\bibfnamefont {T.}~\bibnamefont {Jaouen}}, \bibinfo {author} {\bibfnamefont {T.~K.}\ \bibnamefont {Kim}}, \bibinfo {author} {\bibfnamefont {S.}~\bibnamefont {Mukherjee}}, \bibinfo {author} {\bibfnamefont {K.}~\bibnamefont {Ma}}, \bibinfo {author} {\bibfnamefont {M.}~\bibnamefont {Muntwiler}}, \bibinfo {author} {\bibfnamefont {F.~O.}\ \bibnamefont {von Rohr}}, \bibinfo {author} {\bibfnamefont {C.}~\bibnamefont {Cacho}}, \ and\ \bibinfo {author} {\bibfnamefont {C.}~\bibnamefont {Monney}},\ }\href@noop {} {\bibfield  {journal}
  {\bibinfo  {journal} {Communications Materials}\ }\textbf {\bibinfo {volume} {2}} (\bibinfo {year} {2021})}\BibitemShut {NoStop}%
\bibitem [{\citenamefont {Tokura}(2022)}]{tokura_quantum_2022}%
  \BibitemOpen
  \bibfield  {author} {\bibinfo {author} {\bibfnamefont {Y.}~\bibnamefont {Tokura}},\ }\href {\doibase 10.1038/s41563-022-01339-6} {\bibfield  {journal} {\bibinfo  {journal} {Nature Materials}\ }\textbf {\bibinfo {volume} {21}},\ \bibinfo {pages} {971} (\bibinfo {year} {2022})}\BibitemShut {NoStop}%
\bibitem [{\citenamefont {Kang}\ \emph {et~al.}(2017)\citenamefont {Kang}, \citenamefont {Lee}, \citenamefont {Han}, \citenamefont {Gao}, \citenamefont {Xie}, \citenamefont {Muller},\ and\ \citenamefont {Park}}]{kang_layer_2017}%
  \BibitemOpen
  \bibfield  {author} {\bibinfo {author} {\bibfnamefont {K.}~\bibnamefont {Kang}}, \bibinfo {author} {\bibfnamefont {K.-H.}\ \bibnamefont {Lee}}, \bibinfo {author} {\bibfnamefont {Y.}~\bibnamefont {Han}}, \bibinfo {author} {\bibfnamefont {H.}~\bibnamefont {Gao}}, \bibinfo {author} {\bibfnamefont {S.}~\bibnamefont {Xie}}, \bibinfo {author} {\bibfnamefont {D.~A.}\ \bibnamefont {Muller}}, \ and\ \bibinfo {author} {\bibfnamefont {J.}~\bibnamefont {Park}},\ }\href {\doibase 10.1038/nature23905} {\bibfield  {journal} {\bibinfo  {journal} {Nature}\ }\textbf {\bibinfo {volume} {550}},\ \bibinfo {pages} {229} (\bibinfo {year} {2017})}\BibitemShut {NoStop}%
\end{thebibliography}%


\phantom{xxxx}

\clearpage

\onecolumngrid

\setcounter{figure}{0}
\setcounter{section}{0}
\makeatletter 
\renewcommand{\thefigure}{S\@arabic\c@figure}
\makeatother

\section{Supplemental Materials for: Designer three-dimensional electronic bands in asymmetric transition metal dichalcogenide heterostructures}

\section{Identifying component TMD flakes for heterostructure construction}

\begin{figure}[h]
	\centering
	\includegraphics[width=0.7\textwidth]{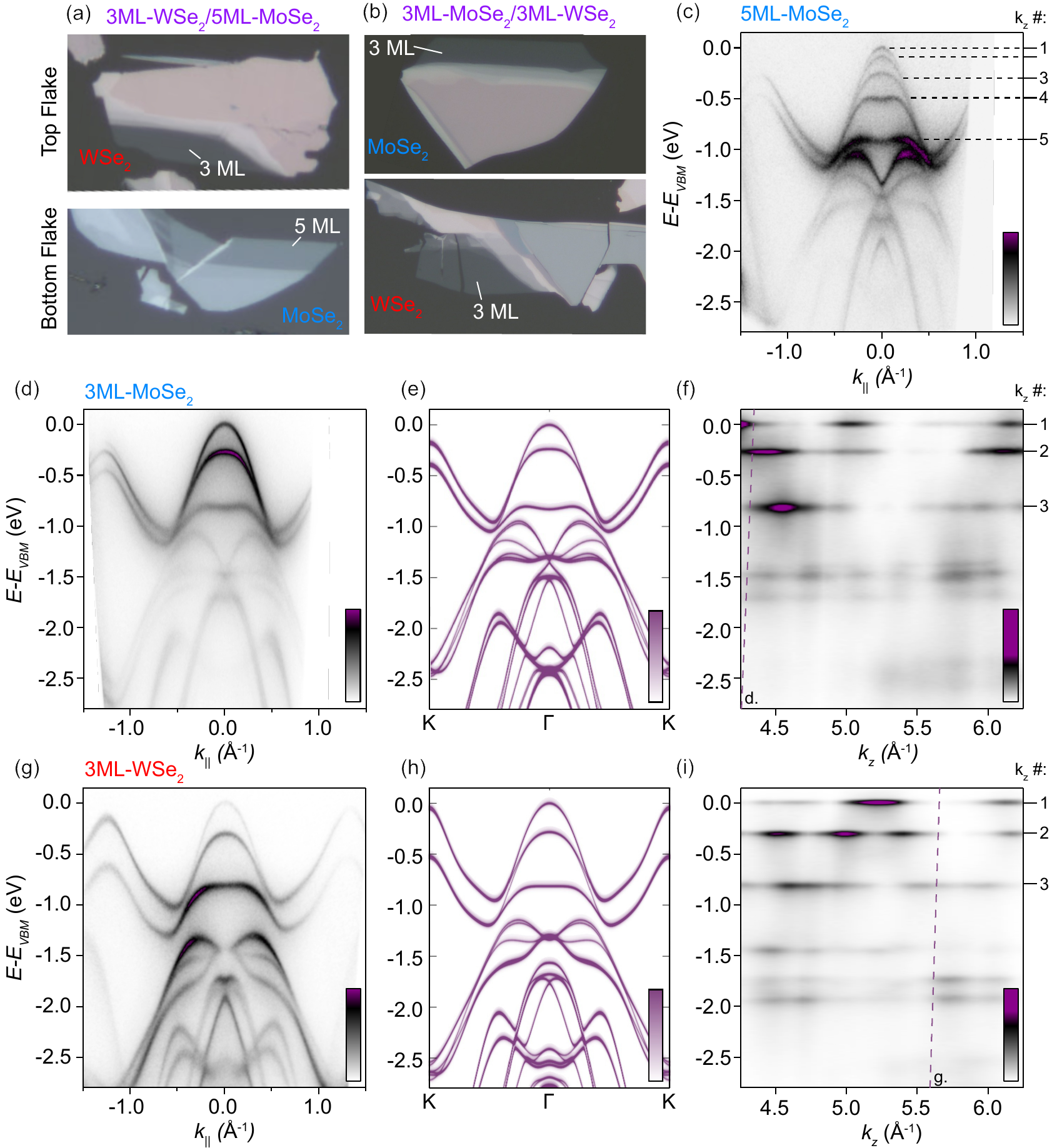}
	\caption{\label{FigS1} (a) Optical images for the two TMD flakes that make up the 8 layer 3ML-WSe$_2$/5ML-MoSe$_2$ heterostructure characterised in this work. The thin regions are indicated with white arrows. (b) Equivalent images for the components of the six-layer 3ML-MoSe$_2$/3ML-WSe$_2$ heterostructure. (c) Band dispersion close to the $\Gamma$-K axis corresponding the 5ML MoSe$_2$ bottom layer of the eight layer heterostructure. $k_z$ subbands for the near-Fermi level band structure at $\Gamma$ are indicated (h$\nu$=81~eV). (d) Near $\Gamma$-K band dispersion extracted from the 3ML MoSe$_2$ top-layer of the six layer heterostructure. (e) Corresponding layer-projected DFT calculations. (f) Corresponding photon energy-dependence ($k_z$ dispersion) for at $k_{\parallel}=0$. The position of the band dispersion in (d) is indicated. (d-i) Equivalent plots but for the 3ML WSe$_2$ top-layer of the eight layer heterostructure. $k_z$ sub-bands and positioning of the band dispersion are again indicated. For the $k_z$ dispersions in (f) and (i), an inner potential of 13~eV was used to match that determined for bulk WSe$_2$~\cite{riley_direct_2014}.} 
\end{figure}

In Sup. Fig.~1(a-b), optical images of the component flakes of (WSe$_2$)$_3$/(MoSe$_2$)$_5$ (a) and (MoSe$_2$)$_3$/(WSe$_2$)$_3$ are presented. These flakes, exfoliated onto PDMS using a standard scotch tape method, have thicknesses determined through comparison to photoluminescence spectra and optical images of reference samples. 

Heterostructures assembled from these flakes are constructed in such a way that regions of individual components remain exposed laterally (see e.g. Fig. 2(a) of the main text), allowing for photoemission spectra to be acquired directly under identical experimental conditions to those of the heterostructures. This procedure  allows for a direct comparison of band structures between component flakes and stacked heterostructures. 

In 2H-TMDs, there is a one-to-one correspondence between the number of distinct quantized $k_z$-subbands of the $\Gamma$-centered $d_{z^2}$ orbital manifolds and the total thickness in monolayers~\cite{splendiani_emerging_2010}. This provides direct \textit{in situ} verification of flake thicknesses through the enumeration of these doubly-degenerate bands in ARPES spectra. In Sup. Fig.~1(c, d, g) band dispersions close to the $\Gamma$-K direction are displayed for 5ML-MoSe$_2$, 3ML-MoSe$_2$ and 3ML-WSe$_2$, with 5, 3 and 3 quantized states, respectively, indicated explicitly with dashed lines in Sup. Fig. 1(c), (f) and (i), respectively. 

The 3ML-MoSe$_2$ and 3ML-WSe$_2$ spectra are compared to density functional theory calculations (see Methods) along the $\Gamma$-K direction. Agreement is very strong across the energy-momentum range shown. 

Finally, in Sup. Fig. 1 (f,i), photon energy dependent spectra are acquired for $k_{\parallel}=0$ for both 3ML-MoSe$_2$ (f) and 3ML-WSe$_2$ (i). The dashed line indicates the $k_z$ values where the corresponding in-plane spectra are taken in (d) and (g). 

The $k_z$ dependent datasets shown here and in Fig.~2 of the main text are acquired by continuously varying the photon energy. The relationship between the photon energy, $h\nu$ and $k_z$ is determined using a free-electron final state approximation~\cite{Dam2004}:

\begin{align}
k_z= \sqrt{\frac{2m_e}{\hbar^2}(V_0+E_k  (\cos^2(\theta) )}, \nonumber
\end{align}

where the inner potential, $V_0$, is taken to be either 13 or 16~eV for all $k_z$ spectra in this work, indicated in the respective captions. These values are chosen to match the value found for photon energy dependent studies on bulk 2H-TMD systems (e.g.~\cite{riley_direct_2014}), or to optimize the match to tight-binding calculations, where appropriate. 

The varying spectral weight between the individually non-dispersive $k_z$ sub-bands (labeled in (f) and (i)) approximates the bulk $k_z$ dispersion of the infinite layer system~\cite{lefevre_two_2024}. With increasing layer number, these quantized states begin to approximate a continuous $k_z$ dispersion, and the in-plane band dispersions of the highly three-dimensional bands become significantly $k_z$-broadened, thereby appearing diffuse, reflecting the small electron mean free path perpendicular to the surface~\cite{Dam2004}.

\clearpage

\section{Lack of rotational disorder in asymmetric heterostructures}

\begin{figure*}[h]
	\centering
	\includegraphics[width=\textwidth]{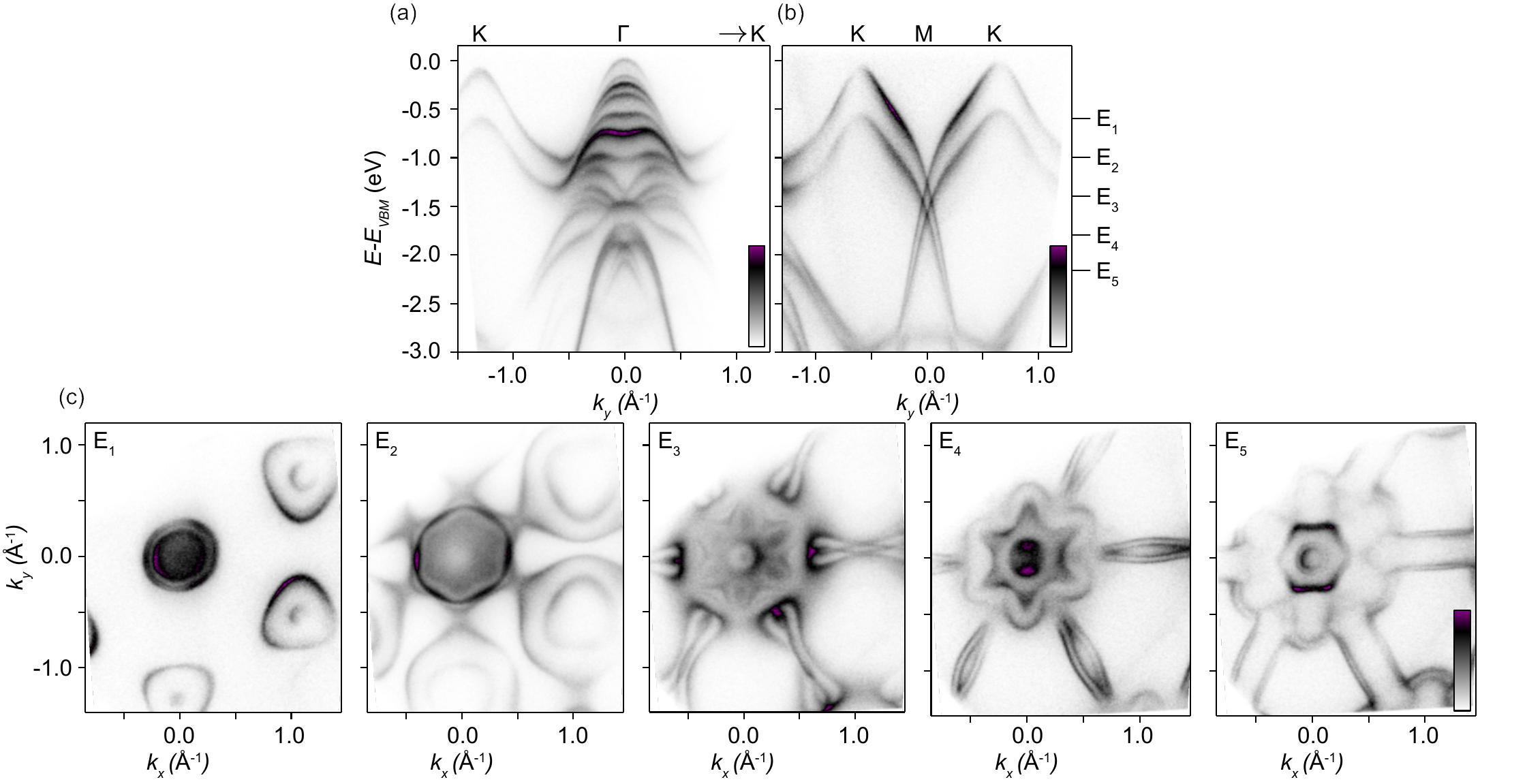}
	\caption{ (a-b) K-$\Gamma$-K and K-M-K band dispersions as extracted from a three-dimensional ($k_x$, $k_y$, $E-E_{\text{F}}$) dataset of the eight-layer 3ML-WSe$_2$/5ML-MoSe$_2$ heterostructure. (c) Constant-energy $k_x$-$k_y$ contours at the energies indicated in (b). 81 eV photons were used for all panels } 
\end{figure*}

In Supplemental Figure 2, constant energy $k_x-k_y$ contours and extracted high symmetry $\Gamma$-K and K-M-K dispersions are shown for the (WSe$_2$)$_3$/(MoSe$_2$)$_5$ heterostructure. Despite the large relative twist angle in both heterostructures between the WSe$_2$ and MoSe$_2$ flakes (see Sup. Fig. 3 for methodology), there is no evidence for rotational disorder. 
Note in particular the absence of a second set of the spin-orbit split local $d_{xy}$ orbital-derived VBM from the K points. The observed splitting at K  is determined by the local crystal structure from the top WSe$_2$ layer, in line with the $d_{xy}$-orbital character and the discussions surrounding Figure 1 of the main text. There are no band features consistent with that of the lower MoSe$_2$ flake. This precludes the scenario that the observed eight-layer spectrum in Sup. Fig.~2(a) and in Figs. 2 and 5 of the main text derive simply from a superposition of spectra from weakly-interacting top and bottom flakes. Indeed, the electron mean free path is on the order of 5.8 ~\AA~ for the 81~eV photons used here,  comparable to the thickness of only a single 1H unit  of a 2H-TMD. Instead all eight visible sub-bands are present in the experimentally accessible top layers. Analagous arguements can be made for other heterostructures in this work.
We note that the quality of the photoemission data throughout our work reflect the high quality of our heterostructure devices and exfoliation methods, though we note the presence of subtle image warping due to instrumentation limitations, explaining both the differing size of the K point valleys and a slight shadowing of the bands at high $k_x$ in Sup. Fig. 2(c).

\clearpage

\section{Twist angle determination between top and bottom flakes}

\begin{figure*}[h]
	\centering
	\includegraphics[width=\textwidth]{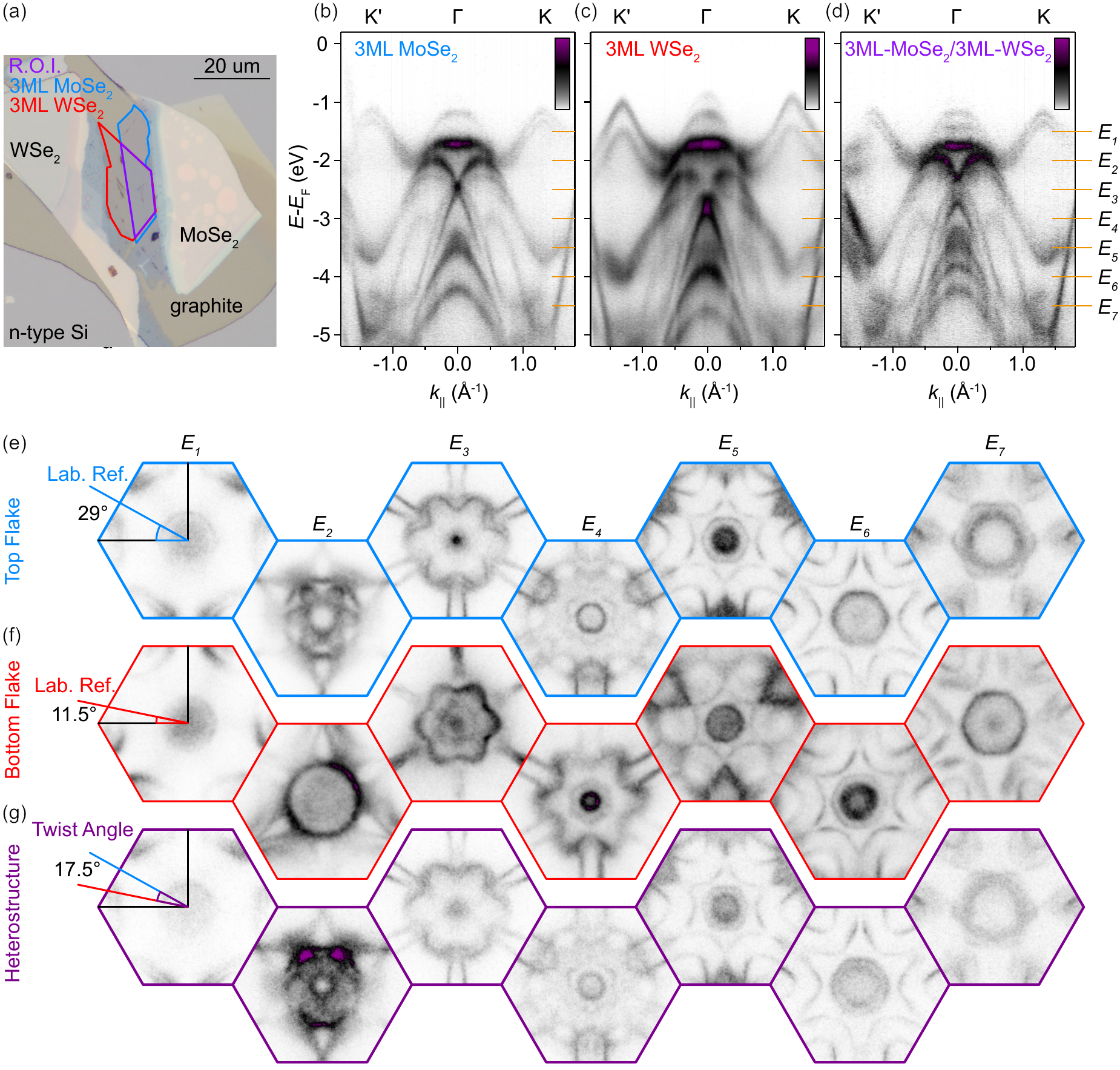}
	\caption{ Three-dimensional ($k_x$, $k_y$, $E-E_{\text{F}}$) datasets acquired from the sample with the six layer (MoSe$_2$)$_3$/(WSe$_2$)$_3$ heterostructure with a nano-ESCA (h$\nu=$21.2~eV) at room temperature. (a) Optical image reproduced from Fig.~2 of the main text. The sample regions with isolated top and bottom flakes are indicated in blue and red, respectively. The overlap region (six layer heterostructure) is indicated in purple.
    (b-d) $\Gamma$-K band dispersions as extracted from the blue (b), red (c) and purple (d) regions. (e-f) Constant energy contours at the binding energies indicated in (b-d) for all three sample regions. A complete Brillouin zone is shown in each case. Angles in the left most panels of (e-g) indicate the angle of rotation of the raw data images (laboratory reference) to display high-symmetry band dispersions. As the sample was not rotated between acquisitions, a twist angle of 17.5 degrees can be determined between the components of the six layer heterostructure. }
\end{figure*}

Supplemental Figure 3 shows three band dispersions from three different regions of the sample housing the (MoSe$_2$)$_3$/(WSe$_2$)$_3$ heterostructure discussed in the main text. Band dispersions and constant energy contours are displayed for three regions selected by moving the sample laterally without rotation: (MoSe$_2$)$_3$ (b,e), (WSe$_2$)$_3$ (c,f) and (MoSe$_2$)$_3$/(WSe$_2$)$_3$(d, g). Data are collected using a nano-ESCA instrument at Flinders University, South Australia, with 21.2~eV photons from a Helium plasma discharge lamp. 

Despite the much lower photon energy and elevated measurement temperature (300~K), these spectra are entirely consistent with those presented in the main text. Note also the continued lack of rotational disorder (as discussed for the eight layer heterostructure in Sup. Fig. 2). Indicated in Sup. Fig. 3 (e-f) are the in-plane azimuthal rotations applied to the raw data to achieve the high symmetry alignment in the Figure. The twist angle can be determined simply by calculating the difference between the azimuthal angle for the top and bottom TMD flakes. The twist angle for this sample is  large at 17.5 degrees, and therefore band hybridisation effects associated with the Moir\'e potential should not be expected in this system. The twist angle for the eight layer WSe$_2$-MoSe$_2$ heterostructure is similarly large.

\clearpage

\section{Fully delocalised bands in stacked combinations of two or three TMD bilayers }

\begin{figure*}[h]
	\centering
	\includegraphics[width=\textwidth]{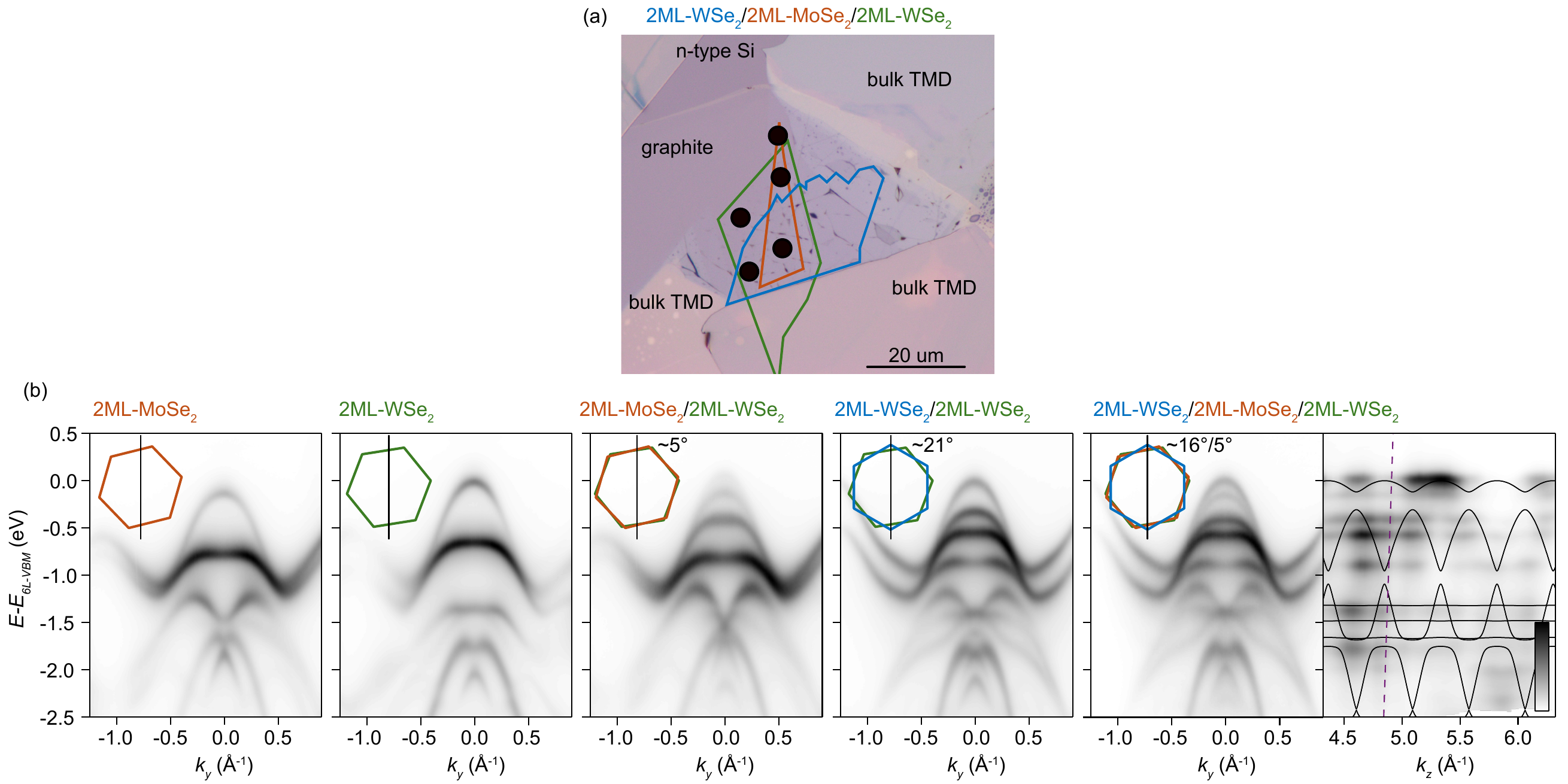}
	\caption{(a) Optical image of a (WSe$_2$)$_2$/(MoSe$_2$)$_2$/(WSe$_2$)$_2$ heterostructure. The three constituent bilayer TMDs are placed with lateral offsets, and therefore regions of pairs and singlets of the three bilayer TMD systems can be found. Black dots indicate the approximate positions from which the ARPES data in (b) were extracted. Each individual bilayer region is indicated with blue (top) red (middle) and green (bottom) outlines. (b) ARPES dispersions for the five distinct sample regions indicated in (a). The insets show how the BZ of the bilayer flake(s) are orientated relative to the $k_y$-axis (black vertical line) of the datasets. The orientations of each flake (and therefore the relative twist angles (indicated inset)) were determined by constant-energy $k_x$-$k_y$ mapping on each bilayer flake. The sample was not rotated between acquisitions of datasets from these five areas, and all energy axes are referenced to the valence band maxima of the stacked six-layer region. 81~eV photons were used for all in-plane datasets. For the six-layer stacked region, a $k_z$ dispersion is also shown, as constructed from photon energy dependent ARPES. Dashed purple line indicates the $k_z$ point accessed with 81~eV photons. An inner potential of 16~eV was used to match the equivalent datasets in Figs.~2 and 4 of the main text. Overlaid solid black lines are a tight binding model for an ordered (MoSe$_2$/WSe$_2$)$_n$ bulk system.}
\end{figure*}

In the main text, we demonstrate how out-of-plane electronic band dispersions form when combining pairs of 3-5 monolayer flakes of MoSe$_2$ and WSe$_2$, despite the fact that the interface results in an asymmetry along the $c-$axis. These out-of-plane band dispersions are discretised due to the finite size of the hybrid system, but crucially reflect its full spatial extent, demonstrating how the $d_{z^2}$-derived bands at $\Gamma$ are fully delocalised across the system. 

In Supplemental Figure 4, the same physics is observed when combining two or three  TMD bilayers, despite large relative twist angles between the three flakes. By comparing the band dispersions from individual 2ML-TMD flakes, two four-monolayer thick heterostructures formed from overlap of two bilayers, and a six-monolayer thick heterostructure formed from the overlap of all three bilayers, it can be seen that the number of $k_z$ sub-bands with the $d_{z^2}$-orbital manifold reflects the total layer number of the heterostructures. This observed delocalisation occurs despite differences in in-plane orientations of the three bilayers (shown inset in Sup. Fig. 4b) and the interfaces between different compounds. 

Although the bands across this sample are broader than those from samples in the main text, full band delocalization can be clearly observed via the increasing numbers of $k_z$ sub-bands. These observations are therefore entirely consistent with those presented in the main text.

\clearpage

\section{Orbital-dependent $c-$axis localisation}

In Supplemental Figures 5 and 6, full layer-resolved DFT calculations for the K-$\Gamma$-K axis is shown for the (MoSe$_2$)$_3$/(WSe$_2$)$_3$ (Sup. Fig. 5) and the (WSe$_2$)$_3$/(MoSe$_2$)$_5$ (Sup. Fig. 6) heterostructures. Consistent with the discussions in the main text and with reference to Fig.~3, bands deriving entirely from out-of-plane orbitals are unchanged through the heterostructure (though intensities do vary), whereas bands deriving  entirely from in-plane orbitals are strongly layer-dependent. 

The topological surface states (TSSs) discussed in the main text and shown in Fig.~5 are visible here only in the surface layers and not close to the interface between the two compound types. This demonstrates how the TSSs are a product of the $k_z$ dispersion of the full heterostructure, and do not arise from one or both of the components in isolation. 

\begin{figure*}[h]
	\centering
	\includegraphics[width=0.7\textwidth]{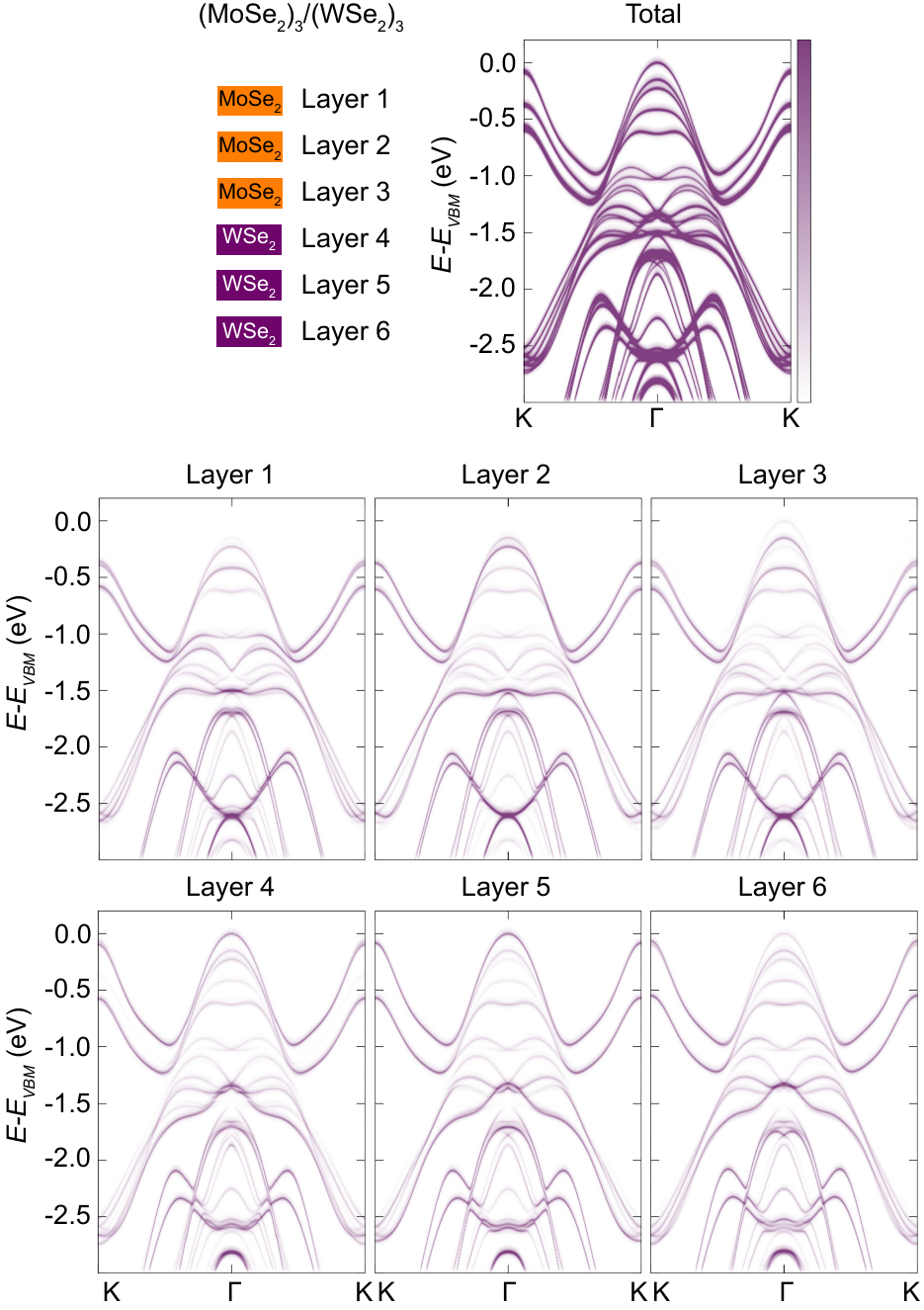}
	\caption{ Layer-integrated and layer-resolved density functional calculations for the (MoSe$_2$)$_3$/(WSe$_2$)$_3$ heterostructure}
\end{figure*}

\begin{figure*}[h]
	\centering
	\includegraphics[width=0.9\textwidth]{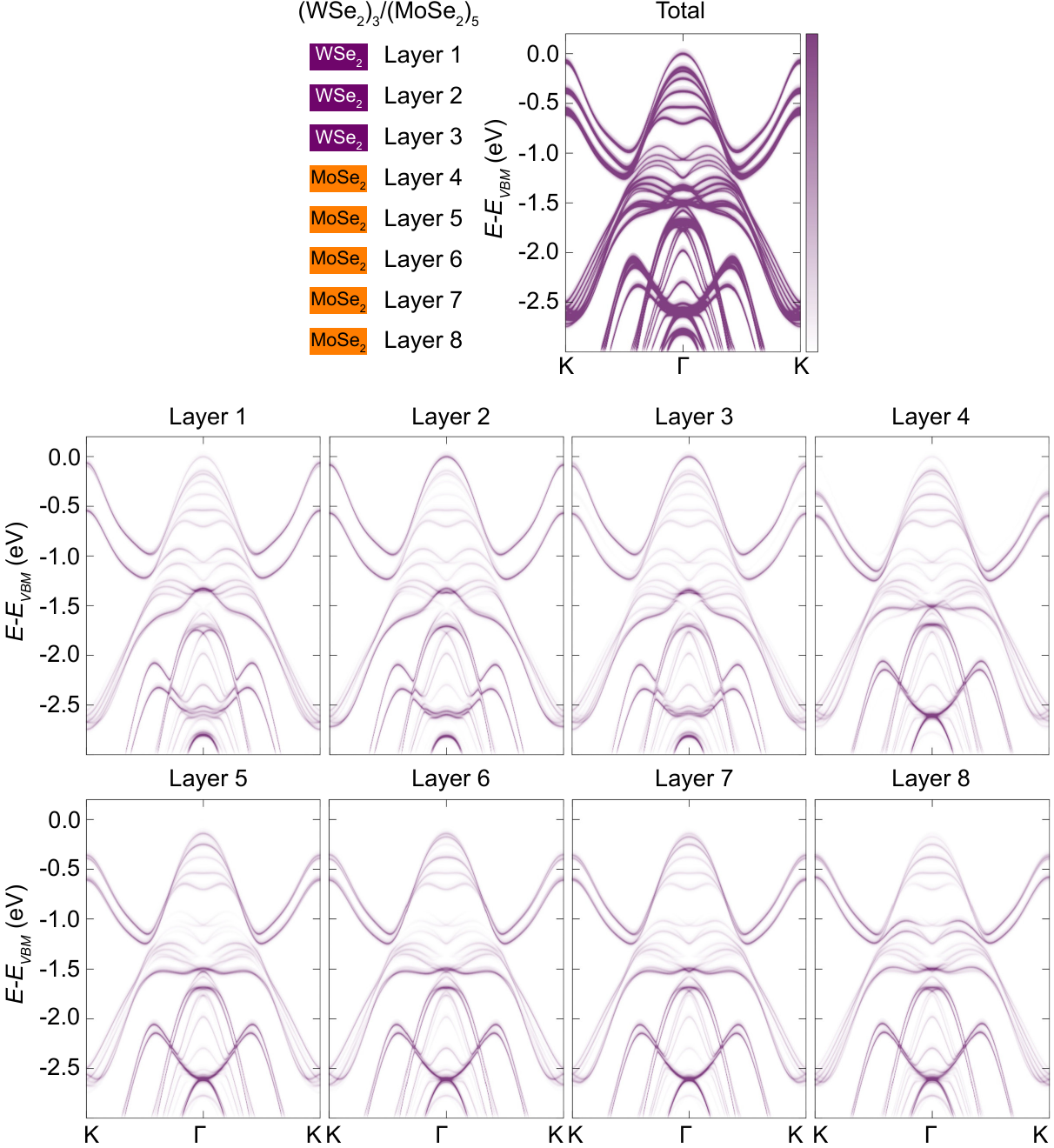}
	\caption{ Layer-integrated and layer-resolved density functional calculations for the (WSe$_2$)$_3$/(MoSe$_2$)$_5$ heterostructure}
\end{figure*}

\clearpage

\section{Supplementary datasets for (WSe$_2$)$_4$/(NbSe$_2$)$_x$}

\begin{figure*}[h]
	\centering
	\includegraphics[width=\textwidth]{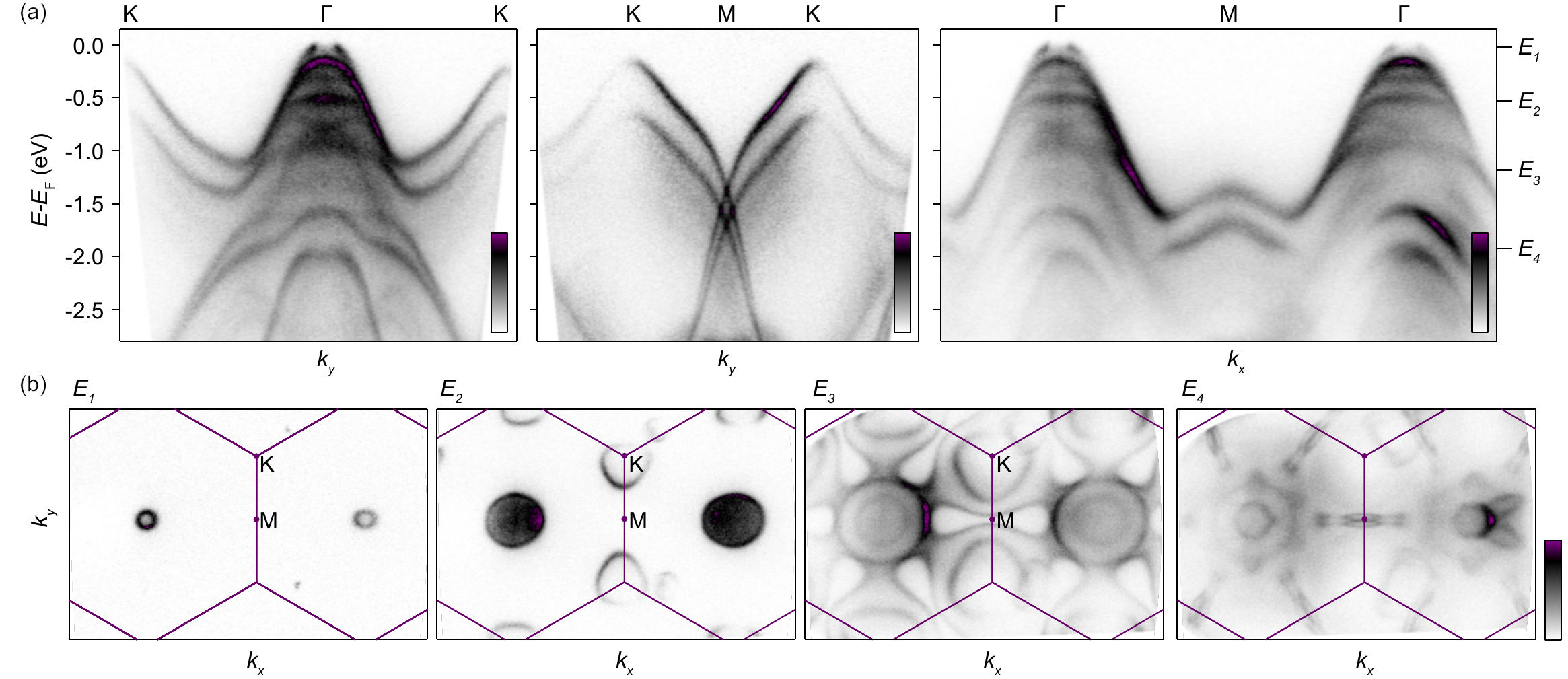}
	\caption{ (a) K-$\Gamma$-K and K-M-K and $\Gamma$-M-$\Gamma$ band dispersions as extracted from a three-dimensional ($k_x$, $k_y$, $E-E_{\text{F}}$) dataset of the six-layer 4ML-WSe$_2$/2ML-NbSe$_2$ heterostructure. (c) Constant-energy $k_x$-$k_y$ contours at the energies indicated in (b). The Brillouin zone is overlaid in purple, and high symmetry points labeled.  91.5 eV photons were used for all panels. } 
\end{figure*}

Sup. Fig. 7 shows constant energy contour $k_x$-$k_y$ mapping and extracted high symmetry dispersions for the (WSe$_2$)$_4$/(NbSe$_2$)$_2$ region of the sample shown in Fig. 4(a) of the main text. 

These data demonstrate how the Fermi pocket created by interfacing the WSe$_2$ and NbSe$_2$ flakes is present in both the first and second Brillouin zones. The constant energy contours also exhibit no rotational disorder and no K valleys consistent with isolated NbSe$_2$, again demonstrating that all observed bands are present within the top layers.

\clearpage

\begin{figure*}[h]
	\centering
	\includegraphics[width=0.7\textwidth]{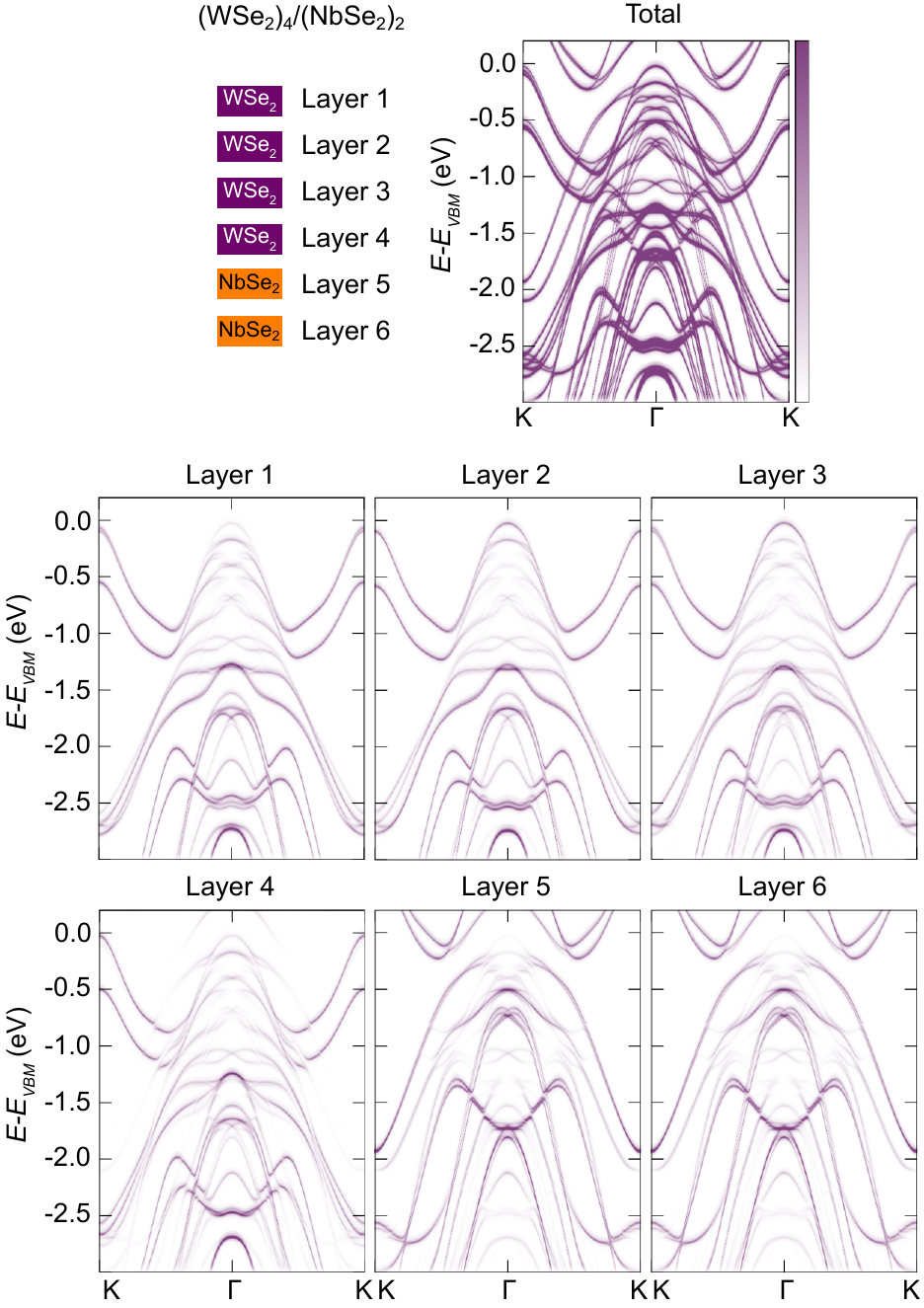}
	\caption{ Layer-integrated and layer-resolved density functional calculations for the (WSe$_2$)$_4$/(NbSe$_2$)$_2$ heterostructure}
\end{figure*}

In Supplemental Figure 8, the full layer-resolved DFT calculations for the K-$\Gamma$-K axis is shown for the (WSe$_2$)$_4$/(NbSe$_2$)$_2$  heterostructure discussed in the main text. As with the previous examples, positions of bands deriving entirely from out-of-plane orbitals are unchanged through the heterostructure (though intensities do vary), whereas bands deriving  entirely from in-plane orbitals are strongly layer-dependent.

\clearpage

\begin{figure*}[h]
	\centering
	\includegraphics[width=0.9\textwidth]{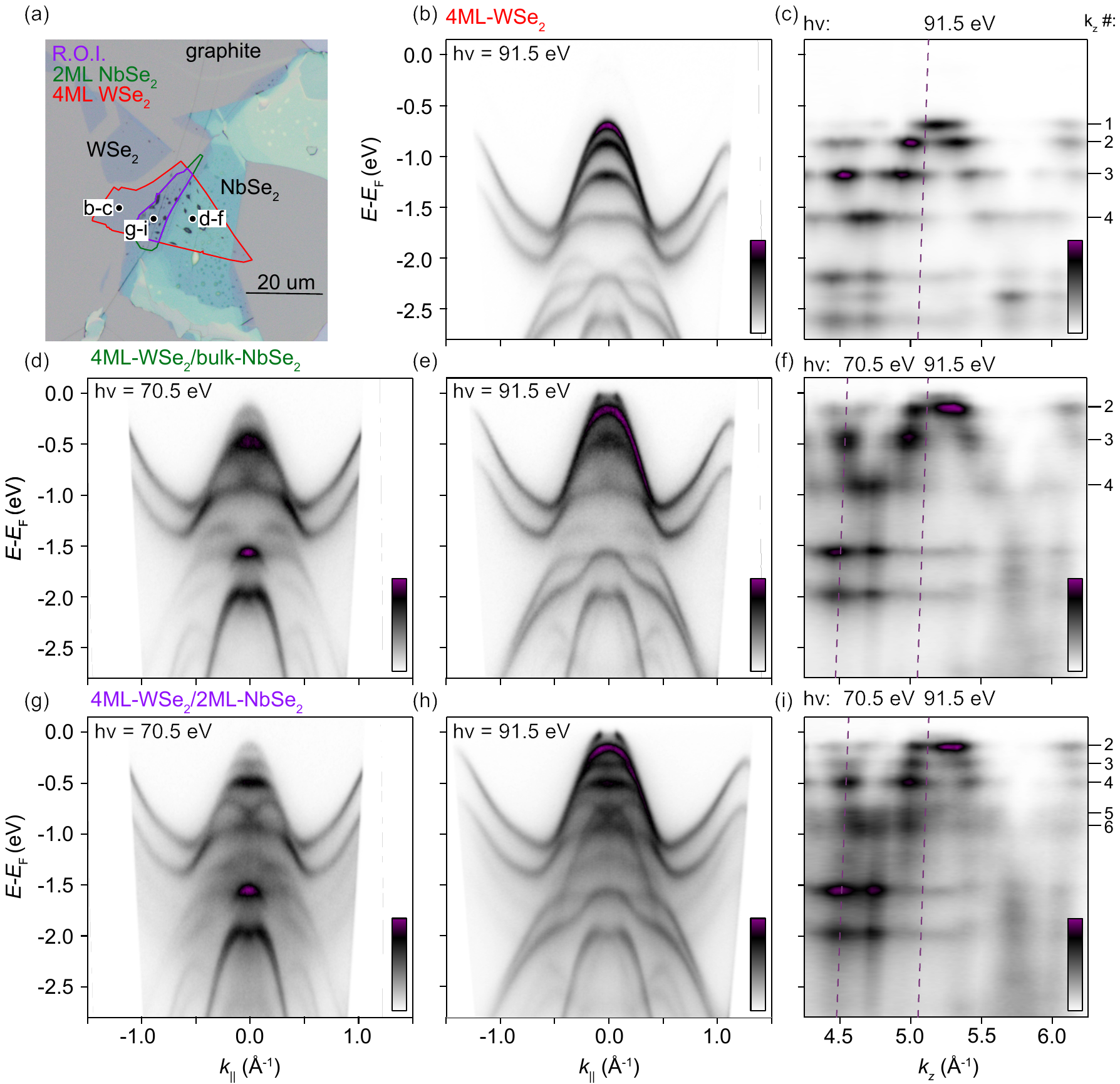}
	\caption{ (a) Optical images for the sample containing the (WSe$_2$)$_4$/(NbSe$_2$)$_2$ heterostructure. All component flakes are indicated. Black dots with white box labels indicate approximate positions from which the ARPES spectra in panels b-i derive. (b) Band dispersion close to $\Gamma$-K as extracted from the 4ML WSe$_2$ top layer of the six layer heterostructure (h$\nu$=91.5~eV). (c) Photon energy dependent dataset of this same sample region for $k_{\parallel}=0$ and h$\nu$=60-144~eV. Dashed lines indicate the photon energy slice from which the data in (b) derive. $k_z$ subbands for the near-Fermi level band structure at $\Gamma$ are indicated. (d-e) near $\Gamma$-K band dispersions as extracted from the region of the sample where 4ML-WSe$_2$ overlays the bulk NbSe$_2$ crystal (h$\nu$=70.5 and 91.5~eV for (d) and (e) respectively). (f) Corresponding photon energy dependent scan over an identical range to that shown in (c). $k_z$ sub-bands and positioning of the band dispersions are again indicated.  (g-i) Equivalent datasets for the (WSe$_2$)$_4$/(NbSe$_2$)$_2$ region of interest. The $k_z$ dispersions in (c, f, i) share an inner potential of 13~eV (matching bulk WSe$_2$~\cite{riley_direct_2014}) to aid comparison.}
\end{figure*}

Supplemental Figure 9 compares the three dimensional electronic structure of (WSe$_2$)$_4$/(NbSe$_2$)$_2$ as shown in Figure 4 of the main text, to equivalent datasets from (WSe$_2$)$_4$ and (WSe$_2$)$_4$/(NbSe$_2$)$_x$, $x>>2$. All spectra are referenced to the Fermi level. 

The in-plane band dispersion (b, close to $\Gamma$-K) and $k_z$ dispersion (c, taken at $k_{\parallel}$=0) for 4ML-WSe$_2$ is consistent with those found for 3ML-WSe$_2$ in Supplemental Fig. 1, with one extra $k_z$ subband in the $d_{z^2}$ manifold, consistent with the increased layer number. These spectra are compared to the case where the same 4ML-WSe$_2$ flake overlaps with a bulk NbSe$_2$ crystal. The number of $k_z$-subbands is unchanged due to the bulk-nature of the bottom component, consistent with the arguments in the main text. Despite this, there are two notable changes relative to free-standing 4ML-WSe$_2$. Firstly, there are additional band features, most notably the conical bands appearing between $k_z$-subbands 3 and 4. In Sup. Fig. 9d, an additional spectra taken with 70.5~eV photons is displayed to increase the spectral weight of this conical feature. The second major change is the shift of all bands to shallower binding energies, with the topmost $k_z$-subband forming a sole Fermi pocket. While tangential to the findings concerning intermediate-regime asymmetric heterostructures, this result suggests yet another potential approach towards band structure engineering of few-layer TMD flakes. With a careful choice of substrate, the fermiology and band structures can be significantly modified. 

Sup. Fig. 9(g-i) again considers the (WSe$_2$)$_4$/(NbSe$_2$)$_2$ case discussed in the main text. An additional band dispersion using 70.5eV photons (g) is displayed for comparison to the bulk case  (d), and shows how a similar conical feature is formed.  

We note that, unlike for the WSe$_2$-MoSe$_2$ heterostructures, one can see that there is a superposition of the stacked region with the top 4ML-WSe$_2$ flake here. This has two possible origins: the probing light spot is at the boundary of the overlap (purple) and WSe$_2$ (red) areas, or the less-than-pristine overlap region, as seen in Sup. Fig. 9(a), precludes full and complete charge transfer between the flakes. 

Regardless of the origin, the significant charge transfer in the overlap region ensures that the $d_{z^2}$-derived bands remain separated from those originating from the top layer alone, thus facilitating an unambiguous enumeration of bands belong to the stacked heterostructure. We also note that the presence of only one extra copy of the WSe$_2$ K point valleys shows that the charge transfer in this region is binary, and thus the observed sub-bands cannot be explained by a series of superimposed band structures with slightly different chemical potentials.



\end{document}